\documentclass[longauth]{aa}

\usepackage{graphicx}

\usepackage{txfonts}
\usepackage[colorlinks=true,citecolor=blue]{hyperref}
\usepackage{lscape}
                   
\usepackage{placeins}
                     
\begin{document}

   \title{TOI-283\,b: A transiting mini-Neptune in a 17.6-day orbit discovered with TESS and ESPRESSO}

   \author{F. Murgas
          \inst{\ref{iac}, \ref{ull}}
          \and
          E. Pall\'e\inst{\ref{iac}, \ref{ull}}
          \and
          A. Su\'{a}rez Mascare\~{n}o\inst{\ref{iac}, \ref{ull}}
          \and
          J. Korth\inst{\ref{ugen}, \ref{lund}}
          \and
          F.\,J.~Pozuelos\inst{\ref{iaa}}
          \and
          M.~J. Hobson\inst{\ref{ugen}}
          \and
          B. Lavie\inst{\ref{ugen}}
          \and
          C. Lovis\inst{\ref{ugen}}
          \and
          S.~G. Sousa\inst{\ref{iaporto}}
          \and
          D. Bossini\inst{\ref{iaporto}}
          \and
          H. Parviainen\inst{\ref{iac}, \ref{ull}}
          \and 
          A.~Castro-Gonz\'{a}lez\inst{\ref{cabcsic}}
          \and
          V. Adibekyan\inst{\ref{iaporto}, \ref{uporto}}
          \and
          C. Allende Prieto\inst{\ref{iac}, \ref{ull}}
          \and
          Y. Alibert\inst{\ref{ubern}}
          \and
          F. Bouchy\inst{\ref{ugen}}
          \and
          C. Brice\~{n}o\inst{\ref{noirlab}}
          \and
          D.~A. Caldwell\inst{\ref{seti}}
          \and
          D. Ciardi\inst{\ref{nasacaltech}}
          \and
          C. Clark\inst{\ref{nasacaltech}}
          \and
          K.~A. Collins\inst{\ref{cfa}}
          \and
          K.~I. Collins\inst{\ref{gmu}}
          \and
          S. Cristiani\inst{\ref{inaftrieste}}
          \and
          X. Dumusque\inst{\ref{ugen}}
          \and
          D. Ehrenreich\inst{\ref{ugen}}
          \and
          P. Figueira\inst{\ref{ugen}, \ref{iaporto}}
          \and
          E. Furlan\inst{\ref{nasacaltech}}
          \and
          R. G\'{e}nova Santos\inst{\ref{iac}, \ref{ull}}
          \and
          C. Gnilka\inst{\ref{nasaames}}
          \and
          J.~I. Gonz\'{a}lez Hern\'{a}ndez\inst{\ref{iac}, \ref{ull}}
          \and
          Z. Hartman\inst{\ref{gemini}}
          \and
          S.~B. Howell\inst{\ref{nasaames}}
          \and
          J.~M. Jenkins\inst{\ref{nasaames}}
          \and
          N. Law\inst{\ref{uncarolina}}
          \and
          C. Littlefield\inst{\ref{bay}, \ref{nasaames}}
          \and
          G. Lo Curto\inst{\ref{esochile}}
          \and
          A.~W. Mann\inst{\ref{uncarolina}}
          \and
          C.~J.~A.~P. Martins\inst{\ref{iaporto}, \ref{caup}}
          \and
          A. Mehner\inst{\ref{esochile}}
          \and
          G. Micela\inst{\ref{inafpalermo}}
          \and
          P. Molaro\inst{\ref{inaftrieste}}
          \and
          N.~J. Nunes\inst{\ref{ialisboa}}
          \and
          F. Pepe\inst{\ref{ugen}}
          \and
          R. Rebolo\inst{\ref{iac}, \ref{ull}, \ref{csic}}
          \and
          H.~M. Relles\inst{\ref{cfa}}
          \and
          N.~C. Santos\inst{\ref{iaporto}, \ref{uporto}}
          \and
          N.~J. Scott\inst{\ref{nasaames}}
          \and
          S.~Seager\inst{\ref{mit_phys}, \ref{mit_earth}, \ref{mit_aero}}
          \and
          A. Sozzetti\inst{\ref{inaftorino}}
          \and
          S. Udry\inst{\ref{ugen}}
          \and
          C.~N. Watkins\inst{\ref{cfa}}
          \and
          J.~N. Winn\inst{\ref{princeton}}
          \and
          M.~R. Zapatero Osorio\inst{\ref{cabcsic}}
          \and
          C. Ziegler\inst{\ref{austin}}
          }

   \institute{
   Instituto de Astrof\'isica de Canarias (IAC), E-38205 La Laguna, Tenerife, Spain \label{iac} \\
              \email{fmurgas@iac.es} 
   \and
   Departamento de Astrof\'isica, Universidad de La Laguna (ULL), E-38206 La Laguna, Tenerife, Spain \label{ull}
   \and
   Observatoire Astronomique de l’Universit\'{e} de Gen\`{e}ve, Chemin Pegasi 51b, 1290 Versoix, Switzerland\label{ugen}
   \and
   Lund Observatory, Division of Astrophysics, Department of Physics, Lund University, Box 118, 22100 Lund, Sweden\label{lund}
   \and
   Instituto de Astrof\'isica de Andaluc\'ia (IAA-CSIC), Glorieta de la Astronom\'ia s/n, 18008 Granada, Spain\label{iaa}
   \and
   Instituto de Astrof\'{\i}sica e Ci\^encias do Espa\c{c}o, Universidade do Porto, CAUP, Rua das Estrelas, 4150-762 Porto, Portugal\label{iaporto}
   \and
   Departamento de F\'{i}sica e Astronomia, Faculdade de Ci\^encias, Universidade do Porto, Rua do Campo Alegre, 4169-007 Porto, Portugal\label{uporto}
   \and
   Centro de Astrobiolog\'{i}a, CSIC-INTA, Camino Bajo del Castillo s/n, 28692 Villanueva de la Ca\~{n}ada, Madrid, Spain\label{cabcsic}
   \and
   Physics Institute of University of Bern, Gesellschafts strasse 6, 3012, Bern, Switzerland\label{ubern}   
   \and
   SOAR Telescope/NSF NOIRLab, Casilla 603, La Serena, Chile\label{noirlab}
   \and
   SETI Institute, Mountain View, CA 94043 USA/NASA Ames Research Center, Moffett Field, CA 94035 USA\label{seti}
   \and
   NASA Exoplanet Science Institute, Caltech/IPAC, Mail Code 100-22, 1200 E. California Blvd., Pasadena, CA 91125, USA\label{nasacaltech}
   \and
   Center for Astrophysics \textbar \ Harvard \& Smithsonian, 60 Garden Street, Cambridge, MA 02138, USA\label{cfa}
   \and
   George Mason University, 4400 University Drive, Fairfax, VA, 22030 USA\label{gmu}
   \and
   INAF – Osservatorio Astronomico di Trieste, Via Tiepolo 11, 34143 Trieste, Italy\label{inaftrieste}
   \and
   NASA Ames Research Center, Moffett Field, CA 94035 USA\label{nasaames}
   \and
   Gemini Observatory/NSF's NOIRLab, 670 A’ohoku Place, Hilo, HI 96720, USA\label{gemini}
   \and
   Department of Physics and Astronomy, The University of North Carolina at Chapel Hill, Chapel Hill, NC 27599-3255, USA\label{uncarolina}
   \and
   Bay Area Environmental Research Institute, Moffett Field, CA 94035, USA\label{bay}
   \and
   European Southern Observatory, Av. Alonso de Cordova, 3107, Vitacura, Santiago de Chile, Chile\label{esochile}
   \and
   Centro de Astrof\'{\i}sica da Universidade do Porto, Rua das Estrelas, 4150-762 Porto, Portugal\label{caup}
   \and
   INAF – Osservatorio Astronomico di Palermo, Piazza del Parlamento 1, 90134 Palermo, Italy\label{inafpalermo}
   \and
   Instituto de Astrof\'{i}sica e Ci\^encias do Espa\c{c}o, Faculdade de Ci\^encias da Universidade de Lisboa, Campo Grande, 1749-016, Lisboa, Portugal\label{ialisboa}
   \and
   Consejo Superior de Investigaciones Cient\'{i}ficas (CSIC), 28006 Madrid, Spain\label{csic}
   \and
   Department of Physics and Kavli Institute for Astrophysics and Space Research, Massachusetts Institute of Technology, Cambridge, MA 02139, USA\label{mit_phys}
   \and
   Department of Earth, Atmospheric and Planetary Sciences, Massachusetts Institute of Technology, Cambridge, MA 02139, USA\label{mit_earth}
   \and
   Department of Aeronautics and Astronautics, MIT, 77 Massachusetts Avenue, Cambridge, MA 02139, USA\label{mit_aero}
   \and
   INAF – Osservatorio Astrofisico di Torino, Strada Osservatorio, 20 10025 Pino Torinese (TO), Italy\label{inaftorino}
   \and
   Department of Astrophysical Sciences, Princeton University, Princeton, NJ 08544, USA\label{princeton}
   \and
   Department of Physics, Engineering and Astronomy, Stephen F. Austin State University, 1936 North St, Nacogdoches, TX 75962, USA\label{austin}
             }

   \date{Received MMMM DD, YYYY; accepted MMMM DD, YYYY}
 
  \abstract 
   {Super-Earths and mini-Neptunes are missing from our Solar System, yet they appear to be the most abundant planetary types in our Galaxy. A detailed characterization of key planets within this population is important for understanding the formation mechanisms of rocky and gas giant planets and the diversity of planetary interior structures.}
   {In 2019, NASA's TESS satellite found a transiting planet candidate in a 17.6-day orbit around the star TOI-283. We started radial velocity (RV) follow-up observations with ESPRESSO to obtain a mass measurement. Mass and radius are measurements critical for planetary classification and internal composition modeling.}
   {We used ESPRESSO spectra to derive the stellar parameters of the planet candidate host star TOI-283. We then performed a joint analysis of the photometric and RV data of this star, using Gaussian processes to model the systematic noise present in both datasets.}
   {We find that the host is a bright K-type star ($d = 82.4$ pc, $\mathrm{T}_\mathrm{eff} = 5213 \pm 70$ K, $V = 10.4$ mag) with a mass and radius of $\mathrm{M}_\star = 0.80 \pm 0.01\; \mathrm{M}_\odot$ and $\mathrm{R}_\star = 0.85 \pm 0.03\; \mathrm{R}_\odot$. The planet has an orbital period of $P = 17.617$ days, a size of $\mathrm{R}_\mathrm{p} = 2.34 \pm 0.09\; \mathrm{R}_\oplus$, and a mass of $\mathrm{M}_\mathrm{p} = 6.54 \pm 2.04\; \mathrm{M}_\oplus$. With an equilibrium temperature of $\sim$600 K and a bulk density of $\rho_\mathrm{p} = 2.81 \pm 0.93$ g cm$^{-3}$, this planet is positioned in the mass-radius diagram where planetary models predict H$_2$O- and H/He-rich envelopes. The ESPRESSO RV data also reveal a long-term trend that is probably related to the star's activity cycle. Further RV observations are required to confirm whether this signal originates from stellar activity or another planetary body in the system.}
   {}


   \maketitle

\section{Introduction}

Ongoing observational efforts, such as those from space telescopes like CoRoT \citep{Baglin2006}, Kepler \citep{Borucki2010}, and TESS \citep[Transiting Exoplanet Survey Satellite,][]{Ricker2015}, have led to the discovery of many transiting exoplanets, allowing for comprehensive statistical population analysis. One of the main results of these population studies is the fact that small exoplanets exhibit a bimodal distribution of radii, characterized by two distinct populations with peaks at $\mathrm{R}_\mathrm{p} \sim 1.3 \; \mathrm{R}_\oplus$ (super-Earths) and $\mathrm{R}_\mathrm{p} \sim 2.4 \; \mathrm{R}_\oplus$ (sub-Neptunes), and separated by a gap referred to as the radius valley \citep{Fulton17}. Various explanations for this phenomenon focus on mechanisms of atmospheric mass loss, such as photoevaporation driven by the host star \citep{Owen2017, Jin2018} or the internal heating of the planet \citep{Ginzburg2018}. Models based on photoevaporation suggest that super-Earth and sub-Neptunian planets share a rocky composition, and their different radii result from whether they retain a primordial hydrogen/helium atmosphere (H/He envelope). In contrast, if these planets had an icy internal composition, the radius valley would manifest at larger planetary radii \citep{Owen2017, Rogers21}. Another possible explanation for the origin of the radius valley is that both populations form in different regions of the disk, with super-Earths forming in drier environments (within the water-ice line of the disk) and sub-Neptunes forming beyond the water-ice line and having more water-rich compositions \citep[e.g.,][]{Venturini2020, Burn2024}. Some observational evidence suggests that, at least for M-dwarf hosts, the radius valley may be a consequence of internal composition rather than an indicator of atmospheric mass loss \citep{Luque2022}.

In terms of radii, sub-Neptunes have sizes ranging from about 1.5 to 4 $\mathrm{R}_{\oplus}$. Sub-Neptunian planets are among the most common types of planet in our Galaxy \citep{Borucki2011, Fulton2018}, and have been a focus of interest because they represent a transition in planetary characteristics \citep{Fulton2018, Christiansen2023}. The presence and composition of the atmosphere play a crucial role in determining the overall properties of sub-Neptune planets. Some may have thick atmospheres, possibly composed of hydrogen and helium, similar to gas giants. However, their bulk composition may be a mixture of rocks and volatiles such as water, methane, and ammonia \citep{Zeng2016, Zeng2019}. Nevertheless, sub-Neptunes can be found at various distances from their host stars, and their orbital characteristics vary widely. There is a limited sample of sub-Neptunes that have relatively long periods ($P > 15$ days) and low stellar irradiances ($S < 30 \; S_\oplus $) and are situated around stars bright enough to allow for atmospheric studies with current instrumentation ($V < 12 $ mag). According to the PlanetS catalog\footnote{Available at: \url{https://dace.unige.ch}} \citep{Otegi2020, Parc2024}, there are only 16 planets that fulfill these characteristics; hence, any addition to this population will have an impact on the study of long-period sub-Neptunes.

The TESS mission is conducting a comprehensive survey of the entire sky to detect transiting planets. Since its launch in mid-2018, TESS has released over 7000 candidate planets, known as TESS objects of interest (TOIs)\footnote{\url{https://tess.mit.edu/toi-releases/}}. The vast majority of these TOIs require ground-based follow-up observations to confirm their planetary nature, in particular radial velocity (RV) measurements to determine planetary masses. Exploiting the synergies between TESS and ESPRESSO \citep{Pepe2021}, an echelle spectrograph mounted on the VLT in Chile has already led to the discovery and characterization of several small transiting planets around different types of stars \citep{Sozzetti2021, Palle2021, Lillo2021, Demangeon2021, Bourrier2022, Barros2022, Lavie2023, Damasso2023}.

Here we use ESPRESSO observations to determine the mass of a transiting candidate around TOI-283, which was part of the guaranteed time observations (GTO) sample designed to cover the parameter space of small planets in and around the radius valley. The planet candidate was validated using a statistical approach in \citet{Giacalone2021}, in which TOI-283\,b was found to have a false positive probability (FPP) of $\mathrm{FFP} = 0.05$ and a nearby false positive probability (NFPP) of $\mathrm{NFPP} = 1.76 \times 10^{-3}$. According to the \citet{Giacalone2021} validation criteria, this puts this candidate on the edge of their “likely planet” category ($\mathrm{FFP} < 0.5$ and $\mathrm{NFPP} < 10^{-3}$). In addition, \cite{Lester2021} obtained high-angular-resolution imaging observations of 517 host stars of TESS exoplanet candidates using the 'Alopeke and Zorro speckle cameras at Gemini North and South. For TOI-283, they found the star to be single and obtained $5\sigma$ $\Delta$mag contrast limits at 0.2$\arcsec$ and 1.0$\arcsec$ in the blue filter (562 nm) of 4.61 and 4.97 magnitudes, respectively. In the red filter (832 nm) the values were 5.52 and 7.51 magnitudes. These ground-based validations triggered the start of ESPRESSO observations for mass determination.

The paper is organized as follows. In Section \ref{Sec:Obs_Space} and Section \ref{Sec:Obs_Ground} we describe the space- and ground-based observations used in this study. Section \ref{Sec:StellarParam} describes our method of obtaining the stellar parameters of TOI-283 from high-resolution spectra. In Section \ref{Sec:Analysis_and_results} we present our analysis and results. Finally, in Section \ref{Sec:Discussion} we present a discussion of the planet, and in Section \ref{Sec:Conclusions} we present our conclusions.

\section{Space-based observations}
\label{Sec:Obs_Space}
\subsection{TESS photometry}
The planet candidate around TOI-283 was discovered by TESS. The star has a TESS Input Catalog \citep{Stassun2018} number of TIC~382626661. Due to its near polar declination ($\delta \sim -65^{\circ}$), TESS observes this target almost continuously for nearly 6 months when the satellite is collecting data from the southern hemisphere. 

The analysis and processing of TESS data is performed by the TESS Science Processing Operations Center \citep[SPOC,][]{Jenkins2016} at the NASA Ames Research Center. The TESS simple aperture photometry \citep[SAP,][]{Morris2020} light curves are analyzed with a reduction process that begins with the removal of systematic effects using the Presearch Data Conditioning (PDC) pipeline module \citep{Smith2012,Stumpe2012,Stumpe2014}, then a search for transit-like signals is performed using a wavelet-based adaptive, noise-compensating matched filter \citep{Jenkins2002,Jenkins2010,JenkinsJM2020}. Finally a transit model that includes the effect of limb darkening \citep{Li2019} is fit to the data. To rule out some false positive scenarios, some validation tests are applied to the time series \citep{Twicken2018}, including the difference image centroid test, which constrains the location of the host star to within $1.7 \pm 2.7 \arcsec$ of the transit source in the analysis of sectors 1 - 69. This result constrains any background source to 1/3 of a pixel at 3$\sigma$ and complements high-resolution imaging, since the difference image centroid test is sensitive to sources within the stamp, which typically subtends $3.85\arcmin$ on each side ($11 \times 11$ pixels). The TESS Science Office (TSO) at the Massachusetts Institute of Technology (MIT) evaluates the target reports, and if the candidate passes the validation tests, it is assigned a TOI number. The TSO announced the detection of a transiting planet candidate around TIC~382626661 on 7 May 2019, and assigned the TOI number TOI 283.01. The detected transit event has a period of $P \sim 17.6$ days and a depth of 722 parts per million (ppm).

To study the transit events of the planet candidate orbiting TOI-283, we used the TESS PDCSAP photometry, covering a total of 36 sectors. Table ~\ref{Tab:TESS_Sectors} summarizes the 2-minute cadence TESS data analyzed in this work. The TESS observations span a baseline of 2449 days, with the flux standard deviation per sector ranging from 810 to 926 ppm. Figure ~\ref{Fig:AllTESS_LightCurves} presents the TESS light curves for each sector.

\begin{table}
\caption{TOI-283 TESS observations with 2-minute cadence used in this work.}
\label{Tab:TESS_Sectors}
\centering
\begin{tabular}{c c c l}
\hline\hline
Sector & Camera & CCD & Observation date \\
\hline
1 & 4 & 3 & 2018-Jul-25 -- 2018-Aug-22 \\
2 & 4 & 3 & 2018-Aug-23 -- 2018-Sep-20 \\
3 & 4 & 4 & 2018-Sep-20 -- 2018-Oct-17 \\
4 & 4 & 4 & 2018-Oct-19 -- 2018-Nov-14 \\
5 & 4 & 4 & 2018-Nov-15 -- 2018-Dec-11 \\
6 & 4 & 4 & 2018-Dec-15 -- 2019-Jan-06 \\
7 & 4 & 1 & 2019-Jan-08 -- 2019-Feb-01 \\
8 & 4 & 1 & 2019-Feb-02 -- 2019-Feb-27 \\
9 & 4 & 1 & 2019-Feb-28 -- 2019-Mar-25 \\
10 & 4 & 2 & 2019-Mar-26 -- 2019-Apr-22 \\
11 & 4 & 2 & 2019-Apr-23 -- 2019-May-20 \\
12 & 4 & 2 & 2019-May-21 -- 2019-Jun-18 \\
13 & 4 & 3 & 2019-Jun-19 -- 2019-Jul-17 \\
27 & 4 & 3 & 2020-Jul-05 -- 2020-Jul-30 \\
28 & 4 & 3 & 2020-Jul-31 -- 2020-Aug-25 \\
29 & 4 & 3 & 2020-Aug-26 -- 2020-Sep-21 \\
30 & 4 & 4 & 2020-Sep-23 -- 2020-Oct-20 \\
31 & 4 & 4 & 2020-Oct-22 -- 2020-Nov-16 \\
32 & 4 & 4 & 2020-Nov-20 -- 2020-Dec-16 \\
33 & 4 & 1 & 2020-Dec-18 -- 2021-Jan-13 \\
34 & 4 & 1 & 2021-Jan-14 -- 2021-Feb-08 \\
35 & 4 & 1 & 2021-Feb-09 -- 2021-Mar-06 \\
36 & 4 & 1 & 2021-Mar-07 -- 2021-Apr-01 \\
37 & 4 & 2 & 2021-Apr-02 -- 2021-Apr-28 \\
38 & 4 & 2 & 2021-Apr-29 -- 2021-May-26 \\
39 & 4 & 2 & 2021-May-27 -- 2021-Jun-24 \\
61 & 4 & 1 & 2023-Jan-18 -- 2023-Feb-12 \\
62 & 4 & 1 & 2023-Feb-12 -- 2023-Mar-10 \\
64 & 3 & 3 & 2023-Apr-06 -- 2023-May-03 \\
65 & 4 & 2 & 2023-May-04 -- 2023-Jun-01 \\
68 & 4 & 3 & 2023-Jul-29 -- 2023-Aug-25 \\
69 & 4 & 3 & 2023-Aug-25 -- 2023-Sep-20 \\
87 & 4 & 1 & 2024-Dec-18 -- 2025-Jan-14 \\
88 & 4 & 1 & 2025-Jan-14 -- 2025-Feb-11 \\
89 & 4 & 1 & 2025-Feb-11 -- 2025-Mar-12 \\
90 & 3 & 3 & 2025-Mar-12 -- 2025-Apr-09 \\
\hline
\end{tabular}
\end{table}

We note that the TESS pixel scale is $\sim$21$\arcsec$, and photometric apertures typically extend to about 1$\arcmin$, which generally results in multiple stars blending into the TESS photometric aperture. Therefore, we used the TESS contamination tool \texttt{TESS-cont}\footnote{Available at \url{https://github.com/castro-gzlz/TESS-cont}} \citep{CastroGonzalez2024} to quantify the flux fraction within the SPOC aperture coming from TOI-283 and nearby sources, and to evaluate whether the observed transit signal could have originated from one of the contaminant stars. We constructed the point response functions (PRFs)\footnote{The PRFs of each Sector, Camera, and CCD were accessed at \url{https://archive.stsci.edu/missions/tess/models/prf_fitsfiles/start_s0004/}} of the 224 nearby \textit{Gaia} DR2 sources \citep{GaiaDR22018} around TOI-283 in a radius of 200$\arcsec$, and the total flux contributions across the TPFs were calculated. In Figure~\ref{Fig:TESS-cont} we summarize the \texttt{TESS-cont} output for Sector~1. We find that 99.5$\,\%$ of the flux comes from TOI-283, indicating a very low degree of contamination. As a sanity check, we also ran \texttt{TESS-cont} based on the DR3 catalog \citep{GaiaDR3} and reached the same conclusion. The five most contaminating stars are TIC 382627415 (Star~1), TIC 382626679 (Star~2), TIC 382626664 (Star~3), TIC 382626686 (Star~4), and TIC 382627418 (Star~5), all of which (except Star 3) are outside the SPOC aperture (see Figure~\ref{Fig:TESS-cont}). We used the \texttt{TESS-cont} \texttt{DILUTION} function to calculate the necessary eclipse depths to generate the observed 722 ppm transit feature, and found eclipses of 71\,\% (Star 1), 102\,\% (Star 2), 123\,\% (Star 3), 126\,\% (Star 4), and 140\,\% (Star 5). Therefore, we conclude that only Star~1 (TIC 382627415) could have produced the observed transit signal, although it would have to present extremely deep ($\sim$70\,\%) eclipses.

Figure~\ref{Fig:TESS_SAP_Periodogram} presents the periodogram analysis of the SAP photometry, aimed at constraining the stellar rotation period. A more detailed discussion is provided in Sect.~\ref{Sec:stellarrot_activ}.

\subsection{Gaia assessment}
We used \textit{Gaia} DR3 \citep{GaiaDR3} to identify potential wide companions associated with TOI-283. Typically, these stars are already cataloged in the TESS Input Catalog, and their influence on the derived transit parameters has been taken into account in the TESS light curves. We searched the \textit{Gaia} DR3 catalog for common proper motion companions of TOI-283 within a radius of 1$^{\circ}$ and applied a parallax constraint, selecting stars with parallaxes within 10\,\% of that of TOI-283. The query returned only TOI-283, indicating that it does not share motion and distance with any other star within one degree, according to the magnitude depth probed by \textit{Gaia}.

\textit{Gaia}'s astrometric data provide additional insights into the possible presence of nearby stellar companions that may have gone undetected by both \textit{Gaia} and high-resolution imaging techniques. The \textit{Gaia} renormalized unit weight error (RUWE) acts as an indicator similar to a reduced $\chi^2$, with values of around $\lesssim 1.4$ indicating that the astrometric solution is consistent with that of a single star. On the other hand, RUWE values of $\gtrsim 1.4$ indicate an excess of astrometric noise, which may suggest the existence of an undetected companion \citep[e.g.,][]{Ziegler2020}. TOI-283, with a RUWE of $1.1$, supports this single-star interpretation, suggesting that there are no hidden companions contributing additional astrometric noise.

\section{Ground-based observations}
\label{Sec:Obs_Ground}
\subsection{LCO photometric follow-up}

To attempt to determine the true source of the TESS detection, we acquired ground-based time-series follow-up photometry of the field around TOI-283 as part of the TESS Follow-up Observing Program \citep[TFOP;][]{collins:2019}\footnote{\url{https://tess.mit.edu/followup}}. We used the {\tt TESS Transit Finder}, which is a customized version of the {\tt Tapir} software package \citep{Jensen:2013}, to schedule our transit observations. 

We observed a partial transit window of TOI-283.01 in Sloan $r'$ band on UTC 26 January 2019 from the Las Cumbres Observatory Global Telescope \citep[LCOGT;][]{Brown:2013} 1\,m network node at the Cerro Tololo Inter-American Observatory in Chile (CTIO). The 1\,m telescope is equipped with a $4096\times4096$ SINISTRO camera that has an image scale of $0.389 \arcsec$ per pixel, resulting in a $26\arcmin\times26\arcmin$ field of view and the images were calibrated by the standard LCOGT {\tt BANZAI} pipeline \citep{McCully:2018}, and differential photometric data were extracted using {\tt AstroImageJ} \citep{Collins:2017}. We used circular photometric apertures of $6.2\arcsec$ that excluded all of the flux from the nearest known neighbor in the \textit{Gaia} DR3 catalog \citep[\textit{Gaia} DR3 5275527777289351040,][]{GaiaDR3} that is $41.4\arcsec$ east of TOI-283. An on-time $\sim$775 ppm ingress was detected on-target (see Fig.~\ref{Fig:LCO_LightCurve} and \ref{Fig:LCO_RpRs_Ditri}). Additionally, we checked for possible nearby eclipsing binaries (NEBs) that could be contaminating the TESS photometric aperture and causing the TESS detection. To account for possible contamination from the wings of neighboring star PSFs, we searched for NEBs out to $2.5\arcmin$ from TOI-283. If fully blended in the SPOC aperture, a neighboring star that is fainter than the target star by 7.9 magnitudes in TESS-band could produce the SPOC-reported flux deficit at mid-transit (assuming a 100\,\% eclipse). To account for possible TESS magnitude uncertainties and possible delta-magnitude differences between TESS-band and Sloan $r'$, we included an additional threshold of 0.5 mag for fainter objects (down to TESS-band magnitude 17.5). We calculated the root mean square (RMS) of each of the 32 nearby star light curves (binned in 10 minute bins) that meet our search criteria and find that the values are smaller by at least a factor of 5 compared to the required NEB depth in each respective star. We then visually inspected each neighboring star's light curve to ensure no obvious eclipse-like signal. Our analysis ruled out an NEB blend as the cause of the SPOC pipeline TOI-283.01 detection in the TESS data. All light curve data are available on the {\tt EXOFOP-TESS} website\footnote{\href{https://exofop.ipac.caltech.edu/tess/target.php?id=382626661}{https://exofop.ipac.caltech.edu/tess/target.php?id=382626661}}. 

\subsection{High-resolution imaging}
As part of our validation process, we observed TOI-283 using high-resolution imaging to place some constraints on the presence of nearby star-bound sources and/or faint background stars not detected by the seeing limited photometric observations. These sources may introduce some unaccounted flux contamination that could affect the TESS photometry, resulting in an underestimated planetary radius, or be the source of astrophysical false positives, such as background eclipsing binaries.

\subsubsection{SOAR/HRCam}

We searched for stellar companions to TOI-283 with speckle imaging on the 4.1 m Southern Astrophysical Research (SOAR) telescope \citep{Tokovinin2018} on 18 February 2019, observing in the Cousins I-band, a similar visible bandpass to TESS. This observation was sensitive with 5-sigma detection to a 5.4-magnitude fainter star at an angular distance of 1 arcsec from the target. More details of the observations within the SOAR TESS survey are available in \cite{Ziegler2020}. The 5$\sigma$ detection sensitivity and speckle autocorrelation functions from the observations are shown in Figure~\ref{Fig:TOI283_HighResImg}. No nearby stars were detected within 3$\arcsec$ of TOI-283 in the SOAR observations.

\subsubsection{Gemini/Zorro}
We observed TOI-283 on two nights (16 March 2020 and 7 April 2023) with the Zorro speckle imager \citep{Scott2021} on the 8.1 m Gemini South telescope. This instrument contains a dichroic that diverts the input beam from the telescope to two electron-multiplying CCD imagers that operate simultaneously in different bandpasses. For our observations, we used filters centered on 562 nm and 832 nm. For the 2020 observation, 5000 frames, each with an exposure time of 60 milliseconds, were obtained in both filters, while for the 2023 observation, 6000 frames were obtained using the same filters. After observing TOI-283, we observed a nearby point source and reduced the data using techniques described by \citet{Howell2011} and \citet{Horch2011}.

\begin{figure*}
    \centering
    \includegraphics[width=\textwidth]{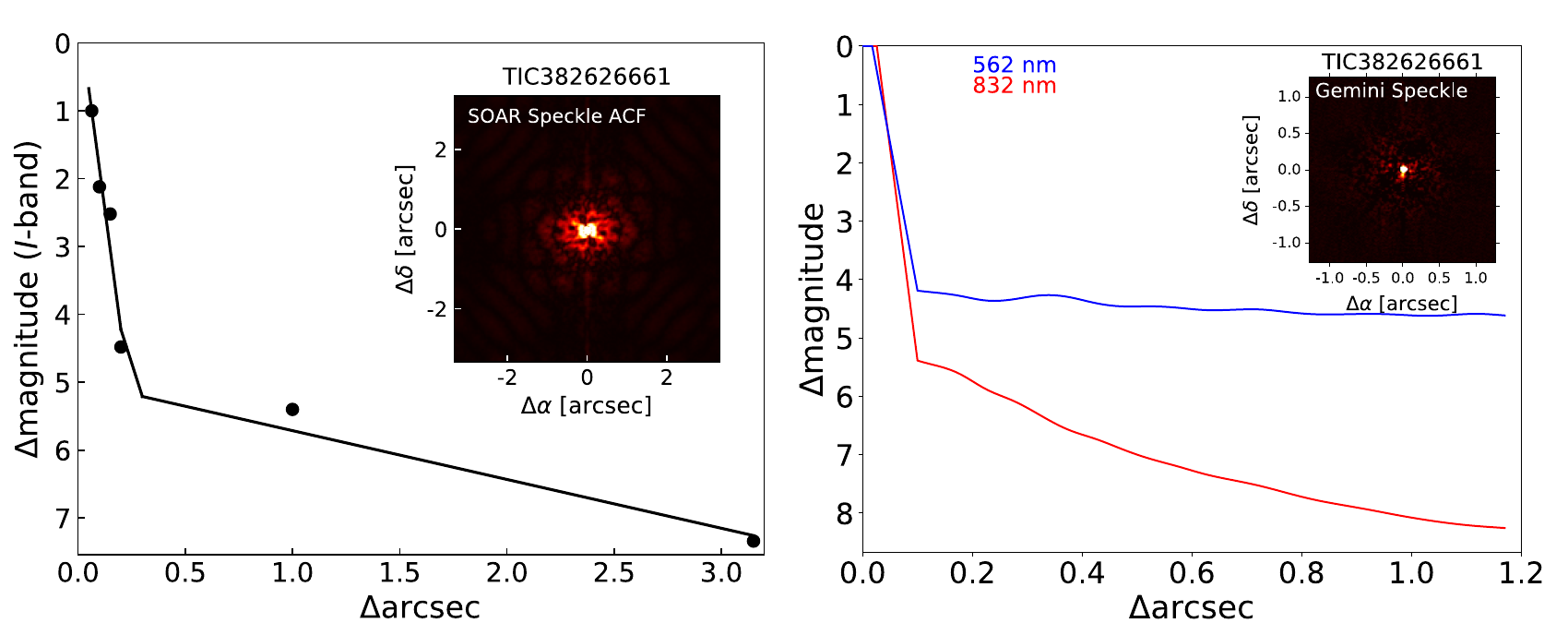}
    \caption{High-resolution images of TOI-283. \textit{Left panel}: SOAR/HRCam observation of TOI-283 taken on 18 February 2019. No nearby stars were detected within 3$\arcsec$. \textit{Inset}: Thumbnail image of TOI-283. \textit{Right panel}: Gemini/Zorro high-resolution image of TOI-283 taken on 7 April 2024. TOI-283 is a single star to a contrast limit of 7 mag at 832 nm. \textit{Inset}: $1.2\arcsec \times 1.2 \arcsec$ thumbnail image of TOI-283.}
    \label{Fig:TOI283_HighResImg}
\end{figure*}

Figure~\ref{Fig:TOI283_HighResImg} presents the results of the observations taken on 7 April 2023. No stellar companions were detected with Zorro between 1.2$\arcsec$ and 0.02$\arcsec$ (the inner limit at 562 nm). The observations have typical 5$\sigma$ limits of 7 mag at 832 nm and 4.8 mag at a separations between 0.1$\arcsec$ and 1.2$\arcsec$ (equivalent to projected separations of 8-99 AU adopting a distance $d = 83$ pc), although the detection limits become increasingly shallow inside of 0.1$\arcsec$.

\subsection{ESPRESSO high-resolution spectroscopy}
\label{Sec:ESPRESSO_HRSpec}
We obtained high-resolution spectroscopic observations of TOI-283 with the ESPRESSO spectrograph \citep{Pepe2014,Pepe2021} installed at the 8.2-m Very Large Telescope (VLT) at Paranal Observatory, Chile. During the observing campaign, we collected a total of 95 measurements, from which we selected a subset to be used in the final fit. Previous works based on ESPRESSO data have shown that instrumental effects can affect the RV measurements. These effects can be traced using telemetry data from the spectrograph (see, for example, \citealp{SuarezMascareno2023,SuarezMascareno2024}). As selection criteria, we retained only those measurements for which the atmospheric dispersion corrector (ADC) operated without technical issues and the temperature of the instrument's optical elements remained within 5$\sigma$ of the median value across the entire dataset. According to these criteria, out of a total of 95 measurements, 11 exhibited ADC-related issues, while none showed temperature-related anomalies. In this work, we analyze a total of 84 high-resolution spectra acquired between 9 February 2019 and 12 February 2022, spanning a time baseline of 1172 days. The data were reduced with the data reduction software (DRS) v.3.0.0\footnote{\url{http://eso.org/sci/software/pipelines/}}. The DRS obtains RV measurements by performing a Gaussian fit to the cross-correlation function (CCF) of the spectrum, which is generated using a binary mask derived from a stellar template \citep{Fellgett1955,Baranne1996,Pepe2000}. The RV measurements exhibit a standard deviation of 7.4 m s$^{-1}$, while the median RV uncertainty is 0.53 m s$^{-1}$ with a standard deviation of 0.11 m s$^{-1}$.

The DRS also provides several stellar activity index values calculated from the CCF, including the full width at half-maximum (FWHM) of the CCF, the bisector inverse slope (BIS) derived from the CCF, and the contrast (CONT) of the CCF \citep{Zechmeister2009}. In addition to the CCF related values, the DRS also delivers some line activity indices such as the Mount Wilson chromospheric $S$ index computed from the \ion{Ca}{II} H \& K lines \citep[$\lambda\lambda$396.8\,nm, 393.3\,nm;][]{Vaughan1978}, \ion{Na}{I} doublet ($\lambda\lambda$589.0\,nm, 589.6\,nm), and H$\alpha$ ($\lambda$656.2\,nm). Table ~\ref{Tab:ESPRESSO_Data_Table}, available at the CDS, presents the RV measurements and activity indices used in this work. Figure~\ref{Fig:RV_vs_ActivityInd} shows the RV measurements versus the activity index values, after subtracting the median from each dataset (E18 and E19) separately, along with the corresponding Pearson correlation coefficient ($r$). No strong correlation (i.e., $|r| > 0.7$) was found between the radial velocities and activity measurements.

\begin{figure*}
    \centering
    \includegraphics[width=\hsize]{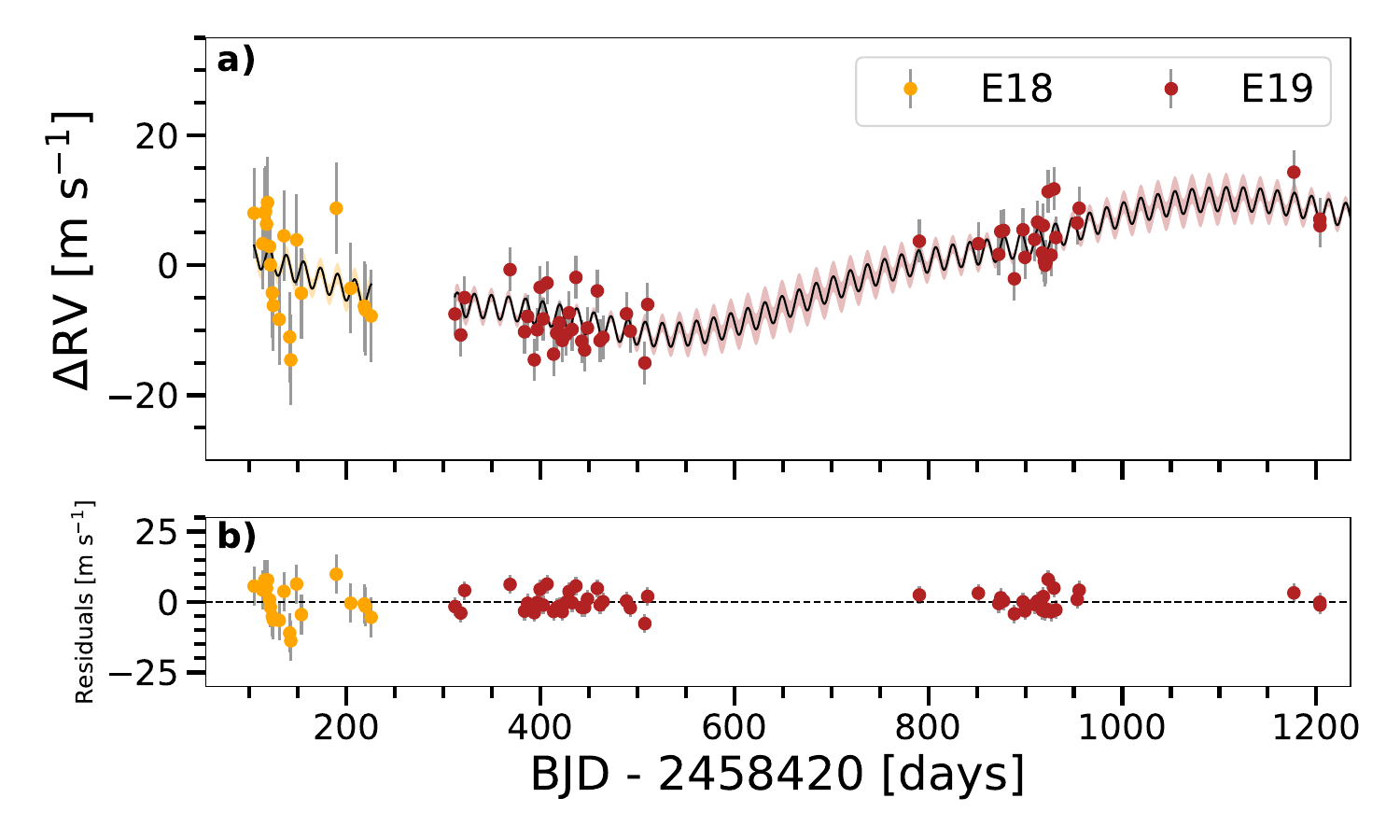}
    \caption{TOI-283\,b RV measurements taken with ESPRESSO. Panel a: RV time series and best-fitting model. The best-fitting model was computed using the median values of the posterior distribution of the fit parameters; the shaded area represents the 1$\sigma$ uncertainty limits of the best-fitting model. Panel b: Residuals of the fit after subtracting the single-planet Keplerian model. The uncertainties shown here include the RV jitter values (added in quadrature) for each set.}
    \label{Fig:TOI283b_ESPRESSO_RV_TimeSeries}
\end{figure*}

During the period in which the RVs were collected, the ESPRESSO spectrograph underwent an intervention in June-July 2019 to replace the fiber link, which improved the efficiency of the instrument by 50\,\% \citep{Pepe2021} and introduced an offset into the measurements. To account for the potential offset caused by the intervention, we considered two datasets in our modeling: the observations before the intervention (E18) and the data after the intervention (E19) (see Section \ref{Sec:JointFit}). Figure~\ref{Fig:TOI283b_ESPRESSO_RV_TimeSeries} shows the observations of the ESPRESSO RV time series.

\section{Stellar properties}
\label{Sec:StellarParam}
The general stellar properties of TOI-283 are presented in Table~\ref{Tab:Star}. The stellar atmospheric parameters ($T_{\mathrm{eff}}$, $\log g$, microturbulence, [Fe/H]) were derived using ARES+MOOG, following the same methodology described in \citet[][]{Santos-13, Sousa-14, Sousa-21}.
The ARES code \footnote{The latest version, ARES v2, can be downloaded from \url{https://github.com/sousasag/ARES}} \citep{Sousa-07, Sousa-15} was used to consistently measure the equivalent widths (EWs) of selected iron lines based on the linelist presented in \citet[][]{Sousa-08}. This was done on a combined ESPRESSO spectrum of TOI-283. We used a minimization process to find the ionization and excitation equilibrium and converge to the best set of spectroscopic parameters. This process makes use of a grid of Kurucz model atmospheres \citep{Kurucz-93} and the radiative transfer code MOOG \citep{Sneden-73}.
In the end, we found the converged parameters that are listed in Table~\ref{Tab:Star}. The trigonometric surface gravity was also derived using \textit{Gaia} DR3 data following the same methodology as described in \citet[][]{Sousa-21}.

The age, mass, and radius were inferred using the Bayesian code PARAM\footnote{\url{http://stev.oapd.inaf.it/cgi-bin/param}} \citep{PARAM,PARAM2,PARAM3}. The code follows a grid-based approach, whereby a well-sampled grid of stellar evolutionary tracks is matched to the observed quantities $T_{\rm eff}$, [Fe/H], and luminosity. Although the first two inputs were taken from this study, the luminosity was computed by converting the bolometric magnitude. Specifically, it was calculated from the observed magnitudes $B$ and $V$ \citep{TYCHO}, $G$ \citep{GAIAEDR3}, $J$, $H$, and $K_s$ \citep{2MASS}, corrected by the corresponding bolometric correction and for the distance. All bolometric corrections were estimated using the online tool YBC\footnote{\url{http://stev.oapd.inaf.it/YBC/index.html}} \citep[PARSEC Bolometric Correction;][]{Chen2019} with the input quantities $T_\mathrm{eff}$, [Fe/H], and $\log{g}$). The distance is instead calculated from the inverse of the parallax (\textit{Gaia} EDR3, \citealt{GAIAEDR3}). The final luminosity ($L=0.456\pm 0.016$ L$_\odot$) was calculated as the weighted average of all the luminosities derived from the photometric bands considered. The grid of stellar evolutionary tracks is from MESA \citep{Paxton2011,Paxton2013,Paxton2015,Paxton2018,Paxton2019}, computed with the physics described in \citet{Moedas22}, grid D1, and with a helium enrichment law of $\Delta Y/\Delta Z=1.23$ (from solar calibration) and an extended mass range of $M=[0.7-1.8]$ M$_\odot$. The final age, mass, and radius are listed in Table~\ref{Tab:Star}. 

Using the activity indices described in Sect. \ref{Sec:ESPRESSO_HRSpec}, we measured a median $S$ index of $0.191 \pm 0.045$; this translates to a $\log_{10}(\mathrm{R'}_{\mathrm{HK}})$ of $-4.92 \pm 0.23$ using the calibration of \citet[][]{Noyes1984}\footnote{\url{https://pyastronomy.readthedocs.io/en/latest/pyaslDoc/aslDoc/sindex.html}}. The rotation period of the star was derived from the time series of the activity indices; the methodology is described in Sect. \ref{Sec:stellarrot_activ}.

\begin{table*}
\centering
\small
\caption{Stellar parameters of TOI-283.} \label{Tab:Star}
\begin{tabular}{lcr}
\hline\hline
\noalign{\smallskip}
Parameter                               & Value                 & Reference \\ 
\hline
\noalign{\smallskip}
\multicolumn{3}{c}{Identifiers}\\
\noalign{\smallskip}
TYC                             & 8919-520-1                    &  {\citet{Hog2000}}.    \\
TOI                             & 283                           & {\it TESS} Alerts      \\  
TIC                             & 382626661                     & {\citet{Stassun2018}}  \\
{\it Gaia} DR3                  & 5275527747228841344           & {\citet{GaiaDR3}}      \\
2MASS                           & J07541669-6526294             & {\citet{2MASS}}        \\  
\noalign{\smallskip}
\multicolumn{3}{c}{Coordinates}\\
\noalign{\smallskip}
$\alpha$ (ICRS, J2000)                        & 07:54:16.71          & {\it Gaia} DR3  \\
$\delta$ (ICRS, J2000)                        &-65:26:29.58          & {\it Gaia} DR3  \\

\noalign{\smallskip}
\multicolumn{3}{c}{Photometry}\\
\noalign{\smallskip}

$B$ [mag]                               & $11.180 \pm 0.010$      & UCAC4      \\
$G_{BP}$ [mag]                          & $10.618 \pm 0.003$      & {\it Gaia} DR3 \\
$V$ [mag]                               & $10.420 \pm 0.120$      & UCAC4      \\
$G$ [mag]                               & $10.199 \pm 0.003$      & {\it Gaia} DR3 \\
$G_{RP}$ [mag]                          & $9.621 \pm 0.004$      & {\it Gaia} DR3 \\
$J$ [mag]                               & $9.001 \pm 0.027$       & 2MASS      \\
$H$ [mag]                               & $8.554 \pm 0.040$       & 2MASS      \\
$K_s$ [mag]                             & $8.517 \pm 0.026$       & 2MASS      \\
$W1$ [mag]                              & $8.423 \pm 0.022$       & AllWISE    \\
$W2$ [mag]                              & $8.483 \pm 0.020$       & AllWISE    \\
$W3$ [mag]                              & $8.456 \pm 0.021$       & AllWISE    \\
$W4$ [mag]                              & $8.480 \pm 0.185$       & AllWISE     \\
\noalign{\smallskip}
\multicolumn{3}{c}{Parallax and kinematics}\\
\noalign{\smallskip}
$\pi$ [mas]                             & $12.12 \pm  0.01$       & {\it Gaia} DR3             \\
$d$ [pc]                                & $82.43^{+0.06}_{-0.07}$ & {\citet{Bailer-Jones2021}}             \\
$\mu_{\alpha}\cos\delta$ [$\mathrm{mas\,yr^{-1}}$] & $-30.84 \pm 0.02$ & {\it Gaia} DR3          \\
$\mu_{\delta}$ [$\mathrm{mas\,yr^{-1}}$]            & $16.02 \pm 0.02$ & {\it Gaia} DR3          \\
$\gamma$ [$\mathrm{km\,s^{-1}}]$        & $-12.64 \pm 0.17$  & {\it Gaia} DR3  \\

\noalign{\smallskip}
\multicolumn{3}{c}{Photospheric parameters}\\
\noalign{\smallskip}
$T_{\mathrm{eff}}$ [K]                      & $5213 \pm 70$         & {This work}   \\
$\log g$                                    & $4.50 \pm 0.03$       & {This work}   \\
$\rm log\,g (cgs)^a$  	                    &  4.42 $\pm$ 0.12      & {This work}   \\
$\rm log\,g_t (cgs)^b$                      &  4.50 $\pm$ 0.03      & {This work}   \\
{[Fe/H]}                                    & $-0.09 \pm 0.05$      & {This work}   \\
$\rm \xi_{\mathrm{t}}$ [km s$^{-1}$]               &  $0.71 \pm 0.06$      & {This work}   \\
$\log_{10}(\mathrm{R'}_{\mathrm{HK}})$      &  $-4.92 \pm 0.23$     & {This work} \\
\noalign{\smallskip}
\multicolumn{3}{c}{Physical parameters}\\
\noalign{\smallskip}
$\mathrm{M}_\star$ [$M_{\odot}$]                       & $0.80 \pm 0.01$     & {This work}       \\
$\mathrm{R}_\star$ [$R_{\odot}$]                       & $0.85 \pm 0.03$     & {This work}       \\
$\rho_\star$ [g cm$^{-3}$]                    & $1.84 \pm 0.20$     & {This work}       \\
$L$ [$L_\odot$]                               & $0.46\pm 0.02$    & {This work}       \\
$P_{\rm rot}$ [days]                          & $30.0^{+5.8}_{-4.1}$ & {This work} \\
Age [Gyr]                                     & 10.4 $\pm$ 3.3 &  {This work} \\
\noalign{\smallskip}
\hline
\end{tabular}
\tablebib{
    {\it Gaia} DR3: \citet{GaiaDR3};
    UCAC: \citet{UCAC4}
    2MASS: \citet{2MASS}.
    AllWISE: \citet{CutriWISECat2014}
. $\rm^a$ From spectral analysis; $\rm^b$ trigonometric surface gravity using \textit{Gaia} DR3 data.\\
}
\end{table*}

\section{Analysis and results}
\label{Sec:Analysis_and_results}
\subsection{TESS light curve analysis}
\label{Sec:tess_tls}
To obtain initial estimates of the period and epoch of the transit event, we conducted an independent analysis of 32 of the 36 available TESS sectors. This subset was selected in order to reduce computational costs, as processing the full dataset is resource-intensive. We began by removing long-term stellar or instrumental trends from the PDCSAP light curves using a \texttt{Python} implementation of the Savitzky-Golay filter \citep{SavitzkyGolay1964}. For each TESS sector, the light curve was normalized by dividing it by a smoothed version of itself, computed with a 48-hour time window. To search for transit events, we used the \texttt{Transit Least Squares} algorithm \citep[\texttt{TLS},][]{Hippke2019}\footnote{\url{https://github.com/hippke/tls}}. The dataset analyzed by \texttt{TLS} contained approximately $510\,700$ data points spanning a baseline of $\sim$1880 days. Within a search range of 0.6 to 940.6 days, \texttt{TLS} detected a periodic transit-like signal with a period of $P = 17.61731 \pm 0.00184$ days and a false alarm probability (FAP) of $8 \times 10^{-5}$, consistent with the period reported by TESS.

\begin{figure}
    \centering
    \includegraphics[width=\hsize]{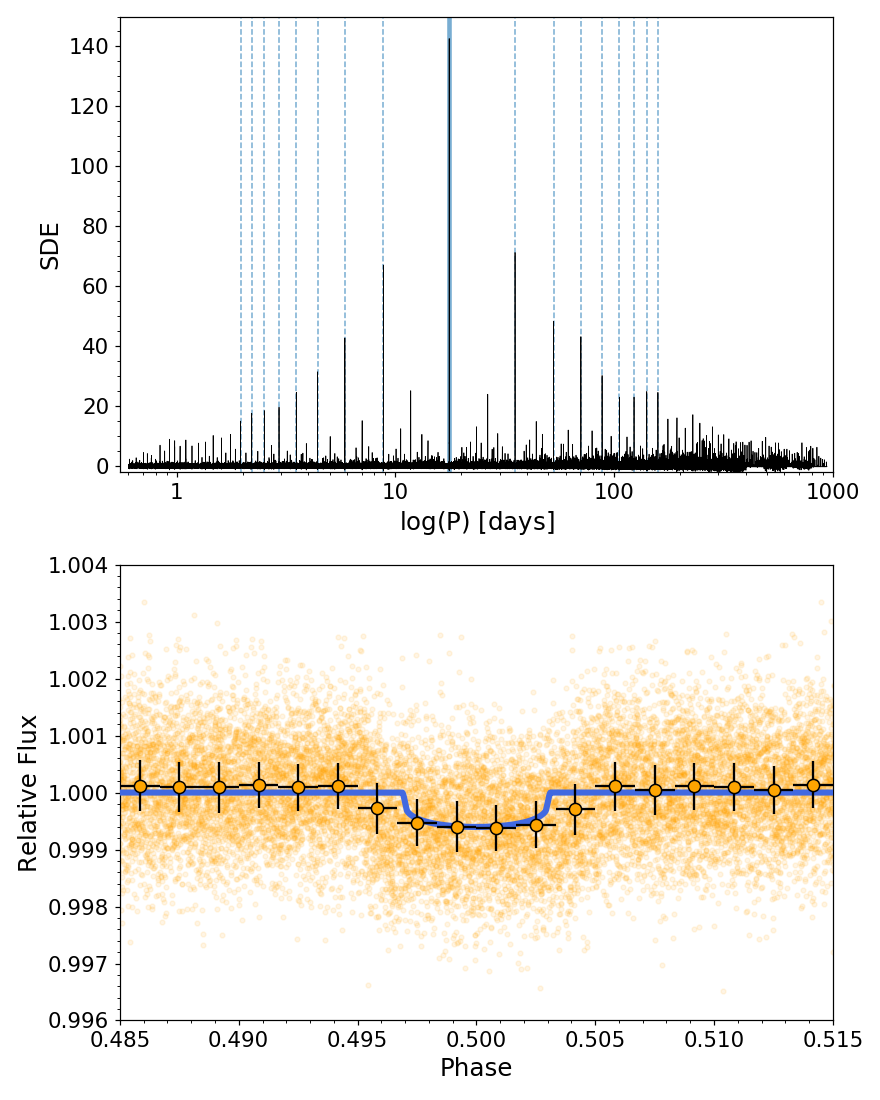}
    \caption{\texttt{TLS} analysis of TESS light curves. \textit{Top panel}: Signal detection efficiency (SDE) versus period. The peak with the maximum SDE corresponds to $P = 17.61731 \pm 0.00184$ days. \textit{Bottom panel}: TESS light curve folded to the maximum SDE period. The blue line corresponds to the \texttt{TLS} transit model.}
    \label{Fig:TLS_TESS}
\end{figure}

Figure~\ref{Fig:TLS_TESS} shows the signal detection efficiency (SDE) as a function of period (top panel) and the phase-folded TESS light curve at the period with the maximum SDE. This result confirms that our reduced-data analysis successfully recovers the known transiting signal. Furthermore, no additional significant periodic signals were found, suggesting the absence of a second transiting planet in the system. These results are in agreement with the more detailed analysis presented in Section~\ref{Sec:TESS_sherlock}.

\subsection{Stellar rotation}
\label{Sec:stellarrot_activ}
We analyzed the ESPRESSO activity indices time series to look for periodic signals that could be caused by the star's rotation. To model the time series of the FWHM, BIS, CONT, $S$ index, Na, H$\alpha$, and Ca indices, we used a combination of a third-degree polynomial to model any long-term trend present in the data and a periodic term using Gaussian processes \citep[GP; e.g.,][]{Rasmussen2006,Gibson2012}. Each individual curve was modeled independently using a time-dependent function described by 

 \begin{equation}
     A(t) = a + b t + c t^2 + d t^3
  \label{Eq:Poly_plus_GP}
 ,\end{equation}
where $a$ is the zero point, and $b$, $c$, and $d$ are the linear, quadratic, and cubic terms of the third-degree polynomial, respectively. To model the systematic noise present in the data, we employed a GP function incorporating a periodic term, specifically the periodic kernel described in \cite{ForemanMackey2017}:

\begin{equation}
    k_{ij\; \mathrm{P}}(t) = \frac{B}{2+C} \exp^{-|t_i-t_j|/L} \left[ \cos \left( \frac{2\pi |t_i-t_j|}{P_\mathrm{rot}} \right) + (1+ C) \right]
 \label{Eq:CeleritePerKernel}
,\end{equation}
where $|t_i-t_j|$ is the difference between two epochs or observations, $P_\mathrm{rot}$ is the period of the variation, and the constants $B$, $C$, and $L$ are all positive and greater than $0$. In addition to the parameters that describe the long-term trend and the GP hyperparameters, we included a jitter term to estimate the white noise present in each time series.

\begin{table}
\caption{Priors for the parameters used to fit the activity indices.}
\label{Tab:StellarActiv_GPfit_Priors}
\centering
\begin{tabular}{c c}
\hline\hline
Parameter & Prior \\
\hline

$a$ & $\mathcal{U}(-10^4, 10^4)$ \\
$b$ & $\mathcal{U}(-50, 50)$ \\
$c$ & $\mathcal{U}(-50, 50)$ \\
$d$ & $\mathcal{U}(-50, 50)$ \\
$\log B$ & $\mathcal{U}(-12, 8)$ \\
$\log L$ & $\mathcal{U}(-12, 12)$ \\
$\log P_\mathrm{rot}$ & $\mathcal{U}(0.0, 5.01)$ \\
$\log C$ & $\mathcal{U}(-12, -5)$ \\
$\sigma_\mathrm{jitter}$ & $\mathcal{U}(10^{-6}, 10)$ \\

\hline
\end{tabular}
\end{table}

We estimated the values of the constants of the third-degree polynomial ($a$, $b$, $c$, and $d$), the GP kernel ($B$, $C$, $L$, and $P$), and the jitter terms using a Bayesian Markov chain Monte Carlo (MCMC) procedure. The priors used for fitting each parameter are presented in Table~\ref{Tab:StellarActiv_GPfit_Priors}. The fitting procedure started with the optimization of a log posterior function using \texttt{PyDE}\footnote{\url{https://github.com/hpparvi/PyDE}}. Once the \texttt{PyDE} converged to a solution, we used this solution as a starting point and ran \texttt{emcee} \citep{ForemanMackey2013} for $25\,000$ iterations using 70 chains (with a thin value of 50). For the fit parameters, we computed the percentiles of the posterior distribution as our final values (median) and 1$\sigma$ uncertainty range.

\begin{figure}
    \centering
    \includegraphics[width=\hsize]{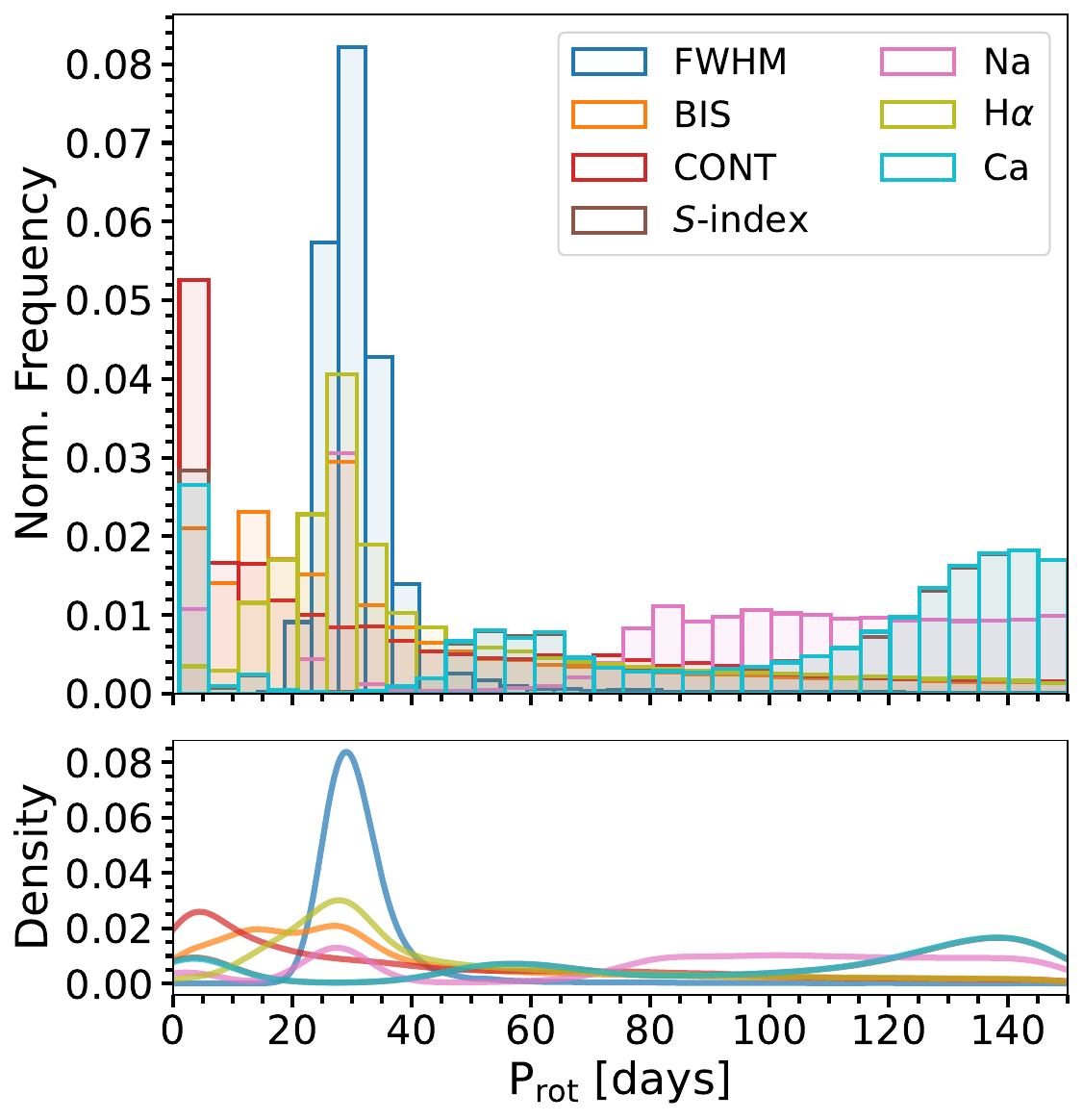}
    \caption{Stellar rotation from activity indices. \textit{Top panel}: Histograms of the posterior distribution of the fit rotation period for the ESPRESSO activity indices. \textit{Bottom panel}: Kernel density estimation of the posterior distributions.}
    \label{Fig:ActivIndices_Prot}
\end{figure}

Figure~\ref{Fig:ActivIndices_Prot} shows the distribution of the fit rotation period ($P_\mathrm{rot}$) derived from each activity index, while Table~\ref{Tab:StellarActiv_GPfit_Prot} summarizes the final rotation period estimates and their associated uncertainties. Of all the activity indicators, only FWHM exhibits a well-defined peak in its posterior distribution, yielding $P_\mathrm{rot} = 30.0^{+5.8}_{-4.1}$ days. Other indices, such as BIS and H$\alpha$, also show peaks near 30 days, though with larger uncertainties. The remaining activity indicators display broader posterior distributions without significant peaks. The rotation period derived from FWHM is consistent with the expected value of $P_\mathrm{rot} = 28 \pm 12$ days predicted by the activity-rotation relationship of \citet{SuarezMascareno2015}.

\begin{table}
\caption{Derived $P_\mathrm{rot}$ values from the activity indices.}
\label{Tab:StellarActiv_GPfit_Prot}
\centering
\begin{tabular}{c c}
\hline\hline
Activity index & $P_\mathrm{rot}$ [days] \\
\hline

FWHM & $30.0^{+5.8}_{-4.1}$ \\
BIS & $27.4^{+46.2}_{-16.9}$ \\
CONT & $22.3^{+59.4}_{-19.1}$ \\
$S$ index & $114.0^{+27.0}_{-91.6}$ \\
Na & $98.2^{+35.2}_{-70.7}$ \\
H$\alpha$ & $31.2^{+48.6}_{-11.1}$ \\
Ca & $115.0^{+25.9}_{-74.8}$ \\

\hline
\end{tabular}
\end{table}

In addition, we analyzed the TESS SAP photometry and searched for long-term photometric archival data that could confirm the rotation period found by our analysis of stellar activity indices. The generalized Lomb-Scargle \citep[GLS,][]{Lomb1976,Scargle1982,Zechmeister2009} periodogram of the TESS SAP data (see Fig.~\ref{Fig:TESS_SAP_Periodogram}) shows that the most significant power peak is around 33 days. This value is consistent with the rotation period found in our analysis, although they are close to the total baseline of a TESS single-sector observation, which is $\sim$27 days, with the spacecraft sending data to Earth and stopping observations every $\sim$13.5 days. Thus, it is possible that the signals detected in the periodogram of the TESS data could be associated with systematics introduced by the orbit of the satellite.

Searching for archival data, we found two TOI-283 time series taken by the All-Sky Automated Survey for Supernovae \citep[ASAS-SN,][]{Shappee2014}. The ASAS-SN Sky Patrol portal\footnote{\url{https://asas-sn.osu.edu/photometry}} provided light curves in the $V$ and $g$ bands with a large number of measurements and baselines greater than $\sim$1600 days, but the data show a large scatter and the GLS periodograms of the light curves do not show a significant peak around $\sim$30 days for both bands.

\subsection{Radial velocity and activity indices frequency analysis}
\label{Sec:rvanalysis}

\begin{figure}
    \centering
    \includegraphics[width=\hsize]{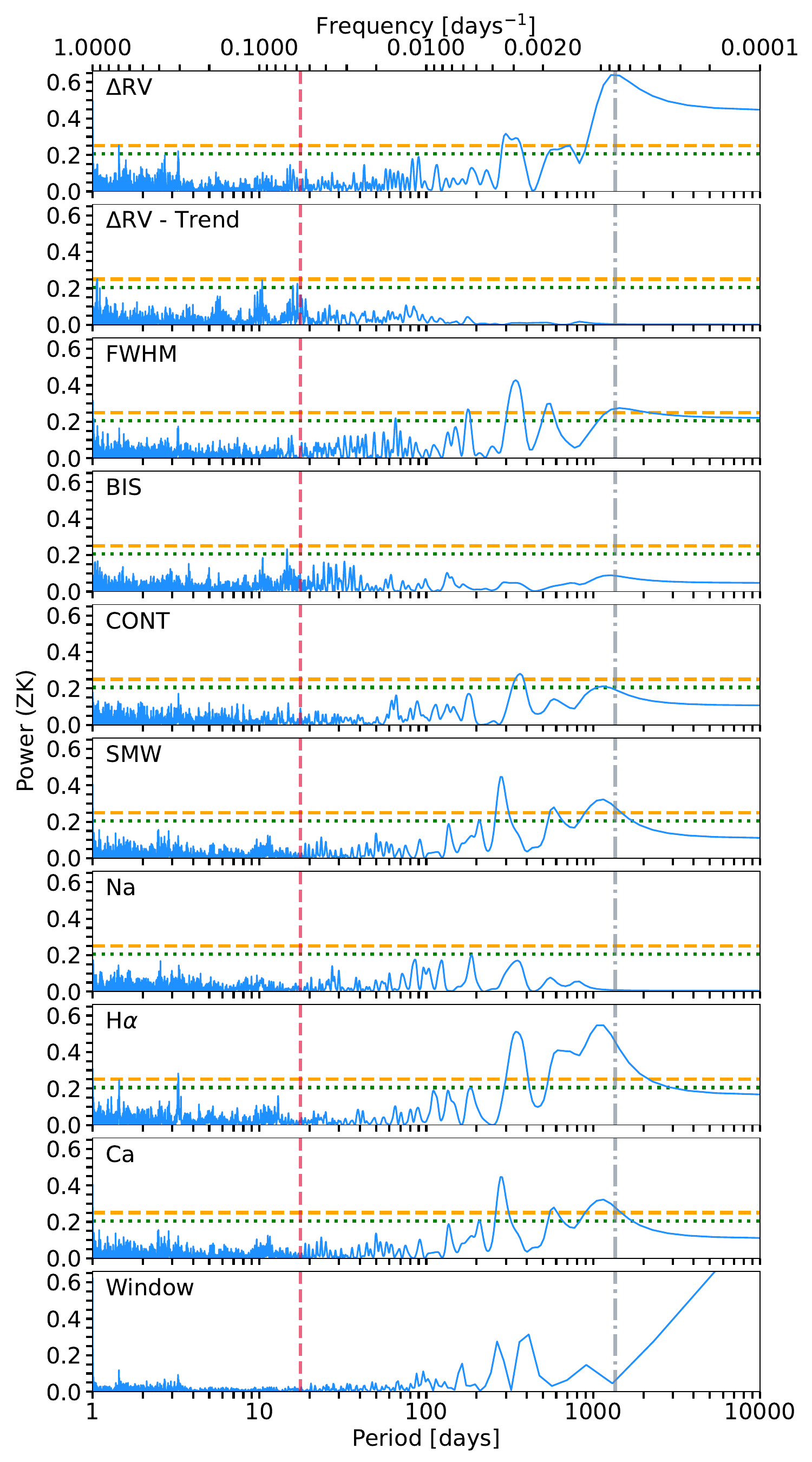}
    \caption{GLS periodograms of the ESPRESSO RVs, activity indices, and the observational window function. The RV and activity index values were median-subtracted for each dataset before (E18) and after (E19) the June–July 2019 intervention. Horizontal lines indicate the FAP levels at 10\% (dash-dotted green line) and 1\% (dashed orange line). The vertical dotted red line marks the orbital period of TOI-283\,b ($P = 17.61745$ days), while the vertical dash-dotted gray line indicates the period of the long-term signal ($P = 1356$ days; see Sect.\ref{Sec:JointFit}).}
    \label{Fig:RVActiv_Periodogram}
\end{figure}

To search for the planetary signal in our ESPRESSO data, we computed the GLS periodogram for the RV measurements and activity indices. To generate the GLS periodograms for each variable, we used the \texttt{Python} implementation of \citet{Zechmeister2009}\footnote{\url{https://github.com/mzechmeister/GLS}}. Figure~\ref{Fig:RVActiv_Periodogram} shows the GLS periodograms, with the orbital period of TOI-283\,b marked with a vertical red line. The periodograms were computed using the RV and activity index values after subtracting the median of each dataset, separately for the pre- (E18) and post- (E19) intervention epochs. The FAP levels shown in the figure were calculated using Eq. (24) from \citet{Zechmeister2009}.

Despite the relatively large number of observations and the long time baseline covered by the ESPRESSO data, the RV periodogram does not show a significant peak at the expected orbital period of the TOI-283\,b transit events. As is shown in Fig.~\ref{Fig:TOI283b_ESPRESSO_RV_TimeSeries}, the RV data may have a long-term trend that could hinder the detection of the transiting planet candidate in the periodogram. As a test, we fit the RV dataset using different approaches to model the presence of the candidate and the long-term trend. We tested four models: 1) a single Keplerian with no trend, 2) a single Keplerian plus a linear trend, 3) a single Keplerian plus a quadratic trend, and 4) a two Keplerian model. The planet-induced RV variation was modeled as a Keplerian assuming a circular orbit, and we used normal priors for the period and central time of the transiting candidate based on TESS estimates. The fitting procedure was similar to that described in Sect.~\ref{Sec:stellarrot_activ}, in which we optimized a posterior function and then applied an MCMC procedure to obtain the posterior probability distributions of the fit parameters. Depending on the complexity of the model we had between seven and ten free parameters and ran the MCMC for $15\,000$ iterations using 80 chains. To evaluate the goodness of fit, we computed the Bayesian information criterion (BIC) metric \citep{Schwarz1978} for each model using the best-fit parameters. The BIC metric is defined as 

\begin{equation}
 \mathrm{BIC} = k \ln(n) - 2\ln(\hat{L})
,\end{equation}
where $\hat{L}$ is the maximized likelihood, $n$ is the number of data points, and $k$ is the number of fit parameters. In this framework, models with lower BIC values are preferred. In our case, the model with the lowest BIC value was model 4, i.e., the two-planet model.

To search for the planetary signal, we fit the long-term variability seen in the RV time series with a second Keplerian function. The second panel from the top in Fig.~\ref{Fig:RVActiv_Periodogram} shows the GLS periodogram of the RV measurements after fitting and removing this long-term trend. The periodogram shows a broad series of discrete peaks around $\sim$8.6 days and $\sim$17.4 days, slightly below the 1\,\% FAP level. While this latter peak periodicity does not match that of the transiting planet, the series of peaks is broad enough to include it.

Similarly, we analyzed the GLS periodograms of several activity indicators derived from the ESPRESSO data. As is shown in Fig.~\ref{Fig:RVActiv_Periodogram}, none of these indices exhibit significant periodicities (even at a false positive alarm level of 10\%) in the short-period regime ($P < 100$ days). However, some indicators display peaks around 200–300 days; for example, the FWHM, $S$ index, H$\alpha$, and Ca. Additionally, several activity indices (e.g., FWHM, $S$ index, H$\alpha$, and Ca) show signs of long-term variability on timescales of $\sim$1000 days; however, these signals are less significant than the peak observed in the RV periodogram.

\begin{figure}
    \centering
    \includegraphics[width=0.45\textwidth]{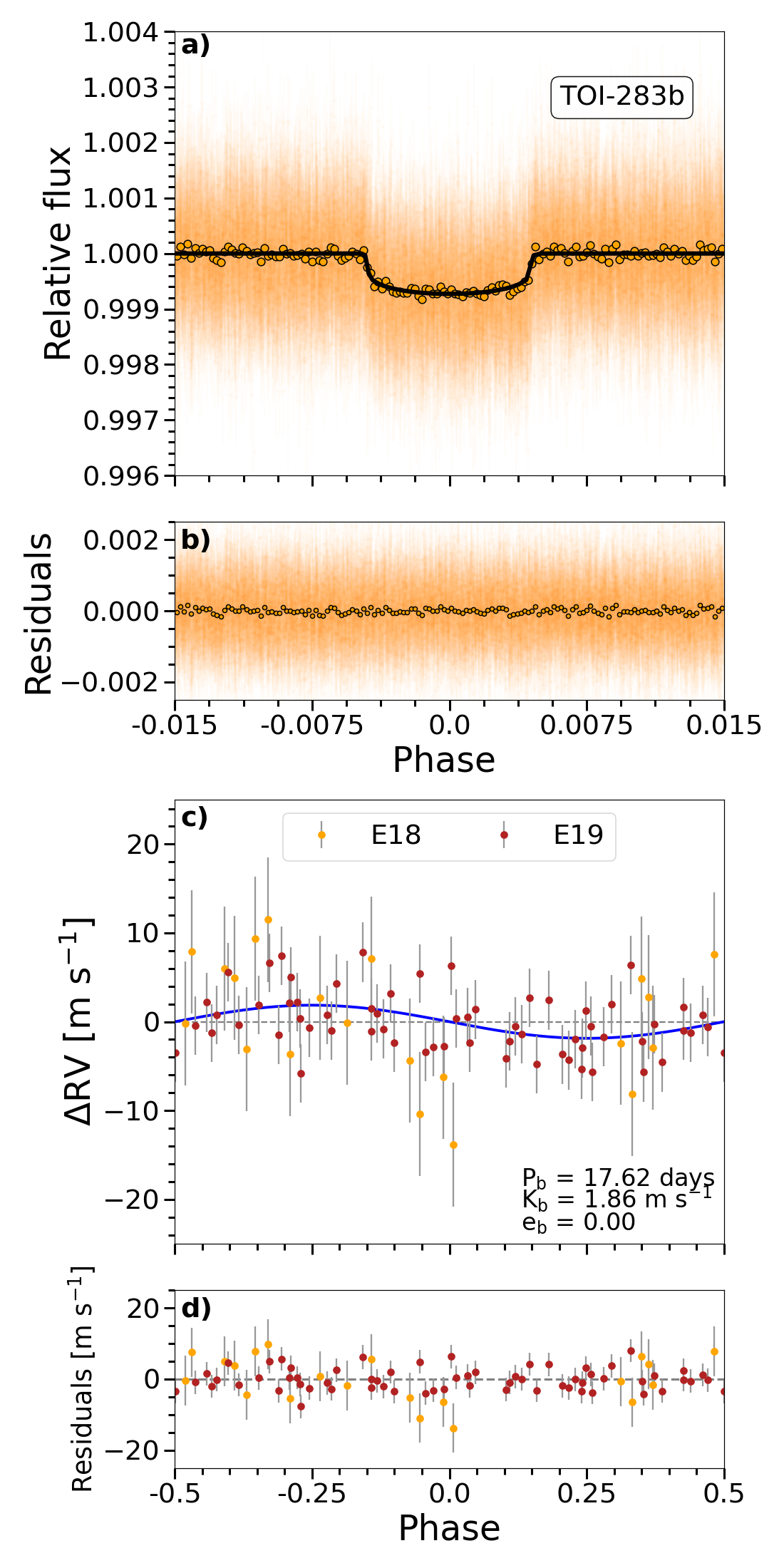}
    \caption{TESS and ESPRESSO phase-folded data of TOI-283\,b. Panel a: TOI-283\,b phase-folded TESS light curve after subtracting the photometric variations from the time series. The best-fit transit model is shown in black. The circles are TESS binned observations. Panel b: Residuals of the transit fit. Panel c: ESPRESSO phase-folded RV measurements and best fit (blue line) after subtracting the quadratic trend and red-noise component. Panel d: Residuals of the RV fit.}
    \label{Fig:TOI283b_TESS_LC_phase}
\end{figure}

\begin{figure}
    \centering
    \includegraphics[width=0.45\textwidth]{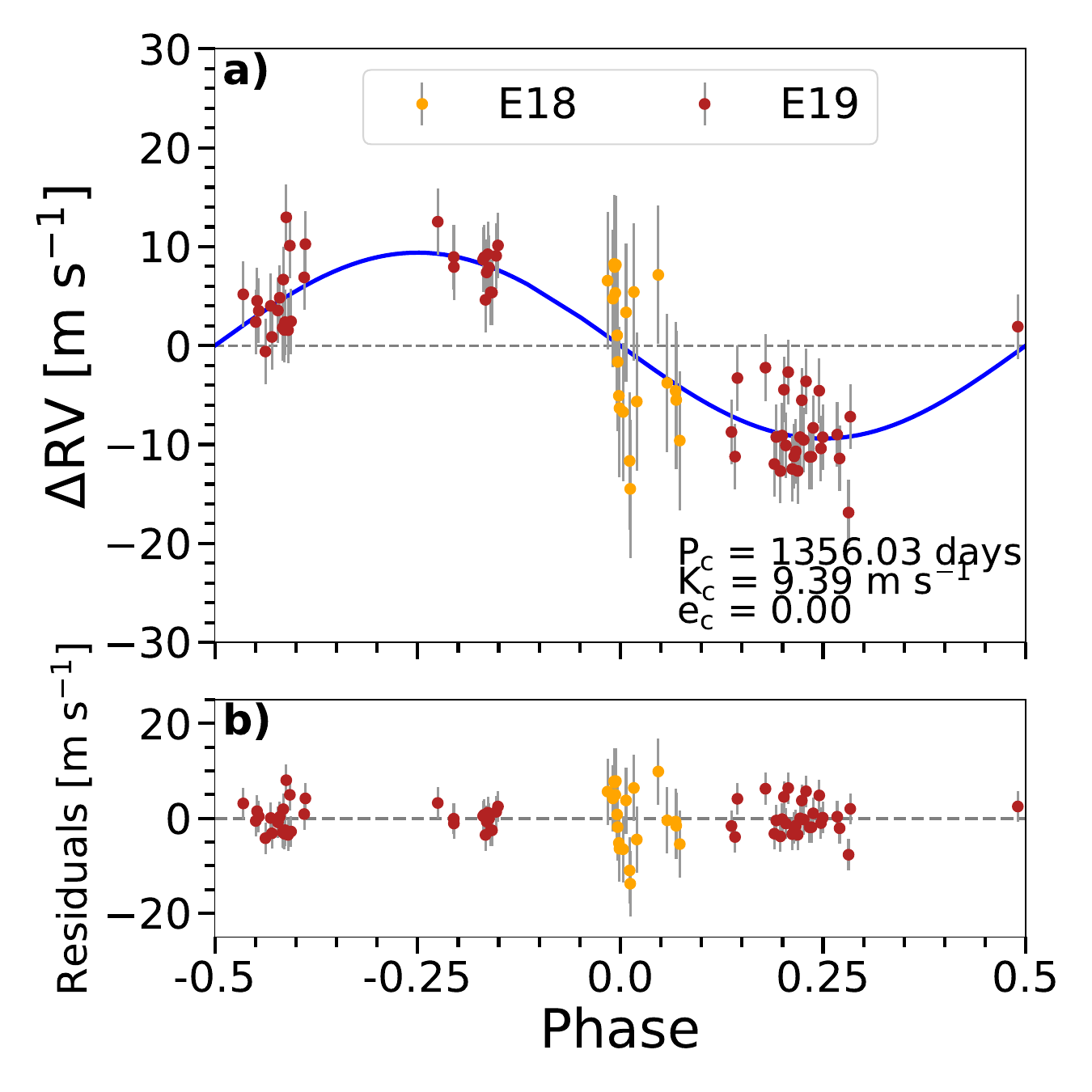}
    \caption{ESPRESSO phase-folded RV time series of the long-term RV trend of TOI-283 (panel a) and the residuals of the fit (panel b).}
    \label{Fig:TOI283_RV_Trend_phase}
\end{figure}

\subsection{Joint fit of TESS and ESPRESSO data}
\label{Sec:JointFit} 

We decided to omit the LCOGT ground-based transit observation from the global fit because the full transit could not be observed and, in addition, the entire event was covered by TESS Sector 7. However, we would like to point out that the LCOGT data provided a transit detection consistent with the TESS results (see Appendix \ref{Sec:Appendix_LCO_transit}). We performed a joint fit of the available data (TESS light curves and ESPRESSO RVs) to obtain the candidate orbital and physical properties. We decided to include the entire TESS time series in the fit, with no binning of the observations. We used a Bayesian MCMC approach to fit the data, as is described in Section \ref{Sec:stellarrot_activ} and in \cite{Murgas2023}. The transit events were modeled using \texttt{PyTransit}\footnote{\url{https://github.com/hpparvi/PyTransit}} \citep{Parviainen2015}. We adopted a quadratic limb darkening (LD) law, during the fit the LD coefficients were compared to the predicted values computed with \texttt{LDTK}\footnote{\url{https://github.com/hpparvi/ldtk}} \citep{Parviainen2015b} using a likelihood function. The predicted \texttt{LDTK} LD values were calculated using the derived stellar parameters presented in Table~\ref{Tab:Star}. The ESPRESSO RV measurements were modeled using \texttt{RadVel}\footnote{\url{https://github.com/California-Planet-Search/radvel}} \citep{Fulton2018}.

Although instrumental effects are removed from the TESS light curves by the PDC algorithm, it is common practice to model the presence of photometric variability and residual systematics with GPs. For the TESS time series, we adopted the commonly used Mat\'ern 3/2 kernel:
\begin{equation}
    k_{ij} = c^2_k \left( 1 + \frac{\sqrt{3} |t_i-t_j|}{\tau_k}\right) \exp\left(-\frac{\sqrt{3} |t_i-t_j|}{\tau_k}\right)
\label{Eq:TESS_GPKernel}
,\end{equation}
where $|t_i-t_j|$ is the time between epochs in the series, and the hyperparameters, $c_k$ and $\tau_k$, were allowed to be free, with $k$ indicating the different light curves analyzed here.

The use of GPs is also applied to the RV measurements to model any velocity variation induced by the stellar activity of the star. We decided to use the \texttt{RadVel} quasi-periodic (QP) kernel described by
\begin{equation}
    k_{ij\; \mathrm{RV}} = \eta_1^2 \exp \left( \frac{-|t_i - t_j|^2}{\eta_2^2} - \frac{1}{2\eta_4^2} \sin^2 \left( \frac{\pi |t_i-t_j|}{\eta_3} \right) \right) 
\label{Eq:RV_GPKernel}
,\end{equation}
where $t_i-t_j$ is the time between the epochs in the series, and the hyperparameters, $\eta_i$ (with $i \in [1,4]$), were set free. For the period of the kernel (i.e., $\eta_3$), we used normal priors centered around the stellar rotation period derived from the activity indices (see Sect. \ref{Sec:stellarrot_activ}).

To model both time series, we used as free parameters the planet-to-star radius ratio ($\mathrm{R}_\mathrm{p}/\mathrm{R}_\star$), the LD coefficients ($q_1$, $q_2$ following \citealp{Kipping2013}), the central time of the transit ($\mathrm{T}_{\mathrm{c}}$), the planetary orbital period ($P$), the stellar density ($\rho_\star$), the transit impact parameter ($\mathrm{b}$), and the RV semi-amplitude ($K$). The long-term trend present in the RVs was modeled using a Keplerian function with its epoch, orbital period, and RV semi-amplitude set as free parameters (see Sect. \ref{Sec:rvanalysis}). Because we split our RV measurements into two epochs, each subset had its own velocity offset ($\gamma$) and RV jitter ($\sigma_{\mathrm{RV\; jitter}}$). We considered two orbital models: one with a circular orbit (eccentricity fixed at $e=0$) and one with an eccentric orbit. For the eccentric model, we used the parameterization $\sqrt{e}\cos(\omega)$ and $\sqrt{e}\sin(\omega)$, where $\omega$ is the argument of the periastron. Considering the orbital parameters and the GP coefficients for each TESS sector and the RV time series, we had a total of 91 and 93 free parameters for the circular and eccentric models, respectively.

We adopted uniform priors for all parameters except the stellar density, for which we used a normal prior based on the mass and radius values derived from our stellar parameters. Since the transit was detected by TSO and confirmed by our independent analysis presented in Sect.~\ref{Sec:tess_tls}, we applied more restrictive uniform priors for $\mathrm{T}_{\mathrm{c}}$ and $P$, spanning a range of $\pm 0.4$ days around the predicted values. For the TESS GPs, we set a uniform prior in logarithmic space for the timescale hyperparameter ($\tau_k$), ranging from one hour to half the length of a TESS sector, i.e., 13 days.

The fitting procedure is the same as the one described in Sect. \ref{Sec:stellarrot_activ}. We optimized a posterior probability function using the differential evolution routine \texttt{PyDE}. The results of the global optimization procedure were used as a starting point of an MCMC procedure using \texttt{emcee}. We ran \texttt{emcee} using 360 chains for 15\,000 iterations as a burn-in phase, and the main MCMC phase was run for 50\,000 iterations, while keeping the parameter values of every 100th iteration to reduce autocorrelation. The final values of the fit parameters were computed using the median and 1$\sigma$ limits of the posterior distributions of the parameters. The prior ranges and the final parameter values with their respective uncertainties are presented in Table~\ref{Tab:Planet_parameters}. The results for the GP hyperparameters are presented in Table~\ref{Tab:GPs_coeffs_jointfit}.

To determine whether the circular or eccentric model provided a better fit to our data, we compared the difference between the BIC values for the circular and eccentric models. We find a difference of $\Delta \mathrm{BIC} = \mathrm{BIC}_{\mathrm{Ecc}} - \mathrm{BIC}_{\mathrm{Circ}} = 28$, which means that the circular model is a better fit to our data. It is also worth noting that our eccentric model also finds an orbit solution with a low eccentricity value of $e = 0.11^{+0.11}_{-0.07}$, which could be considered consistent with a circular orbit according to the criterion of \cite{Lucy1971}. Given the results of the BIC comparison, we adopted the results of the circular orbit fit as our final values for the remainder of the paper.

Figure~\ref{Fig:TOI283b_ESPRESSO_RV_TimeSeries} shows the RV time series including the results for the best-fit circular orbit model, Figure~\ref{Fig:TOI283b_TESS_LC_phase} shows the final light curve fit to the TESS data and the radial velocities phase-folded to the planet's orbital period. The RV residuals RMSs are 6.5 m s$^{-1}$ for the pre-intervention data (E18) and 3.1 m s$^{-1}$ for the post-intervention measurements (E19); these values were calculated including the fit RV jitter value (added in quadrature) for each set. The relatively large scatter of the residuals could be explained by stellar activity, although no strong correlations between the RVs and the activity indices were found (see Fig.~\ref{Fig:RV_vs_ActivityInd}). Figure~\ref{Fig:MCMC_CornerPlot} shows the posterior distributions for the fit transit and orbital parameters of TOI-283\,b. 

Figure~\ref{Fig:TOI283_RV_Trend_phase} presents the phase-folded fit to the long-term RV trend. Our Keplerian model fit finds a period of $P = 1350^{+489}_{-176}$ days and a RV semi-amplitude of $K = 9.4^{+2.2}_{-1.7}$ m s$^{-1}$. If the signal originated from an additional companion in the system, its minimum mass would be $\mathrm{M}\mathrm{p} \sin i = 0.44^{+0.12}_{-0.08} \, \mathrm{M}_\mathrm{Jup}$, placing it within the planetary regime. However, a compelling alternative explanation is that the long-term trend arises from stellar activity. The GLS periodogram of the activity indicators shown in Fig.~\ref{Fig:RVActiv_Periodogram} reveals significant power at periods around 300–400 days in the FWHM, $S$ index, H$\alpha$, and Ca indices. In particular, many of these indices also show peaks above the 1\,\% FAP threshold near the period derived for the long-term RV signal. This alignment strengthens the interpretation that stellar activity may be responsible for the observed trend. Moreover, our RV time span of $\sim$1170 days only marginally covers a single orbit of the proposed companion, complicating any robust planetary interpretation. Additional long-term RV monitoring will be essential to determine whether the trend detected in the ESPRESSO data is indeed caused by a planetary companion or is instead stellar in origin.

In addition to the results presented above, we explored alternative joint fitting strategies by varying the GP treatment of the RV data. These tests were conducted to determine whether it was possible to reduce the level of RV jitter. Specifically, we tested: (i) a fit without including GPs; (ii) a fit with a non-periodic GP kernel (squared exponential); and (iii) a fit including a third Keplerian signal with orbital periods ranging from 0.5 to 15 days and the same QP kernel that was described previously. Across all these tests, the RV jitter values remain consistent, likely because the GP parameters, particularly those describing the exponential timescales, are only weakly constrained. Crucially, the inferred orbital parameters of the transiting planet are unchanged within uncertainties in all cases, demonstrating that the planet’s properties are robustly estimated despite the relatively high levels of RV jitter.

\begin{table*}[t]
\centering
\caption{Prior functions, and fit and derived parameters of TOI-283\,b.}
\label{Tab:Planet_parameters}
\begin{tabular}{l c c c}
\hline 
\hline
\noalign{\smallskip}
Parameter & Prior & Circular model & Eccentric model \\
\noalign{\smallskip}
\hline
\noalign{\smallskip}
\multicolumn{4}{c}{Fit transit and orbital parameters} \\
\noalign{\smallskip}

$\rho_*$ [g cm$^{-3}$] & $\mathcal{N}(1.84,0.2)$ & $1.83 \pm 0.04$ & $1.83 \pm 0.03$ \\
$\mathrm{R}_\mathrm{p}/\mathrm{R}_\star$ & $\mathcal{U}(0.005, 0.035)$ & $0.0252 \pm 0.0003$ & $0.0252^{+0.0007}_{-0.0005}$ \\
$T_{c}$ - 2\,457\,000  [BJD] & $\mathcal{U}(2548.685,2549.485)$ & $2549.0822^{+0.0007}_{-0.0008}$ & $2549.0823 \pm 0.0007$  \\
$P$ [days] & $\mathcal{U}(17.218,18.018)$ & $17.61745^{+0.00002}_{-0.00001}$ & $17.61745^{+0.00002}_{-0.00001}$ \\
$b$ & $\mathcal{U}(0,1)$ & $0.476 \pm 0.016$ & $0.467^{+0.136}_{-0.192}$ \\
$\sqrt{e}\cos(\omega)$  & $\mathcal{U}(-1,1)$ &  0  (fixed) & $0.17^{+0.23}_{-0.28}$ \\
$\sqrt{e}\sin(\omega)$  & $\mathcal{U}(-1,1)$ &  0  (fixed) & $0.00^{+0.24}_{-0.28}$ \\
$K$ [m s$^{-1}$] & $\mathcal{U}(0,80)$ & $1.86^{+0.58}_{-0.56}$ & $1.89 \pm 0.60$ \\
$\gamma_1$ [m s$^{-1}$] & $\mathcal{U}(-12900, -12200)$ & $-12588.65^{+6.95}_{-5.28}$ & $-12589.41^{+6.43}_{-4.92}$ \\
$\sigma_{1}$ [m s$^{-1}$] & $\mathcal{U}(0,20)$ & $6.95^{+1.35}_{-1.06}$ & $6.89^{+1.35}_{-1.05}$ \\
$\gamma_2$ [m s$^{-1}$] & $\mathcal{U}(-12900, -12200)$ & $-12590.84^{+1.71}_{-1.28}$ & $-12590.88^{+1.46}_{-1.19}$ \\
$\sigma_{2}$ [m s$^{-1}$] & $\mathcal{U}(0,20)$ & $3.26^{+0.35}_{-0.31}$ & $3.28^{+0.37}_{-0.32}$ \\

\noalign{\smallskip}
\multicolumn{4}{c}{Derived orbital parameters} \\
\noalign{\smallskip}

$a/R_*$ & & $31.12 \pm 0.18$ & $31.11 \pm 0.19$ \\
$e$  & & 0 (fixed) & $0.11^{+0.11}_{-0.07}$ \\
$\omega$ [deg] & & 90 (fixed) & $0.6^{+81.9}_{-81.8}$ \\
$i$ [deg] & & $89.12 \pm 0.03$ & $89.14^{+0.35}_{-0.26}$ \\

\multicolumn{4}{c}{Derived planet parameters} \\
\noalign{\smallskip}

$\mathrm{R}_\mathrm{p}$ [R$_{\oplus}$] & & $2.34 \pm 0.09$ & $2.33 \pm 0.10$ \\ 
$\mathrm{M}_\mathrm{p}$ [M$_{\oplus}$] & & $6.54 \pm 2.04$ & $6.57 \pm 2.12$ \\
$\rho_\mathrm{p}$ [g cm$^{-3}$] & & $2.81 \pm 0.93$ & $2.84 \pm 0.99$ \\
$g_\mathrm{p}$ [m s$^{-2}$] & & $11.7 \pm 3.8$ & $11.8 \pm 3.9$ \\
$a$ [au] & & $0.123 \pm 0.004$ & $0.123 \pm 0.004$ \\
$\langle F_\mathrm{p} \rangle$ [$10^3$\,W\,m$^{-2}$] & & $41.4 \pm 3.4$ & $41.4 \pm 3.4$ \\
$S_\mathrm{p}$ [$S_\oplus$] & & $30.4 \pm 2.5$ & $30.4 \pm 2.5$ \\
$T_{\rm eq}$ ($A_{\rm Bond} = 0.0$) [K] & & $661 \pm 9$ & $661 \pm 9$ \\

\noalign{\smallskip}
\multicolumn{4}{c}{Long-term trend fit parameters} \\
\noalign{\smallskip}

$T_{c\; \mathrm{trend}}$ - 2\,457\,000  [BJD] & $\mathcal{U}(611.15,3611.15)$ & $1546.1^{+102.2}_{-238.0}$ & $1568.6^{+87.7}_{-194.9}$  \\
$P_\mathrm{trend}$ [days] & $\mathcal{U}(50,2000)$ & $1356.0^{+497.8}_{-178.9}$ & $1315.8^{+388.2}_{-151.7}$ \\
$K_\mathrm{trend}$ [m s$^{-1}$] & $\mathcal{U}(0,80)$ & $9.39^{+2.27}_{-1.64}$ & $9.37^{+2.02}_{-1.57}$ \\

\noalign{\smallskip}
\multicolumn{4}{c}{Fit LD coefficients} \\
\noalign{\smallskip}

$q_{1\;TESS}$ & $\mathcal{U}(0,1)$ & $0.362 \pm 0.002$ & $0.363 \pm 0.002$ \\
$q_{2\;TESS}$ & $\mathcal{U}(0,1)$ & $0.374 \pm 0.003$ & $0.375 \pm 0.003$ \\

\noalign{\smallskip}
\multicolumn{4}{c}{Derived LD coefficients} \\
\noalign{\smallskip}

$u_{1\;TESS}$ & & $0.450 \pm 0.003$ & $0.451 \pm 0.003$ \\
$u_{2\;TESS}$ & & $0.152 \pm 0.004$ & $0.151 \pm 0.004$ \\

\noalign{\smallskip}
\hline
\end{tabular}
\tablefoot{$\mathcal{U}$, $\mathcal{N}$ represent uniform and normal prior functions, respectively. The equilibrium temperature was computed as $T_{\rm eq} = T_\star \sqrt{\frac{R_\star}{2a}} \left( 1 - A_{\rm Bond} \right)^{1/4}$, where $A_{\rm Bond}$ is the Bond albedo. }
\end{table*}

\subsection{Transit timing variations}
\begin{figure}
    \centering
    \includegraphics[width=\hsize]{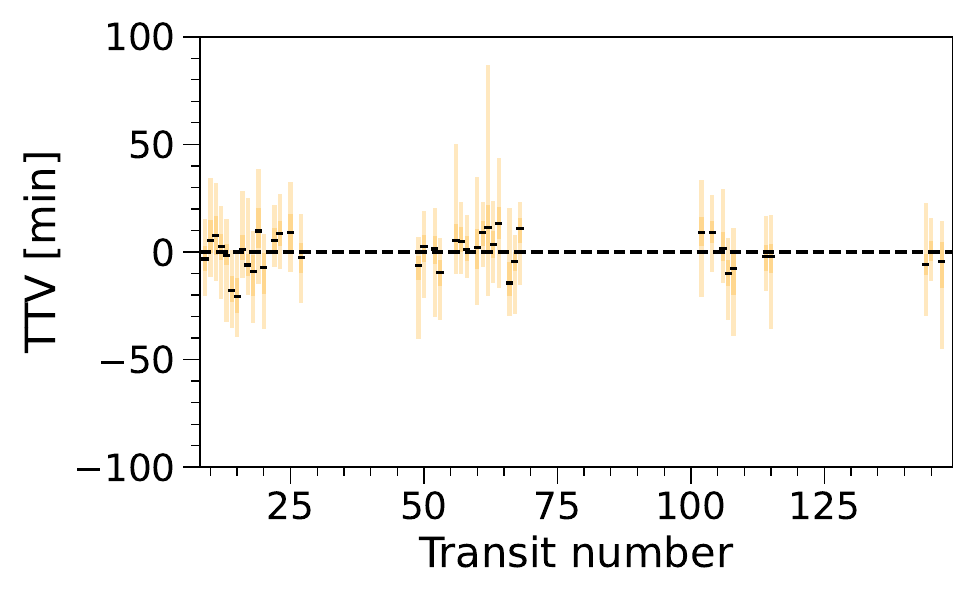}
    \caption{TOI-283\,b TTVs from TESS. No significant TTVs are detected in a time baseline of $\sim$2449 days.}
    \label{Fig:TTVs}
\end{figure}

We checked the presence of measurable transit timing variations (TTVs) using TESS photometric data. We utilized \texttt{PyTTV} to fit the transits, treating each central transit time as a free parameter following the method described in \cite{Korth2023}. Our search for TTVs involved jointly fitting the TESS photometry of all sectors with \texttt{PyTTV}. The results showed no significant deviations in the transit centers from the linear ephemeris. The TTVs are illustrated in Figure~\ref{Fig:TTVs}.

\subsection{Planet searches and detection limits from TESS photometry}
\label{Sec:TESS_sherlock}
We analyzed the TESS 120\,s cadence data in the search for hints of potential planetary candidates that remained unnoticed by the SPOC and Quick-Look Pipeline (QLP) pipelines. To this end, we used the \texttt{SHERLOCK} package \citep{pozuelos2020,demory2020} by combining all available sectors exploring the orbital periods from 1 to 30\,d, using ten detrended scenarios corresponding to window sizes ranging from 0.2 to 1.2\,d. We refer the reader to \cite{pozuelos2023} and \cite{devora2024} for further details about different search strategies.

In the first run, we found a strong signal corresponding to TOI-283\,b, which allowed us to independently confirm the detectability of this candidate. In the subsequent runs, \texttt{SHERLOCK} found other weaker signals, all attributable to noise or spurious detections, and hence we did not find any other signal hinting at extra transiting planets in the orbital periods explored. 

We then conducted injection and retrieval experiments to establish detection limits. In this context, we used the \texttt{MATRIX} package \citep{matrix}, which produces a set of synthetic planets by integrating various orbital periods, planetary radii, and orbital phases that were inserted into the dataset \citep[see, e.g.,][]{delrez2022}. Due to the vast number of sectors available, this experiment was computationally costly, generating an extremely dense grid of scenarios. Instead, our strategy explored some illustrative combinations of radii, periods, and phases, following \cite{vangrootel2021}. In particular, we used planetary sizes of 1, 2, and 3\,R$_{\oplus}$, with the orbital periods of 1, 5, 10, 15, 20, 25, and 30\,d. Then, each pair of radius-period was evaluated in five different orbital phases. Hence, in total, we explored 105 scenarios. We found that Earth-size planets are 100\,\% recovered when residing in short orbital periods ($<$10\,d), allowing us to rule out the presence of any transiting planet in close-in orbits. However, for orbital periods equal to or larger than 15\,d, the recovery of these small planets rapidly decreases to 0\,\%, making them invisible in the data. In contrast, larger planets (2 and 3\,R$_{\oplus}$) were easily recovered for the full set of periods explored, with recovery rates larger than 85\,\%, allowing us to conclude that they are likely not present in the system.  

\section{Discussion}
\label{Sec:Discussion}

\begin{figure}
    \centering
    \includegraphics[width=\hsize]{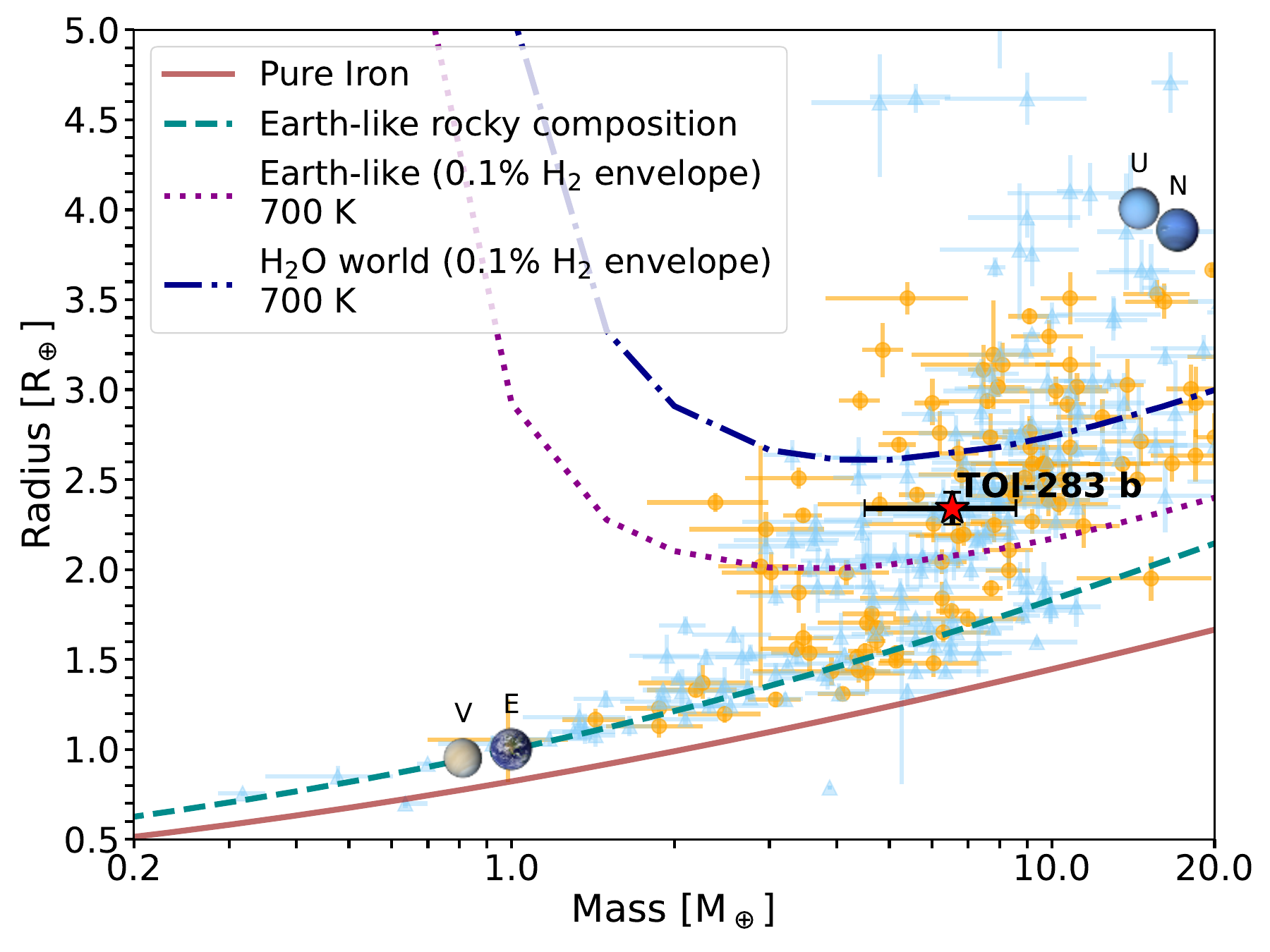}
    \caption{Mass-radius diagram for TOI-283\,b (red star) and known transiting planets with mass determinations with a precision better than 30\,\% (parameters taken from the TEPCat database; \citealp{Southworth2011}). Planets orbiting K-type stars ($4000 \leq T_{\mathrm{eff}} \leq 5300$ K) are marked with orange circles. The lines in the mass-radius diagram represent the composition models of \cite{Zeng2016, Zeng2019} for planets with pure iron cores (100\,\% Fe, brown line), Earth-like rocky compositions (32.5\,\% Fe plus 67.5\,\% MgSiO$_3$, dashed green line), Earth-like compositions with a 0.1\% H$_2$ envelope (dotted purple line), and a water world with a 0.1\,\% H$_2$ gas envelope (dash-dotted blue line). We show some Solar System planets for reference (Venus, Earth, Uranus, and Neptune).}
    \label{Fig:MassRadius}
\end{figure}

From our analysis, we determined that TOI-283\,b has a radius of $\mathrm{R}_\mathrm{p} = 2.34 \pm 0.09 \; \mathrm{R}_\oplus$, a mass of $\mathrm{M}_\mathrm{p} = 6.54 \pm 2.04 \; M_\oplus$, and an orbital period of 17.61 days. The planet orbits its star with a separation of $a = 0.123 \pm 0.004$ au. Assuming a Bond albedo of 0, this leads to an equilibrium temperature of $T_\mathrm{eq} = 661 \pm 9$ K, while we derive a stellar insolation of $S_\mathrm{p} = 30.4 \pm 2.5 \; S_\oplus$. Figure~\ref{Fig:MassRadius} places TOI-283\,b in a mass-radius diagram of known transiting planets found in the TEPCat\footnote{\url{https://www.astro.keele.ac.uk/jkt/tepcat/}} catalog \citep{Southworth2011}. For illustration purposes, we also include the composition models from \cite{Zeng2016, Zeng2019} with the closest temperature to TOI-283\,b ($700$ K). TOI-283\,b occupies a densely populated area of the diagram, where several bulk compositions are plausible, including the presence of significant H/He-rich atmospheres and water-rich planets.

\begin{figure}
   \centering
    \includegraphics[width=\hsize]{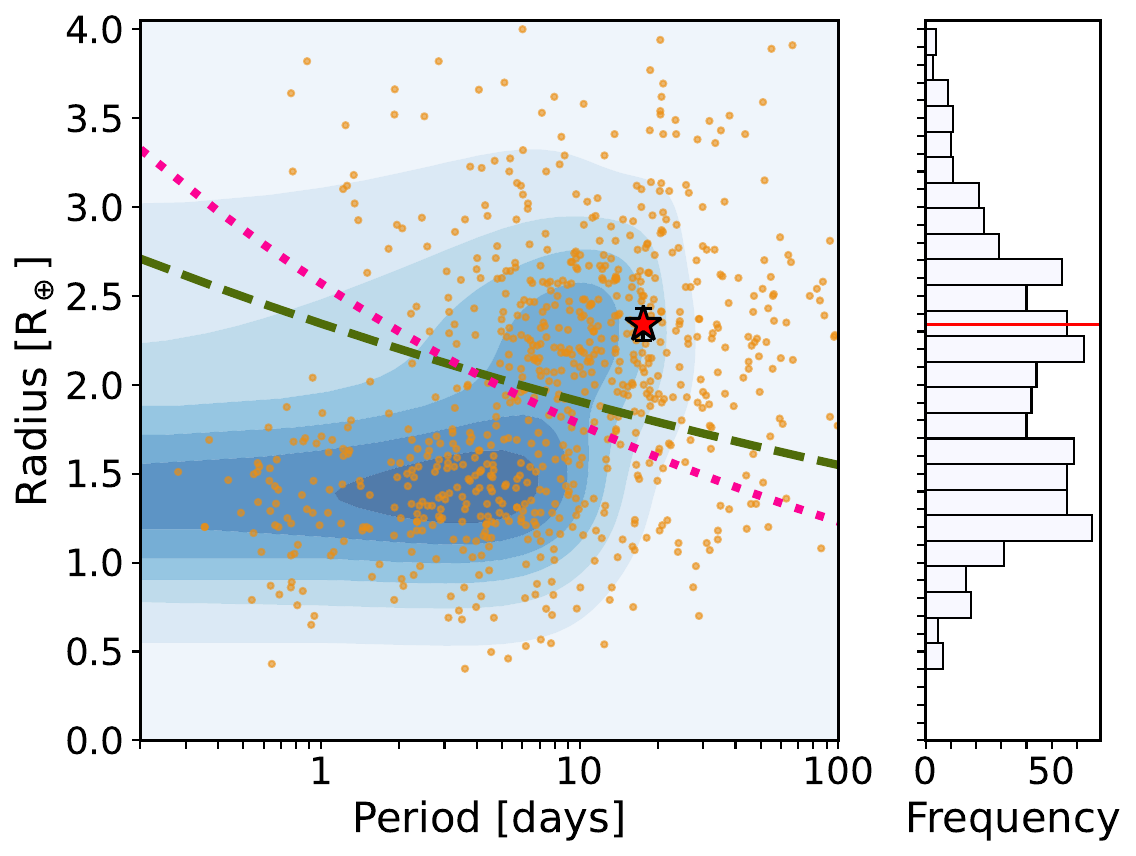}
    \caption{Period-radius diagram for small planets ($\mathrm{R}_\mathrm{p} < 4 \; R_{\oplus}$) around K-type stars ($4000 \; \mathrm{K} < T_\mathrm{eff} < 5300 \; \mathrm{K}$). The shaded blue area represents a Gaussian kernel density estimation of the points. The position of TOI-283\,b is marked by the red star. The dashed olive line shows the position of the radius gap from \cite{VanEylen2018}, while the dotted pink line shows the position of the gap from \cite{Venturini2024}. The data were taken from the NASA Exoplanet Archive (\url{https://exoplanetarchive.ipac.caltech.edu/}).}
    \label{Fig:TOI283b_RadGap}
\end{figure}

\cite{Fulton17} and \cite{Fulton2018b} show that the radii of small planets in short-period orbits ($P < 100$ days) have a bimodal distribution, with a gap at $\mathrm{R}_\mathrm{p}\sim 1. 8 \; \mathrm{R}_\oplus$ separating smaller super-Earths ($\mathrm{R}_\mathrm{p}\sim 1.3 \; \mathrm{R}_\oplus$) from larger sub-Neptunes ($\mathrm{R}_\mathrm{p}\sim 2.4 \; \mathrm{R}_\oplus$). With a planetary radius of $\mathrm{R}_\mathrm{p} = 2.33 \pm 0.09 \; \mathrm{R}_\oplus$, this planet is closer to the sub-Neptune population than to the accepted super-Earth regime. Figure~\ref{Fig:TOI283b_RadGap} shows the period-radius diagram for known transiting planets around K-type stars ($4000 \; \mathrm{K} < T_\mathrm{eff} < 5300 \; \mathrm{K}$). We only show planets with radii smaller than $4 \; \mathrm{R}_\oplus$. From the figure is clear that TOI-283\,b is above the gap measured by \cite{VanEylen2018} and the gap position derived by \cite{Venturini2024}. Our derived planetary radius is consistent with the majority of the population of sub-Neptunes around K stars. 

\begin{figure*}
    \centering
    \includegraphics[width=\hsize]{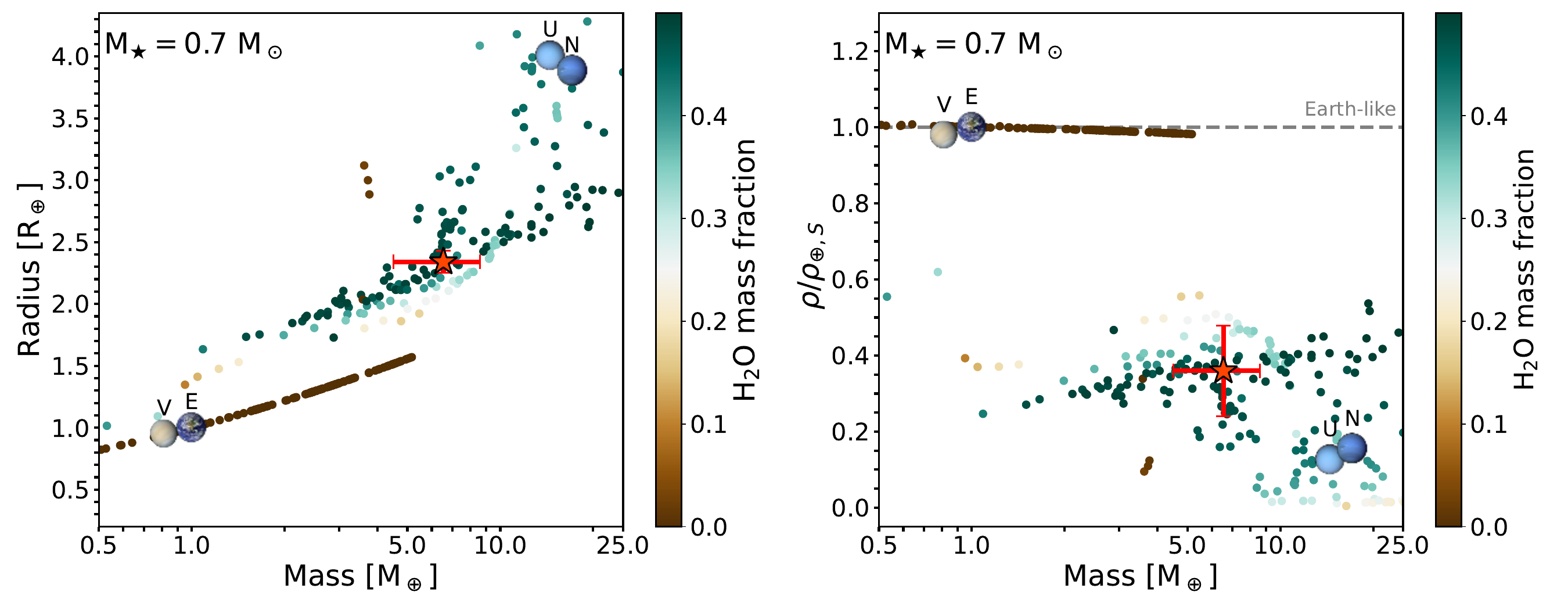}
    \caption{Comparison of TOI-283\,b to synthetic planets with short orbital periods ($P< 100$ days) and an age of 2 Gyr around a 0.7 $\mathrm{M}_\odot$ star from the simulations of \cite{Venturini2024}. The position of TOI-283\,b is marked by the orange star. The planets are marked by dots with the color indicating the fraction of water. \textit{Left panel}: Mass-radius diagram. \textit{Right panel}: Mass-density diagram (density normalized by an Earth-like composition).}
    \label{Fig:MR_MRho_Venturini24}
\end{figure*}

\subsection{Planet composition}

To better constrain the properties of TOI-283\,b, we compared it to the synthetic exoplanet population presented by \cite{Venturini2024}\footnote{\url{https://zenodo.org/records/10719523}}. In their work, \cite{Venturini2024} model planet formation around stars with masses of 0.1, 0.4, 0.7, 1.0, 1.3, and 1.5 $\mathrm{M}_\odot$ using a pebble-accretion formation framework \citep[see also][]{Venturini2020}. Their simulations assume the formation of a single planetary embryo per disk and include planetary migration processes. Following the dispersal of the protoplanetary disk, the models track the planets’ thermal evolution, including cooling and photoevaporation, up to an age of 2 Gyr. Most of the resulting synthetic planets have orbital periods of $\sim$11–18 days. In the left panel of Fig.~\ref{Fig:MR_MRho_Venturini24}, we present the mass–radius diagram for synthetic planets formed around a 0.7 $\mathrm{M}_\odot$ star. TOI-283\,b is located in a region populated by planets with water mass fractions exceeding $0.3$, suggesting that TOI-283\,b may be predominantly water-rich. However, this interpretation must be treated with caution, as the synthetic population models are limited to 2 Gyr. As our stellar age estimates suggest that the TOI-283 system is much older, evolutionary processes such as prolonged atmospheric loss may have significantly altered the planet’s composition beyond the time span considered in these simulations.

In the right panel of Fig.~\ref{Fig:MR_MRho_Venturini24}, we present a mass-density diagram with the density normalized by the Earth-like composition model from \cite{Zeng2016, Zeng2019}. The simulations are able to reproduce the distinct density populations found in M dwarfs by \cite{Luque2022}. In this case there is a noticeable lack of planets with normalized densities between $\rho/\rho_{\oplus,S} \sim 0.6-0.8$ in agreement with the limit proposed by \cite{Luque2022} of $\rho/\rho_{\oplus,S} \sim 0.65$. As with the mass-radius diagram, TOI-283\,b's position places it in the water-rich composition and away from the rocky planet normalized density region. This could indicate that TOI-283\,b is a mini-Neptune with a significant amount of water, but \cite{Venturini2024} point out that for planets with $T_\mathrm{eq} > 400$ K, as is the case for TOI-283\,b with $T_\mathrm{eq} \approx 660$ K, the water layer is in the form of water vapor and not a liquid ocean. This has the effect of increasing the radius of the planet compared to the radii of condensed water worlds.

\begin{figure*}
    \centering
    \includegraphics[width=0.49\hsize]{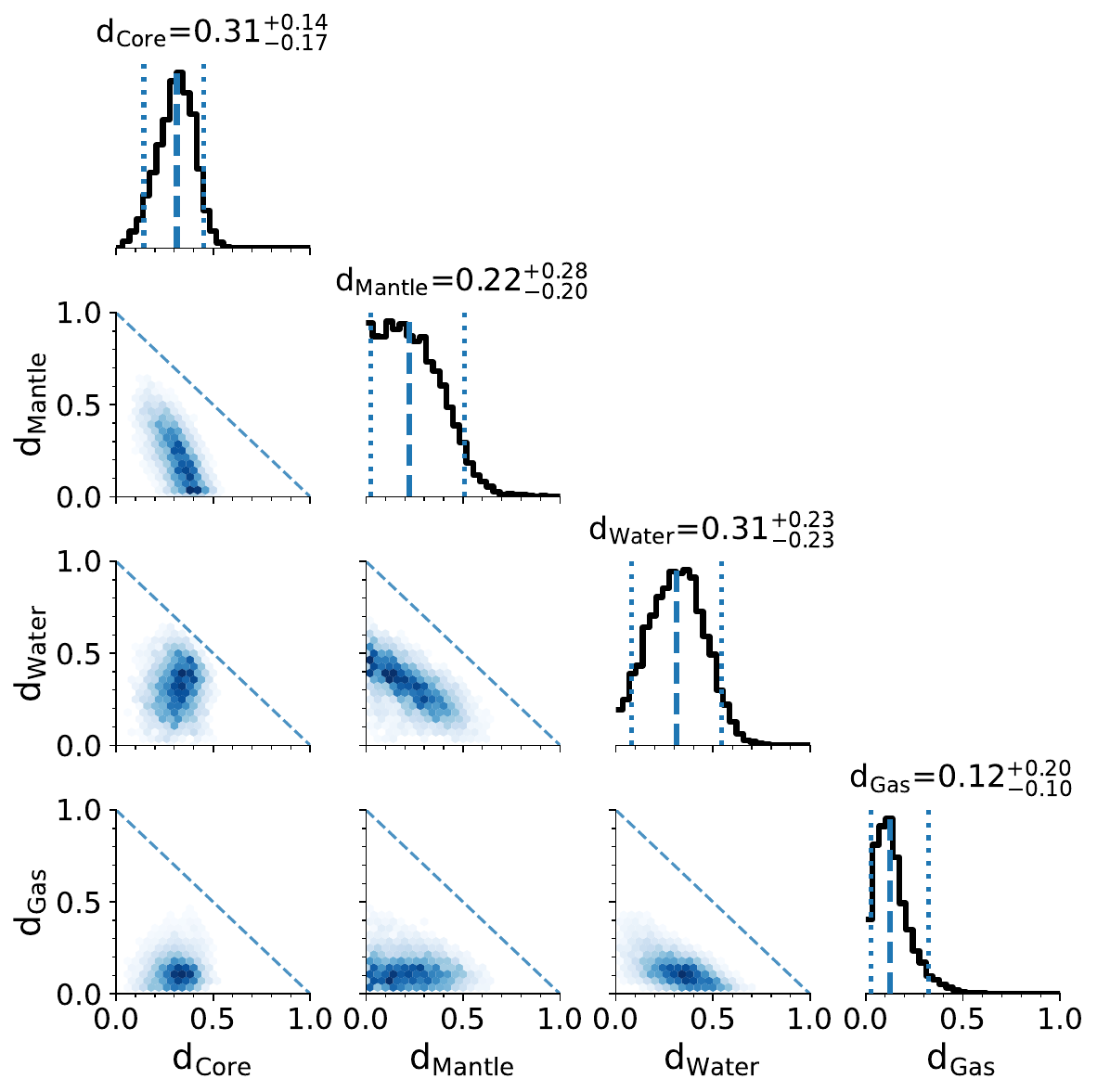}
    \includegraphics[width=0.49\hsize]{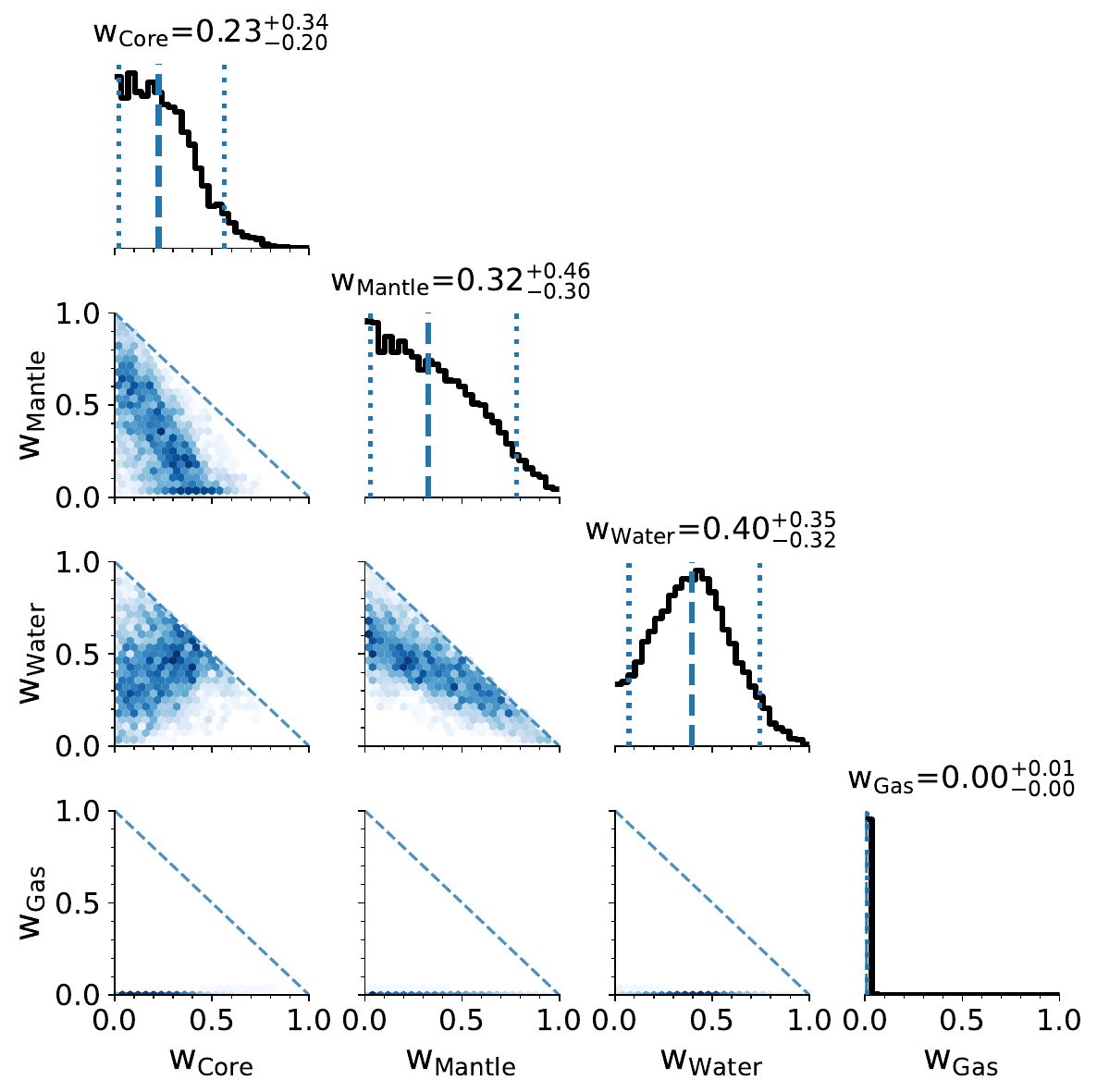}
    \caption{\texttt{ExoMDN} interior modeling results. \textit{Left}: Predicted thickness of the interior layers of TOI-283\,b. \textit{Right}: Predicted mass fraction of the interior layers of TOI-283\,b.}
    \label{Fig:TOI283b_ExoMDN_RadiusFraction}
\end{figure*}

We used \texttt{ExoMDN}\footnote{\url{https://github.com/philippbaumeister/ExoMDN}} \citep{BaumeisterTosi2023} to investigate the interior structure of TOI-283\,b. \texttt{ExoMDN} is an inference model for planetary interiors based on a mixture density network (MDN) trained on 5.6 million synthetic planet models computed with the \texttt{TATOOINE} code \citep{Baumeister2020,MacKenzie2023}. The \texttt{TATOOINE} interior models assume compositions consisting of an iron core, a silicate mantle, a water and high-pressure ice layer, and an H/He atmosphere. The training set planets have masses below 25 $M_\oplus$ and equilibrium temperatures in the range of 100--1000 K. To produce the \texttt{ExoMDN} interior models, we used as input parameters the mass, radius, and equilibrium temperature of TOI-283\,b. In Figure~\ref{Fig:TOI283b_ExoMDN_RadiusFraction} we show the \texttt{ExoMDN} predicted thickness and mass fraction of the interior layers of TOI-283\,b. The models predict that the planet consists of an interior structure comprising a $31^{+14}_{-17}$\,\% core, a $22^{+28}_{-20}$\,\% mantle, a $31^{+23}_{-23}$\,\% water layer, and a $12^{+20}_{-10}$\,\% gas envelope, where all percentages refer to the respective thicknesses relative to the planetary radius. In terms of mass fraction, TOI-283\,b could be dominated by a water envelope with $40^{+35}_{-32}$\,\% of its mass budget occupied by this molecule. This implies that TOI-283\,b could be a water-rich planet, in agreement with the composition models shown in Figure~\ref{Fig:MassRadius} and the synthetic population planets of \cite{Venturini2024}.

\subsection{Atmospheric characterization}

\begin{figure}
    \centering
    \includegraphics[width=\hsize]{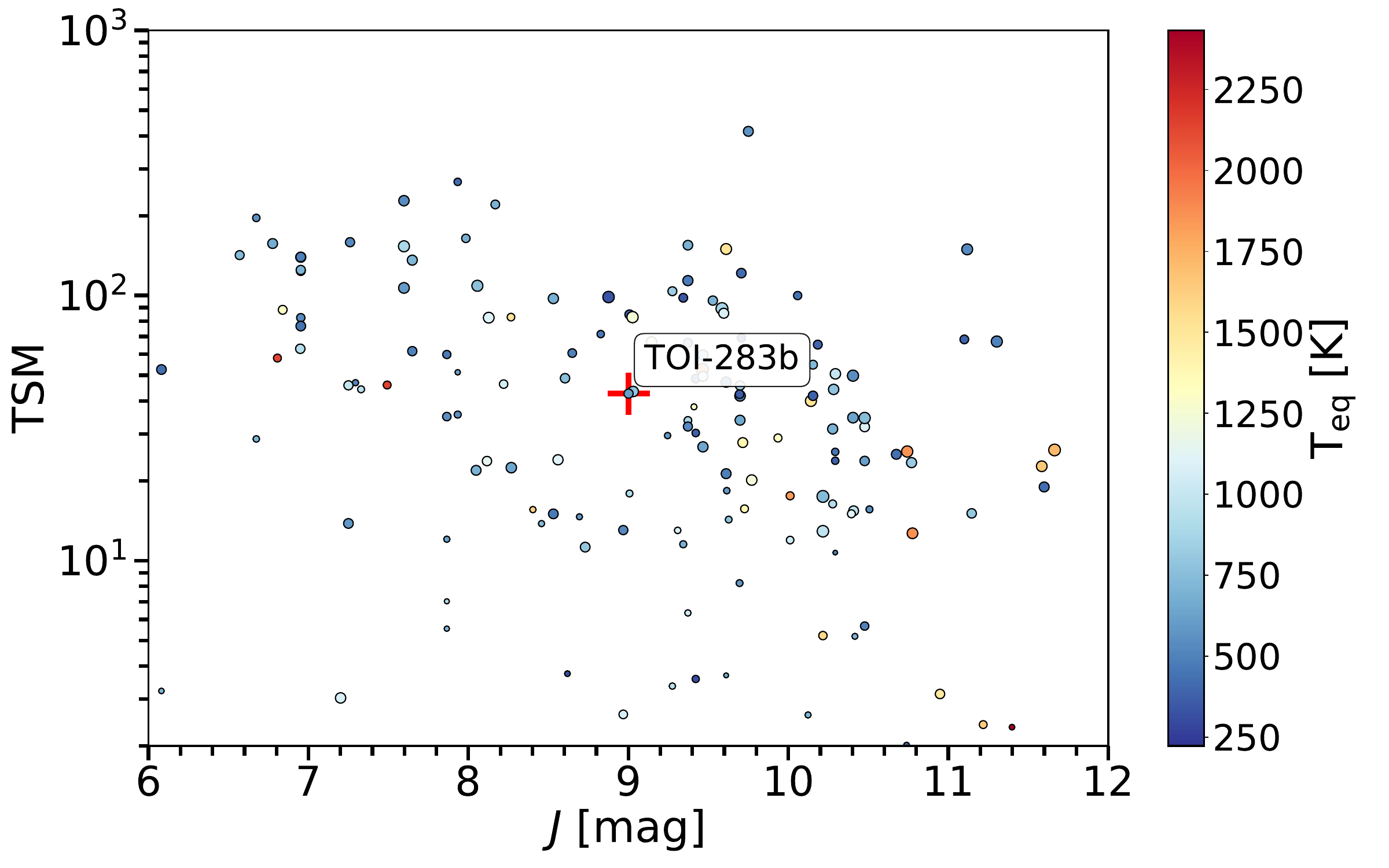}
    \caption{Transmission spectroscopy metric (TSM) versus $J$ band apparent magnitude of known transiting planets with radii of $\mathrm{R}_\mathrm{p} < 4 \; \mathrm{R}_\oplus$.}
    \label{Fig:TSmmetrics}
\end{figure}

To evaluate the potential for atmospheric characterization of TOI-283\,b, we calculated the transmission spectroscopy metric (TSM; \citealt{Kempton2018}), which is proportional to the expected signal-to-noise ratio based on the strength of spectral features. We retrieved a TSM value of 35, which is below the recommended threshold value of 90 for prioritizing planets in the parameter space of TOI-283\,b. For completeness the emission spectroscopy metric (ESM) is 1.8, very low, as is expected for a small warm planet. In Figure~\ref{Fig:TSmmetrics} we plot the TSM value against the J magnitude of the host star, from all known small ($\mathrm{R}_\mathrm{p} < 4 \; \mathrm{R}_\oplus$) planets with radius measurements, including TOI-283\,b. This means that transmission spectroscopy with JWST is possible but challenging.

Although JWST observations can be challenging, if this space-based telescope were able to probe the atmosphere of TOI-283\,b through transmission spectroscopy, the resulting molecular detections could offer valuable insights into its interior structure and water content. Planets with radii between 1 and 2.6~$\mathrm{R}_\oplus$, masses between 1 and 10~$\mathrm{M}_\oplus$, and H$_2$-rich atmospheres containing significant amounts of water in the form of oceans -- the so-called Hycean planets \citep{Madhusudhan2021} -- may exhibit distinctive atmospheric signatures. Several studies suggest that Hycean atmospheres can show enhancements in CO$_2$, H$_2$O, and/or CH$_4$, depletion of NH$_3$ due to dissolution in a liquid-water ocean, and enhanced CH$_3$OH abundances \citep[e.g.,][]{Hu2021, Tsai2021, Madhusudhan2023, Holmberg2024}. The detection (or non-detection) of these species could help differentiate between a water-rich world and one with an Earth-like composition. An additional caveat for TOI-283\,b is its high equilibrium temperature ($\mathrm{T_{eq}} \sim 660\,\mathrm{K}$), which suggests that any water present is likely to be predominantly in the gas phase rather than in liquid form (see previous section). Therefore, predictions developed for Hycean planets near the habitable zone may not be fully applicable to TOI-283\,b.

\section{Conclusions}
\label{Sec:Conclusions}
We report the discovery of TOI-283\,b, a sub-Neptunian planet orbiting a K star with an orbital period of 17.6 days. The planet candidate was first detected by NASA's TESS mission; ground-based observations confirmed that the star has no detectable stellar companions and that the transits occur on the star TOI-283. High-precision RV measurements made with the VLT's ESPRESSO instrument enabled the mass of the planet candidate to be determined.

We have obtained the stellar parameters of the planet's host star by analyzing a combined high-resolution ESPRESSO stellar spectrum. Our analysis indicates that TOI-283 has an effective temperature of $T_\mathrm{eff} = 5213 \pm 70$ K, an iron abundance of $[\mathrm{Fe/H}] = -0.09 \pm 0.05$ dex, and a derived stellar mass and radius of $\mathrm{M}_\star = 0.80 \pm 0.01 \; \mathrm{M}_\odot$ and $\mathrm{R}_\star = 0.85 \pm 0.03 \; \mathrm{R}_\odot$, respectively. From the stellar activity indicators derived from the spectroscopic observations, we estimate a stellar rotation period of about 30 days, with the FWHM indicator having the smallest uncertainty at $P_\mathrm{rot} = 30.0^{+5.8}_{-4.1}$ days.

To determine the main orbital parameters of the planet, we jointly fit the light curves of TESS (36 sectors) and 84 RV measurements taken with the ESPRESSO spectrograph. The TESS and ESPRESSO datasets were fit simultaneously using an MCMC procedure, with both time series taking into account the contribution of red noise sources modeled by Gaussian processes. From our best-fit model, we measured an orbital period of $P = 17.61745 \pm 0.00002$ days and derived a planetary radius of $\mathrm{R}_\mathrm{p} = 2.34 \pm 0.09 \; \mathrm{R}_\oplus$ and a mass of $\mathrm{M}_\mathrm{p} = 6.54 \pm 2.04 \; M_\oplus$. These values allow us to classify TOI-283\,b as a mini-Neptune. The position in the mass-radius diagram of this planet and theoretical compositional models indicate that this planet has a significant contribution from light elements and is consistent with an extended H/He-rich atmosphere and/or significant water content. Given the possibility that TOI-283\,b is a water-rich planet, atmospheric characterization with JWST could offer valuable insights into the presence of water in its atmosphere.

The data and model selection criteria indicate the presence of a long-term trend in the RV time series. During the fitting procedure, we modeled this signal with a second Keplerian component. For this long-term trend, we derived a period of $P = 1356^{+498}_{-179}$ days and a RV semi-amplitude of $K = 9.4^{+2.3}_{-1.6}$ m s$^{-1}$. The GLS periodograms of the stellar activity indices show that several indicators exhibit peaks above the 1\,\% FAP threshold at periods compatible with that of the long-term RV signal, suggesting that the trend could be driven by a stellar activity cycle. Conversely, if the signal originated from an additional companion in the system, the object would have a minimum mass of $\mathrm{M}\mathrm{p} \sin i = 0.44^{+0.12}_{-0.08} \; \mathrm{M}_\mathrm{Jup}$, placing it within the planetary regime. Further RV observations are required to determine whether this signal is of planetary origin or induced by stellar variability.

\begin{acknowledgements}
    
    Based on observations collected at the European Organisation for Astronomical Research in the Southern Hemisphere under ESO programme 1102.C-0744.
    
    This paper includes data collected by the TESS mission, which are publicly available from the Mikulski Archive for Space Telescopes (MAST). Funding for the TESS mission is provided by NASA's Science Mission Directorate. Resources supporting this work were provided by the NASA High-End Computing (HEC) Program through the NASA Advanced Supercomputing (NAS) Division at Ames Research Center for the production of the SPOC data products. We acknowledge the use of public TESS data from pipelines at the TESS Science Office and at the TESS Science Processing Operations Center. This research has made use of the Exoplanet Follow-up Observation Program website, which is operated by the California Institute of Technology, under contract with the National Aeronautics and Space Administration under the Exoplanet Exploration Program. 
    
    This research has made use of the Exoplanet Follow-up Observation Program (ExoFOP; DOI: 10.26134/ExoFOP5) website, which is operated by the California Institute of Technology, under contract with the National Aeronautics and Space Administration under the Exoplanet Exploration Program.
    
    This work makes use of observations from the LCOGT network. Part of the LCOGT telescope time was granted by NOIRLab through the Mid-Scale Innovations Program (MSIP). MSIP is funded by NSF.
    
    Based in part on observations obtained at the Southern Astrophysical Research (SOAR) telescope, which is a joint project of the Minist\'{e}rio da Ci\^{e}ncia, Tecnologia e Inova\c{c}\~{o}es (MCTI/LNA) do Brasil, the US National Science Foundation’s NOIRLab, the University of North Carolina at Chapel Hill (UNC), and Michigan State University (MSU).
    
    Some of the observations in the paper made use of the High-Resolution Imaging instrument Zorro. Zorro was funded by the NASA Exoplanet Exploration Program and built at the NASA Ames Research Center by Steve B. Howell, Nic Scott, Elliott P. Horch, and Emmett Quigley. Data were reduced using a software pipeline originally written by Elliott Horch and Mark Everett. Zorro was mounted on the Gemini South telescope of the international Gemini Observatory, a program of NSF NOIRLab, which is managed by the Association of Universities for Research in Astronomy (AURA) under a cooperative agreement with the U.S. National Science Foundation on behalf of the Gemini partnership: the U.S. National Science Foundation (United States), National Research Council (Canada), Agencia Nacional de Investigaci\'{o}n y Desarrollo (Chile), Ministerio de Ciencia, Tecnolog\'{i}a e Innovaci\'{o}n (Argentina), Minist\'{e}rio da Ci\^{e}ncia, Tecnologia, Inova\c{c}\~{o}es e Comunica\c{c}\~{o}es (Brazil), and Korea Astronomy and Space Science Institute (Republic of Korea).
    
    This work has made use of data from the European Space Agency (ESA) mission {\it Gaia} (\url{https://www.cosmos.esa.int/gaia}), processed by the {\it Gaia} Data Processing and Analysis Consortium (DPAC,
\url{https://www.cosmos.esa.int/web/gaia/dpac/consortium}). Funding for the DPAC has been provided by national institutions, in particular the institutions participating in the {\it Gaia} Multilateral Agreement.
    
    We thank the Swiss National Science Foundation (SNSF) and the Geneva University for their continuous support to our planet low-mass companion search programmes. This work has been carried out within the framework of the National Centre of Competence in Research PlanetS supported by the Swiss National Science Foundation.

    The authors acknowledge support from the Swiss NCCR PlanetS and the Swiss National Science Foundation. This work has been carried out within the framework of the NCCR PlanetS supported by the Swiss National Science Foundation under grants 51NF40182901 and 51NF40205606.
    
    F. M. acknowledges financial support from the Agencia Estatal de Investigaci\'{o}n del Ministerio de Ciencia, Innovaci\'{o}n y Universidades (MCIU/AEI) through grant PID2023-152906NA-I00.
    
    J.K. acknowledges support from the Swedish Research Council (Project Grant 2022-04043) and of the Swiss National Science Foundation under grant number TMSGI2\_211697.
    
    F.J.P acknowledges financial support from the Severo Ochoa grant CEX2021-001131-S funded by MCIN/AEI/10.13039/501100011033 and Ministerio de Ciencia e Innovaci\'on through the project PID2022-137241NB-C43.
    
    A.C.-G. is funded by the Spanish Ministry of Science through MCIN/AEI/10.13039/501100011033 grant PID2019-107061GB-C61.
    
    KAC and CNW acknowledge support from the TESS mission via subaward s3449 from MIT.
    
    The INAF authors acknowledge financial support of the Italian Ministry of Education, University, and Research with PRIN 201278X4FL and the "Progetti Premiali" funding scheme.
    
    X.D acknowledges the support from the European Research Council (ERC) under the European Union’s Horizon 2020 research and innovation programme (grant agreement SCORE No 851555) and from the Swiss National Science Foundation under the grant SPECTRE (No 200021\textunderscore215200).
    
    This work of CJM was financed by Portuguese funds through FCT (Funda\c c\~ao para a Ci\^encia e a Tecnologia) in the framework of the project 2022.04048.PTDC (Phi in the Sky, DOI 10.54499/2022.04048.PTDC). CJM also acknowledges FCT and POCH/FSE (EC) support through Investigador FCT Contract 2021.01214.CEECIND/CP1658/CT0001 (DOI 10.54499/2021.01214.CEECIND/CP1658/CT0001).
    
    Co-funded by the European Union (ERC, FIERCE, 101052347). Views and opinions expressed are however those of the author(s) only and do not necessarily reflect those of the European Union or the European Research Council. Neither the European Union nor the granting authority can be held responsible for them. This work was supported by FCT - Funda\c{c}\~{a}o para a Ci\^encia e a Tecnologia through national funds and by FEDER through COMPETE2020 - Programa Operacional Competitividade e Internacionaliza\c{c}\~{a}o by these grants: UIDB/04434/2020; UIDP/04434/2020.
    
    JIGH, ASM, CAP, and RR acknowledge financial support from the Spanish Ministry of Science, Innovation and Universities (MICIU) projects PID2020-117493GB-I00 and PID2023-149982NB-I00.
    
\end{acknowledgements}

\bibliographystyle{aa}
\bibliography{references}

\begin{thebibliography}{124}
\expandafter\ifx\csname natexlab\endcsname\relax\def\natexlab#1{#1}\fi

\bibitem[{{Baglin} {et~al.}(2006){Baglin}, {Auvergne}, {Boisnard}, {Lam-Trong}, {Barge}, {Catala}, {Deleuil}, {Michel}, \& {Weiss}}]{Baglin2006}
{Baglin}, A., {Auvergne}, M., {Boisnard}, L., {et~al.} 2006, in 36th COSPAR Scientific Assembly, Vol.~36, 3749

\bibitem[{{Bailer-Jones} {et~al.}(2021){Bailer-Jones}, {Rybizki}, {Fouesneau}, {Demleitner}, \& {Andrae}}]{Bailer-Jones2021}
{Bailer-Jones}, C.~A.~L., {Rybizki}, J., {Fouesneau}, M., {Demleitner}, M., \& {Andrae}, R. 2021, \aj, 161, 147

\bibitem[{{Baranne} {et~al.}(1996){Baranne}, {Queloz}, {Mayor}, {Adrianzyk}, {Knispel}, {Kohler}, {Lacroix}, {Meunier}, {Rimbaud}, \& {Vin}}]{Baranne1996}
{Baranne}, A., {Queloz}, D., {Mayor}, M., {et~al.} 1996, \aaps, 119, 373

\bibitem[{{Barros} {et~al.}(2022){Barros}, {Demangeon}, {Alibert}, {Leleu}, {Adibekyan}, {Lovis}, {Bossini}, {Sousa}, {Hara}, {Bouchy}, {Lavie}, {Rodrigues}, {Gomes da Silva}, {Lillo-Box}, {Pepe}, {Tabernero}, {Zapatero Osorio}, {Sozzetti}, {Su{\'a}rez Mascare{\~n}o}, {Micela}, {Allende Prieto}, {Cristiani}, {Damasso}, {Di Marcantonio}, {Ehrenreich}, {Faria}, {Figueira}, {Gonz{\'a}lez Hern{\'a}ndez}, {Jenkins}, {Lo Curto}, {Martins}, {Micela}, {Nunes}, {Pall{\'e}}, {Santos}, {Rebolo}, {Seager}, {Twicken}, {Udry}, {Vanderspek}, \& {Winn}}]{Barros2022}
{Barros}, S.~C.~C., {Demangeon}, O.~D.~S., {Alibert}, Y., {et~al.} 2022, \aap, 665, A154

\bibitem[{{Baumeister} {et~al.}(2020){Baumeister}, {Padovan}, {Tosi}, {Montavon}, {Nettelmann}, {MacKenzie}, \& {Godolt}}]{Baumeister2020}
{Baumeister}, P., {Padovan}, S., {Tosi}, N., {et~al.} 2020, \apj, 889, 42

\bibitem[{{Baumeister} \& {Tosi}(2023)}]{BaumeisterTosi2023}
{Baumeister}, P. \& {Tosi}, N. 2023, \aap, 676, A106

\bibitem[{{Borucki} {et~al.}(2010){Borucki}, {Koch}, {Basri}, {Batalha}, {Brown}, {Caldwell}, {Caldwell}, {Christensen-Dalsgaard}, {Cochran}, {DeVore}, {Dunham}, {Dupree}, {Gautier}, {Geary}, {Gilliland}, {Gould}, {Howell}, {Jenkins}, {Kondo}, {Latham}, {Marcy}, {Meibom}, {Kjeldsen}, {Lissauer}, {Monet}, {Morrison}, {Sasselov}, {Tarter}, {Boss}, {Brownlee}, {Owen}, {Buzasi}, {Charbonneau}, {Doyle}, {Fortney}, {Ford}, {Holman}, {Seager}, {Steffen}, {Welsh}, {Rowe}, {Anderson}, {Buchhave}, {Ciardi}, {Walkowicz}, {Sherry}, {Horch}, {Isaacson}, {Everett}, {Fischer}, {Torres}, {Johnson}, {Endl}, {MacQueen}, {Bryson}, {Dotson}, {Haas}, {Kolodziejczak}, {Van Cleve}, {Chandrasekaran}, {Twicken}, {Quintana}, {Clarke}, {Allen}, {Li}, {Wu}, {Tenenbaum}, {Verner}, {Bruhweiler}, {Barnes}, \& {Prsa}}]{Borucki2010}
{Borucki}, W.~J., {Koch}, D., {Basri}, G., {et~al.} 2010, Science, 327, 977

\bibitem[{{Borucki} {et~al.}(2011){Borucki}, {Koch}, {Basri}, {Batalha}, {Brown}, {Bryson}, {Caldwell}, {Christensen-Dalsgaard}, {Cochran}, {DeVore}, {Dunham}, {Gautier}, {Geary}, {Gilliland}, {Gould}, {Howell}, {Jenkins}, {Latham}, {Lissauer}, {Marcy}, {Rowe}, {Sasselov}, {Boss}, {Charbonneau}, {Ciardi}, {Doyle}, {Dupree}, {Ford}, {Fortney}, {Holman}, {Seager}, {Steffen}, {Tarter}, {Welsh}, {Allen}, {Buchhave}, {Christiansen}, {Clarke}, {Das}, {D{\'e}sert}, {Endl}, {Fabrycky}, {Fressin}, {Haas}, {Horch}, {Howard}, {Isaacson}, {Kjeldsen}, {Kolodziejczak}, {Kulesa}, {Li}, {Lucas}, {Machalek}, {McCarthy}, {MacQueen}, {Meibom}, {Miquel}, {Prsa}, {Quinn}, {Quintana}, {Ragozzine}, {Sherry}, {Shporer}, {Tenenbaum}, {Torres}, {Twicken}, {Van Cleve}, {Walkowicz}, {Witteborn}, \& {Still}}]{Borucki2011}
{Borucki}, W.~J., {Koch}, D.~G., {Basri}, G., {et~al.} 2011, \apj, 736, 19

\bibitem[{{Bourrier} {et~al.}(2022){Bourrier}, {Zapatero Osorio}, {Allart}, {Attia}, {Cretignier}, {Dumusque}, {Lovis}, {Adibekyan}, {Borsa}, {Figueira}, {Gonz{\'a}lez Hern{\'a}ndez}, {Mehner}, {Santos}, {Schmidt}, {Seidel}, {Sozzetti}, {Alibert}, {Casasayas-Barris}, {Ehrenreich}, {Lo Curto}, {Martins}, {Di Marcantonio}, {M{\'e}gevand}, {Nunes}, {Palle}, {Poretti}, \& {Sousa}}]{Bourrier2022}
{Bourrier}, V., {Zapatero Osorio}, M.~R., {Allart}, R., {et~al.} 2022, \aap, 663, A160

\bibitem[{{Brown} {et~al.}(2013){Brown}, {Baliber}, {Bianco}, {Bowman}, {Burleson}, {Conway}, {Crellin}, {Depagne}, {De Vera}, {Dilday}, {Dragomir}, {Dubberley}, {Eastman}, {Elphick}, {Falarski}, {Foale}, {Ford}, {Fulton}, {Garza}, {Gomez}, {Graham}, {Greene}, {Haldeman}, {Hawkins}, {Haworth}, {Haynes}, {Hidas}, {Hjelstrom}, {Howell}, {Hygelund}, {Lister}, {Lobdill}, {Martinez}, {Mullins}, {Norbury}, {Parrent}, {Paulson}, {Petry}, {Pickles}, {Posner}, {Rosing}, {Ross}, {Sand}, {Saunders}, {Shobbrook}, {Shporer}, {Street}, {Thomas}, {Tsapras}, {Tufts}, {Valenti}, {Vander Horst}, {Walker}, {White}, \& {Willis}}]{Brown:2013}
{Brown}, T.~M., {Baliber}, N., {Bianco}, F.~B., {et~al.} 2013, \pasp, 125, 1031

\bibitem[{{Burn} {et~al.}(2024){Burn}, {Mordasini}, {Mishra}, {Haldemann}, {Venturini}, {Emsenhuber}, \& {Henning}}]{Burn2024}
{Burn}, R., {Mordasini}, C., {Mishra}, L., {et~al.} 2024, Nature Astronomy, 8, 463

\bibitem[{{Castro-Gonz{\'a}lez} {et~al.}(2024){Castro-Gonz{\'a}lez}, {Lillo-Box}, {Armstrong}, {Acu{\~n}a}, {Aguichine}, {Bourrier}, {Gandhi}, {Sousa}, {Delgado-Mena}, {Moya}, {Adibekyan}, {Correia}, {Barrado}, {Damasso}, {Winn}, {Santos}, {Barkaoui}, {Barros}, {Benkhaldoun}, {Bouchy}, {Brice{\~n}o}, {Caldwell}, {Collins}, {Essack}, {Ghachoui}, {Gillon}, {Hounsell}, {Jehin}, {Jenkins}, {Keniger}, {Law}, {Mann}, {Nielsen}, {Pozuelos}, {Schanche}, {Seager}, {Tan}, {Timmermans}, {Villase{\~n}or}, {Watkins}, \& {Ziegler}}]{CastroGonzalez2024}
{Castro-Gonz{\'a}lez}, A., {Lillo-Box}, J., {Armstrong}, D.~J., {et~al.} 2024, \aap, 691, A233

\bibitem[{{Chen} {et~al.}(2019){Chen}, {Girardi}, {Fu}, {Bressan}, {Aringer}, {Dal Tio}, {Pastorelli}, {Marigo}, {Costa}, \& {Zhang}}]{Chen2019}
{Chen}, Y., {Girardi}, L., {Fu}, X., {et~al.} 2019, \aap, 632, A105

\bibitem[{Christiansen {et~al.}(2023)Christiansen, Zink, Hardegree-Ullman, Fernandes, Hopkins, Rebull, Boley, Bergsten, \& Bhure}]{Christiansen2023}
Christiansen, J.~L., Zink, J.~K., Hardegree-Ullman, K.~K., {et~al.} 2023, The Astronomical Journal, 166, 248

\bibitem[{{Collins}(2019)}]{collins:2019}
{Collins}, K. 2019, in American Astronomical Society Meeting Abstracts, Vol. 233, American Astronomical Society Meeting Abstracts \#233, 140.05

\bibitem[{{Collins} {et~al.}(2017){Collins}, {Kielkopf}, {Stassun}, \& {Hessman}}]{Collins:2017}
{Collins}, K.~A., {Kielkopf}, J.~F., {Stassun}, K.~G., \& {Hessman}, F.~V. 2017, \aj, 153, 77

\bibitem[{{Cutri} {et~al.}(2021){Cutri}, {Wright}, {Conrow}, {Fowler}, {Eisenhardt}, {Grillmair}, {Kirkpatrick}, {Masci}, {McCallon}, {Wheelock}, {Fajardo-Acosta}, {Yan}, {Benford}, {Harbut}, {Jarrett}, {Lake}, {Leisawitz}, {Ressler}, {Stanford}, {Tsai}, {Liu}, {Helou}, {Mainzer}, {Gettngs}, {Gonzalez}, {Hoffman}, {Marsh}, {Padgett}, {Skrutskie}, {Beck}, {Papin}, \& {Wittman}}]{CutriWISECat2014}
{Cutri}, R.~M., {Wright}, E.~L., {Conrow}, T., {et~al.} 2021, VizieR Online Data Catalog, II/328

\bibitem[{{da Silva} {et~al.}(2006){da Silva}, {Girardi}, {Pasquini}, {Setiawan}, {von der L{\"u}he}, {de Medeiros}, {Hatzes}, {D{\"o}llinger}, \& {Weiss}}]{PARAM}
{da Silva}, L., {Girardi}, L., {Pasquini}, L., {et~al.} 2006, A\&A, 458, 609

\bibitem[{{Damasso} {et~al.}(2023){Damasso}, {Rodrigues}, {Castro-Gonz{\'a}lez}, {Lavie}, {Davoult}, {Zapatero Osorio}, {Dou}, {Sousa}, {Owen}, {Sossi}, {Adibekyan}, {Osborn}, {Leinhardt}, {Alibert}, {Lovis}, {Delgado Mena}, {Sozzetti}, {Barros}, {Bossini}, {Ziegler}, {Ciardi}, {Matthews}, {Carter}, {Lillo-Box}, {Su{\'a}rez Mascare{\~n}o}, {Cristiani}, {Pepe}, {Rebolo}, {Santos}, {Allende Prieto}, {Benatti}, {Bouchy}, {Brice{\~n}o}, {Di Marcantonio}, {D'Odorico}, {Dumusque}, {Egger}, {Ehrenreich}, {Faria}, {Figueira}, {G{\'e}nova Santos}, {Gonzales}, {Gonz{\'a}lez Hern{\'a}ndez}, {Law}, {Lo Curto}, {Mann}, {Martins}, {Mehner}, {Micela}, {Molaro}, {Nunes}, {Palle}, {Poretti}, {Schlieder}, \& {Udry}}]{Damasso2023}
{Damasso}, M., {Rodrigues}, J., {Castro-Gonz{\'a}lez}, A., {et~al.} 2023, \aap, 679, A33

\bibitem[{{Delrez} {et~al.}(2022){Delrez}, {Murray}, {Pozuelos}, {Narita}, {Ducrot}, {Timmermans}, {Watanabe}, {Burgasser}, {Hirano}, {Rackham}, {Stassun}, {Van Grootel}, {Aganze}, {Cointepas}, {Howell}, {Kaltenegger}, {Niraula}, {Sebastian}, {Almenara}, {Barkaoui}, {Baycroft}, {Bonfils}, {Bouchy}, {Burdanov}, {Caldwell}, {Charbonneau}, {Ciardi}, {Collins}, {Daylan}, {Demory}, {de Wit}, {Dransfield}, {Fajardo-Acosta}, {Fausnaugh}, {Fukui}, {Furlan}, {Garcia}, {Gnilka}, {G{\'o}mez Maqueo Chew}, {G{\'o}mez-Mu{\~n}oz}, {G{\"u}nther}, {Harakawa}, {Heng}, {Hooton}, {Hori}, {Ikoma}, {Jehin}, {Jenkins}, {Kagetani}, {Kawauchi}, {Kimura}, {Kodama}, {Kotani}, {Krishnamurthy}, {Kudo}, {Kunovac}, {Kusakabe}, {Latham}, {Littlefield}, {McCormac}, {Melis}, {Mori}, {Murgas}, {Palle}, {Pedersen}, {Queloz}, {Ricker}, {Sabin}, {Schanche}, {Schroffenegger}, {Seager}, {Shiao}, {Sohy}, {Standing}, {Tamura}, {Theissen}, {Thompson}, {Triaud}, {Vanderspek}, {Vievard}, {Wells}, {Winn}, {Zou}, {Z{\'u}{\~n}iga-Fern{\'a}ndez}, \&
  {Gillon}}]{delrez2022}
{Delrez}, L., {Murray}, C.~A., {Pozuelos}, F.~J., {et~al.} 2022, \aap, 667, A59

\bibitem[{{Demangeon} {et~al.}(2021){Demangeon}, {Zapatero Osorio}, {Alibert}, {Barros}, {Adibekyan}, {Tabernero}, {Antoniadis-Karnavas}, {Camacho}, {Su{\'a}rez Mascare{\~n}o}, {Oshagh}, {Micela}, {Sousa}, {Lovis}, {Pepe}, {Rebolo}, {Cristiani}, {Santos}, {Allart}, {Allende Prieto}, {Bossini}, {Bouchy}, {Cabral}, {Damasso}, {Di Marcantonio}, {D'Odorico}, {Ehrenreich}, {Faria}, {Figueira}, {G{\'e}nova Santos}, {Haldemann}, {Hara}, {Gonz{\'a}lez Hern{\'a}ndez}, {Lavie}, {Lillo-Box}, {Lo Curto}, {Martins}, {M{\'e}gevand}, {Mehner}, {Molaro}, {Nunes}, {Pall{\'e}}, {Pasquini}, {Poretti}, {Sozzetti}, \& {Udry}}]{Demangeon2021}
{Demangeon}, O.~D.~S., {Zapatero Osorio}, M.~R., {Alibert}, Y., {et~al.} 2021, \aap, 653, A41

\bibitem[{{Demory} {et~al.}(2020){Demory}, {Pozuelos}, {G{\'o}mez Maqueo Chew}, {Sabin}, {Petrucci}, {Schroffenegger}, {Grimm}, {Sestovic}, {Gillon}, {McCormac}, {Barkaoui}, {Benz}, {Bieryla}, {Bouchy}, {Burdanov}, {Collins}, {de Wit}, {Dressing}, {Garcia}, {Giacalone}, {Guerra}, {Haldemann}, {Heng}, {Jehin}, {Jofr{\'e}}, {Kane}, {Lillo-Box}, {Maign{\'e}}, {Mordasini}, {Morris}, {Niraula}, {Queloz}, {Rackham}, {Savel}, {Soubkiou}, {Srdoc}, {Stassun}, {Triaud}, {Zambelli}, {Ricker}, {Latham}, {Seager}, {Winn}, {Jenkins}, {Calvario-Vel{\'a}squez}, {Franco Herrera}, {Colorado}, {Cadena Zepeda}, {Figueroa}, {Watson}, {Lugo-Ibarra}, {Carigi}, {Guisa}, {Herrera}, {Sierra D{\'\i}az}, {Su{\'a}rez}, {Barrado}, {Batalha}, {Benkhaldoun}, {Chontos}, {Dai}, {Essack}, {Ghachoui}, {Huang}, {Huber}, {Isaacson}, {Lissauer}, {Morales-Calder{\'o}n}, {Robertson}, {Roy}, {Twicken}, {Vanderburg}, \& {Weiss}}]{demory2020}
{Demory}, B.~O., {Pozuelos}, F.~J., {G{\'o}mez Maqueo Chew}, Y., {et~al.} 2020, \aap, 642, A49

\bibitem[{{D{\'e}vora-Pajares} \& {Pozuelos}(2022)}]{matrix}
{D{\'e}vora-Pajares}, M. \& {Pozuelos}, F.~J. 2022, {MATRIX: Multi-phAse Transits Recovery from Injected eXoplanets}

\bibitem[{{D{\'e}vora-Pajares} {et~al.}(2024){D{\'e}vora-Pajares}, {Pozuelos}, {Thuillier}, {Timmermans}, {Van Grootel}, {Bonidie}, {Mota}, \& {Su{\'a}rez}}]{devora2024}
{D{\'e}vora-Pajares}, M., {Pozuelos}, F.~J., {Thuillier}, A., {et~al.} 2024, \mnras, 532, 4752

\bibitem[{{Fellgett}(1955)}]{Fellgett1955}
{Fellgett}, P. 1955, Optica Acta, 2, 9

\bibitem[{{Foreman-Mackey} {et~al.}(2017){Foreman-Mackey}, {Agol}, {Ambikasaran}, \& {Angus}}]{ForemanMackey2017}
{Foreman-Mackey}, D., {Agol}, E., {Ambikasaran}, S., \& {Angus}, R. 2017, \aj, 154, 220

\bibitem[{{Foreman-Mackey} {et~al.}(2013){Foreman-Mackey}, {Hogg}, {Lang}, \& {Goodman}}]{ForemanMackey2013}
{Foreman-Mackey}, D., {Hogg}, D.~W., {Lang}, D., \& {Goodman}, J. 2013, \pasp, 125, 306

\bibitem[{{Fulton} \& {Petigura}(2018)}]{Fulton2018b}
{Fulton}, B.~J. \& {Petigura}, E.~A. 2018, \aj, 156, 264

\bibitem[{{Fulton} {et~al.}(2018){Fulton}, {Petigura}, {Blunt}, \& {Sinukoff}}]{Fulton2018}
{Fulton}, B.~J., {Petigura}, E.~A., {Blunt}, S., \& {Sinukoff}, E. 2018, \pasp, 130, 044504

\bibitem[{{Fulton} {et~al.}(2017){Fulton}, {Petigura}, {Howard}, {Isaacson}, {Marcy}, {Cargile}, {Hebb}, {Weiss}, {Johnson}, {Morton}, {Sinukoff}, {Crossfield}, \& {Hirsch}}]{Fulton17}
{Fulton}, B.~J., {Petigura}, E.~A., {Howard}, A.~W., {et~al.} 2017, \aj, 154, 109

\bibitem[{{Gaia Collaboration} {et~al.}(2018){Gaia Collaboration}, {Brown}, {Vallenari}, {Prusti}, {de Bruijne}, {Babusiaux}, {Bailer-Jones}, {Biermann}, {Evans}, {Eyer}, \& et~al.}]{GaiaDR22018}
{Gaia Collaboration}, {Brown}, A.~G.~A., {Vallenari}, A., {et~al.} 2018, \aap, 616, A1

\bibitem[{{Gaia Collaboration} {et~al.}(2021){Gaia Collaboration}, {Brown}, {Vallenari}, {Prusti}, {de Bruijne}, {Babusiaux}, {Biermann}, {Creevey}, {Evans}, {Eyer}, {Hutton}, {Jansen}, {Jordi}, {Klioner}, {Lammers}, {Lindegren}, {Luri}, {Mignard}, {Panem}, {Pourbaix}, {Randich}, {Sartoretti}, {Soubiran}, {Walton}, {Arenou}, {Bailer-Jones}, {Bastian}, {Cropper}, {Drimmel}, {Katz}, {Lattanzi}, {van Leeuwen}, {Bakker}, {Cacciari}, {Casta{\~n}eda}, {De Angeli}, {Ducourant}, {Fabricius}, {Fouesneau}, {Fr{\'e}mat}, {Guerra}, {Guerrier}, {Guiraud}, {Jean-Antoine Piccolo}, {Masana}, {Messineo}, {Mowlavi}, {Nicolas}, {Nienartowicz}, {Pailler}, {Panuzzo}, {Riclet}, {Roux}, {Seabroke}, {Sordo}, {Tanga}, {Th{\'e}venin}, {Gracia-Abril}, {Portell}, {Teyssier}, {Altmann}, {Andrae}, {Bellas-Velidis}, {Benson}, {Berthier}, {Blomme}, {Brugaletta}, {Burgess}, {Busso}, {Carry}, {Cellino}, {Cheek}, {Clementini}, {Damerdji}, {Davidson}, {Delchambre}, {Dell'Oro}, {Fern{\'a}ndez-Hern{\'a}ndez}, {Galluccio}, {Garc{\'\i}a-Lario},
  {Garcia-Reinaldos}, {Gonz{\'a}lez-N{\'u}{\~n}ez}, {Gosset}, {Haigron}, {Halbwachs}, {Hambly}, {Harrison}, {Hatzidimitriou}, {Heiter}, {Hern{\'a}ndez}, {Hestroffer}, {Hodgkin}, {Holl}, {Jan{\ss}en}, {Jevardat de Fombelle}, {Jordan}, {Krone-Martins}, {Lanzafame}, {L{\"o}ffler}, {Lorca}, {Manteiga}, {Marchal}, {Marrese}, {Moitinho}, {Mora}, {Muinonen}, {Osborne}, {Pancino}, {Pauwels}, {Petit}, {Recio-Blanco}, {Richards}, {Riello}, {Rimoldini}, {Robin}, {Roegiers}, {Rybizki}, {Sarro}, {Siopis}, {Smith}, {Sozzetti}, {Ulla}, {Utrilla}, {van Leeuwen}, {van Reeven}, {Abbas}, {Abreu Aramburu}, {Accart}, {Aerts}, {Aguado}, {Ajaj}, {Altavilla}, {{\'A}lvarez}, {{\'A}lvarez Cid-Fuentes}, {Alves}, {Anderson}, {Anglada Varela}, {Antoja}, {Audard}, {Baines}, {Baker}, {Balaguer-N{\'u}{\~n}ez}, {Balbinot}, {Balog}, {Barache}, {Barbato}, {Barros}, {Barstow}, {Bartolom{\'e}}, {Bassilana}, {Bauchet}, {Baudesson-Stella}, {Becciani}, {Bellazzini}, {Bernet}, {Bertone}, {Bianchi}, {Blanco-Cuaresma}, {Boch}, {Bombrun}, {Bossini},
  {Bouquillon}, {Bragaglia}, {Bramante}, {Breedt}, {Bressan}, {Brouillet}, {Bucciarelli}, {Burlacu}, {Busonero}, {Butkevich}, {Buzzi}, {Caffau}, {Cancelliere}, {C{\'a}novas}, {Cantat-Gaudin}, {Carballo}, {Carlucci}, {Carnerero}, {Carrasco}, {Casamiquela}, {Castellani}, {Castro-Ginard}, {Castro Sampol}, {Chaoul}, {Charlot}, {Chemin}, {Chiavassa}, {Cioni}, {Comoretto}, {Cooper}, {Cornez}, {Cowell}, {Crifo}, {Crosta}, {Crowley}, {Dafonte}, {Dapergolas}, {David}, {David}, {de Laverny}, {De Luise}, {De March}, {De Ridder}, {de Souza}, {de Teodoro}, {de Torres}, {del Peloso}, {del Pozo}, {Delbo}, {Delgado}, {Delgado}, {Delisle}, {Di Matteo}, {Diakite}, {Diener}, {Distefano}, {Dolding}, {Eappachen}, {Edvardsson}, {Enke}, {Esquej}, {Fabre}, {Fabrizio}, {Faigler}, {Fedorets}, {Fernique}, {Fienga}, {Figueras}, {Fouron}, {Fragkoudi}, {Fraile}, {Franke}, {Gai}, {Garabato}, {Garcia-Gutierrez}, {Garc{\'\i}a-Torres}, {Garofalo}, {Gavras}, {Gerlach}, {Geyer}, {Giacobbe}, {Gilmore}, {Girona}, {Giuffrida}, {Gomel}, {Gomez},
  {Gonzalez-Santamaria}, {Gonz{\'a}lez-Vidal}, {Granvik}, {Guti{\'e}rrez-S{\'a}nchez}, {Guy}, {Hauser}, {Haywood}, {Helmi}, {Hidalgo}, {Hilger}, {H{\l}adczuk}, {Hobbs}, {Holland}, {Huckle}, {Jasniewicz}, {Jonker}, {Juaristi Campillo}, {Julbe}, {Karbevska}, {Kervella}, {Khanna}, {Kochoska}, {Kontizas}, {Kordopatis}, {Korn}, {Kostrzewa-Rutkowska}, {Kruszy{\'n}ska}, {Lambert}, {Lanza}, {Lasne}, {Le Campion}, {Le Fustec}, {Lebreton}, {Lebzelter}, {Leccia}, {Leclerc}, {Lecoeur-Taibi}, {Liao}, {Licata}, {Lindstr{\o}m}, {Lister}, {Livanou}, {Lobel}, {Madrero Pardo}, {Managau}, {Mann}, {Marchant}, {Marconi}, {Marcos Santos}, {Marinoni}, {Marocco}, {Marshall}, {Martin Polo}, {Mart{\'\i}n-Fleitas}, {Masip}, {Massari}, {Mastrobuono-Battisti}, {Mazeh}, {McMillan}, {Messina}, {Michalik}, {Millar}, {Mints}, {Molina}, {Molinaro}, {Moln{\'a}r}, {Montegriffo}, {Mor}, {Morbidelli}, {Morel}, {Morris}, {Mulone}, {Munoz}, {Muraveva}, {Murphy}, {Musella}, {Noval}, {Ord{\'e}novic}, {Orr{\`u}}, {Osinde}, {Pagani}, {Pagano},
  {Palaversa}, {Palicio}, {Panahi}, {Pawlak}, {Pe{\~n}alosa Esteller}, {Penttil{\"a}}, {Piersimoni}, {Pineau}, {Plachy}, {Plum}, {Poggio}, {Poretti}, {Poujoulet}, {Pr{\v{s}}a}, {Pulone}, {Racero}, {Ragaini}, {Rainer}, {Raiteri}, {Rambaux}, {Ramos}, {Ramos-Lerate}, {Re Fiorentin}, {Regibo}, {Reyl{\'e}}, {Ripepi}, {Riva}, {Rixon}, {Robichon}, {Robin}, {Roelens}, {Rohrbasser}, {Romero-G{\'o}mez}, {Rowell}, {Royer}, {Rybicki}, {Sadowski}, {Sagrist{\`a} Sell{\'e}s}, {Sahlmann}, {Salgado}, {Salguero}, {Samaras}, {Sanchez Gimenez}, {Sanna}, {Santove{\~n}a}, {Sarasso}, {Schultheis}, {Sciacca}, {Segol}, {Segovia}, {S{\'e}gransan}, {Semeux}, {Shahaf}, {Siddiqui}, {Siebert}, {Siltala}, {Slezak}, {Smart}, {Solano}, {Solitro}, {Souami}, {Souchay}, {Spagna}, {Spoto}, {Steele}, {Steidelm{\"u}ller}, {Stephenson}, {S{\"u}veges}, {Szabados}, {Szegedi-Elek}, {Taris}, {Tauran}, {Taylor}, {Teixeira}, {Thuillot}, {Tonello}, {Torra}, {Torra}, {Turon}, {Unger}, {Vaillant}, {van Dillen}, {Vanel}, {Vecchiato}, {Viala}, {Vicente},
  {Voutsinas}, {Weiler}, {Wevers}, {Wyrzykowski}, {Yoldas}, {Yvard}, {Zhao}, {Zorec}, {Zucker}, {Zurbach}, \& {Zwitter}}]{GAIAEDR3}
{Gaia Collaboration}, {Brown}, A.~G.~A., {Vallenari}, A., {et~al.} 2021, \aap, 649, A1

\bibitem[{{Gaia Collaboration} {et~al.}(2023){Gaia Collaboration}, {Vallenari}, {Brown}, {Prusti}, {de Bruijne}, {Arenou}, {Babusiaux}, {Biermann}, {Creevey}, {Ducourant}, {Evans}, {Eyer}, {Guerra}, {Hutton}, {Jordi}, {Klioner}, {Lammers}, {Lindegren}, {Luri}, {Mignard}, {Panem}, {Pourbaix}, {Randich}, {Sartoretti}, {Soubiran}, {Tanga}, {Walton}, {Bailer-Jones}, {Bastian}, {Drimmel}, {Jansen}, {Katz}, {Lattanzi}, {van Leeuwen}, {Bakker}, {Cacciari}, {Casta{\~n}eda}, {De Angeli}, {Fabricius}, {Fouesneau}, {Fr{\'e}mat}, {Galluccio}, {Guerrier}, {Heiter}, {Masana}, {Messineo}, {Mowlavi}, {Nicolas}, {Nienartowicz}, {Pailler}, {Panuzzo}, {Riclet}, {Roux}, {Seabroke}, {Sordo}, {Th{\'e}venin}, {Gracia-Abril}, {Portell}, {Teyssier}, {Altmann}, {Andrae}, {Audard}, {Bellas-Velidis}, {Benson}, {Berthier}, {Blomme}, {Burgess}, {Busonero}, {Busso}, {C{\'a}novas}, {Carry}, {Cellino}, {Cheek}, {Clementini}, {Damerdji}, {Davidson}, {de Teodoro}, {Nu{\~n}ez Campos}, {Delchambre}, {Dell'Oro}, {Esquej},
  {Fern{\'a}ndez-Hern{\'a}ndez}, {Fraile}, {Garabato}, {Garc{\'\i}a-Lario}, {Gosset}, {Haigron}, {Halbwachs}, {Hambly}, {Harrison}, {Hern{\'a}ndez}, {Hestroffer}, {Hodgkin}, {Holl}, {Jan{\ss}en}, {Jevardat de Fombelle}, {Jordan}, {Krone-Martins}, {Lanzafame}, {L{\"o}ffler}, {Marchal}, {Marrese}, {Moitinho}, {Muinonen}, {Osborne}, {Pancino}, {Pauwels}, {Recio-Blanco}, {Reyl{\'e}}, {Riello}, {Rimoldini}, {Roegiers}, {Rybizki}, {Sarro}, {Siopis}, {Smith}, {Sozzetti}, {Utrilla}, {van Leeuwen}, {Abbas}, {{\'A}brah{\'a}m}, {Abreu Aramburu}, {Aerts}, {Aguado}, {Ajaj}, {Aldea-Montero}, {Altavilla}, {{\'A}lvarez}, {Alves}, {Anders}, {Anderson}, {Anglada Varela}, {Antoja}, {Baines}, {Baker}, {Balaguer-N{\'u}{\~n}ez}, {Balbinot}, {Balog}, {Barache}, {Barbato}, {Barros}, {Barstow}, {Bartolom{\'e}}, {Bassilana}, {Bauchet}, {Becciani}, {Bellazzini}, {Berihuete}, {Bernet}, {Bertone}, {Bianchi}, {Binnenfeld}, {Blanco-Cuaresma}, {Blazere}, {Boch}, {Bombrun}, {Bossini}, {Bouquillon}, {Bragaglia}, {Bramante}, {Breedt},
  {Bressan}, {Brouillet}, {Brugaletta}, {Bucciarelli}, {Burlacu}, {Butkevich}, {Buzzi}, {Caffau}, {Cancelliere}, {Cantat-Gaudin}, {Carballo}, {Carlucci}, {Carnerero}, {Carrasco}, {Casamiquela}, {Castellani}, {Castro-Ginard}, {Chaoul}, {Charlot}, {Chemin}, {Chiaramida}, {Chiavassa}, {Chornay}, {Comoretto}, {Contursi}, {Cooper}, {Cornez}, {Cowell}, {Crifo}, {Cropper}, {Crosta}, {Crowley}, {Dafonte}, {Dapergolas}, {David}, {David}, {de Laverny}, {De Luise}, {De March}, {De Ridder}, {de Souza}, {de Torres}, {del Peloso}, {del Pozo}, {Delbo}, {Delgado}, {Delisle}, {Demouchy}, {Dharmawardena}, {Di Matteo}, {Diakite}, {Diener}, {Distefano}, {Dolding}, {Edvardsson}, {Enke}, {Fabre}, {Fabrizio}, {Faigler}, {Fedorets}, {Fernique}, {Fienga}, {Figueras}, {Fournier}, {Fouron}, {Fragkoudi}, {Gai}, {Garcia-Gutierrez}, {Garcia-Reinaldos}, {Garc{\'\i}a-Torres}, {Garofalo}, {Gavel}, {Gavras}, {Gerlach}, {Geyer}, {Giacobbe}, {Gilmore}, {Girona}, {Giuffrida}, {Gomel}, {Gomez}, {Gonz{\'a}lez-N{\'u}{\~n}ez},
  {Gonz{\'a}lez-Santamar{\'\i}a}, {Gonz{\'a}lez-Vidal}, {Granvik}, {Guillout}, {Guiraud}, {Guti{\'e}rrez-S{\'a}nchez}, {Guy}, {Hatzidimitriou}, {Hauser}, {Haywood}, {Helmer}, {Helmi}, {Sarmiento}, {Hidalgo}, {Hilger}, {H{\l}adczuk}, {Hobbs}, {Holland}, {Huckle}, {Jardine}, {Jasniewicz}, {Jean-Antoine Piccolo}, {Jim{\'e}nez-Arranz}, {Jorissen}, {Juaristi Campillo}, {Julbe}, {Karbevska}, {Kervella}, {Khanna}, {Kontizas}, {Kordopatis}, {Korn}, {K{\'o}sp{\'a}l}, {Kostrzewa-Rutkowska}, {Kruszy{\'n}ska}, {Kun}, {Laizeau}, {Lambert}, {Lanza}, {Lasne}, {Le Campion}, {Lebreton}, {Lebzelter}, {Leccia}, {Leclerc}, {Lecoeur-Taibi}, {Liao}, {Licata}, {Lindstr{\o}m}, {Lister}, {Livanou}, {Lobel}, {Lorca}, {Loup}, {Madrero Pardo}, {Magdaleno Romeo}, {Managau}, {Mann}, {Manteiga}, {Marchant}, {Marconi}, {Marcos}, {Marcos Santos}, {Mar{\'\i}n Pina}, {Marinoni}, {Marocco}, {Marshall}, {Martin Polo}, {Mart{\'\i}n-Fleitas}, {Marton}, {Mary}, {Masip}, {Massari}, {Mastrobuono-Battisti}, {Mazeh}, {McMillan}, {Messina}, {Michalik},
  {Millar}, {Mints}, {Molina}, {Molinaro}, {Moln{\'a}r}, {Monari}, {Mongui{\'o}}, {Montegriffo}, {Montero}, {Mor}, {Mora}, {Morbidelli}, {Morel}, {Morris}, {Muraveva}, {Murphy}, {Musella}, {Nagy}, {Noval}, {Oca{\~n}a}, {Ogden}, {Ordenovic}, {Osinde}, {Pagani}, {Pagano}, {Palaversa}, {Palicio}, {Pallas-Quintela}, {Panahi}, {Payne-Wardenaar}, {Pe{\~n}alosa Esteller}, {Penttil{\"a}}, {Pichon}, {Piersimoni}, {Pineau}, {Plachy}, {Plum}, {Poggio}, {Pr{\v{s}}a}, {Pulone}, {Racero}, {Ragaini}, {Rainer}, {Raiteri}, {Rambaux}, {Ramos}, {Ramos-Lerate}, {Re Fiorentin}, {Regibo}, {Richards}, {Rios Diaz}, {Ripepi}, {Riva}, {Rix}, {Rixon}, {Robichon}, {Robin}, {Robin}, {Roelens}, {Rogues}, {Rohrbasser}, {Romero-G{\'o}mez}, {Rowell}, {Royer}, {Ruz Mieres}, {Rybicki}, {Sadowski}, {S{\'a}ez N{\'u}{\~n}ez}, {Sagrist{\`a} Sell{\'e}s}, {Sahlmann}, {Salguero}, {Samaras}, {Sanchez Gimenez}, {Sanna}, {Santove{\~n}a}, {Sarasso}, {Schultheis}, {Sciacca}, {Segol}, {Segovia}, {S{\'e}gransan}, {Semeux}, {Shahaf}, {Siddiqui}, {Siebert},
  {Siltala}, {Silvelo}, {Slezak}, {Slezak}, {Smart}, {Snaith}, {Solano}, {Solitro}, {Souami}, {Souchay}, {Spagna}, {Spina}, {Spoto}, {Steele}, {Steidelm{\"u}ller}, {Stephenson}, {S{\"u}veges}, {Surdej}, {Szabados}, {Szegedi-Elek}, {Taris}, {Taylor}, {Teixeira}, {Tolomei}, {Tonello}, {Torra}, {Torra}, {Torralba Elipe}, {Trabucchi}, {Tsounis}, {Turon}, {Ulla}, {Unger}, {Vaillant}, {van Dillen}, {van Reeven}, {Vanel}, {Vecchiato}, {Viala}, {Vicente}, {Voutsinas}, {Weiler}, {Wevers}, {Wyrzykowski}, {Yoldas}, {Yvard}, {Zhao}, {Zorec}, {Zucker}, \& {Zwitter}}]{GaiaDR3}
{Gaia Collaboration}, {Vallenari}, A., {Brown}, A.~G.~A., {et~al.} 2023, \aap, 674, A1

\bibitem[{{Giacalone} {et~al.}(2021){Giacalone}, {Dressing}, {Jensen}, {Collins}, {Ricker}, {Vanderspek}, {Seager}, {Winn}, {Jenkins}, {Barclay}, {Barkaoui}, {Cadieux}, {Charbonneau}, {Collins}, {Conti}, {Doyon}, {Evans}, {Ghachoui}, {Gillon}, {Guerrero}, {Hart}, {Jehin}, {Kielkopf}, {McLean}, {Murgas}, {Palle}, {Parviainen}, {Pozuelos}, {Relles}, {Shporer}, {Socia}, {Stockdale}, {Tan}, {Torres}, {Twicken}, {Waalkes}, \& {Waite}}]{Giacalone2021}
{Giacalone}, S., {Dressing}, C.~D., {Jensen}, E. L.~N., {et~al.} 2021, \aj, 161, 24

\bibitem[{{Gibson} {et~al.}(2012){Gibson}, {Aigrain}, {Roberts}, {Evans}, {Osborne}, \& {Pont}}]{Gibson2012}
{Gibson}, N.~P., {Aigrain}, S., {Roberts}, S., {et~al.} 2012, \mnras, 419, 2683

\bibitem[{{Ginzburg} {et~al.}(2018){Ginzburg}, {Schlichting}, \& {Sari}}]{Ginzburg2018}
{Ginzburg}, S., {Schlichting}, H.~E., \& {Sari}, R. 2018, \mnras, 476, 759

\bibitem[{{Hippke} \& {Heller}(2019)}]{Hippke2019}
{Hippke}, M. \& {Heller}, R. 2019, \aap, 623, A39

\bibitem[{{H{\o}g} {et~al.}(2000{\natexlab{a}}){H{\o}g}, {Fabricius}, {Makarov}, {Urban}, {Corbin}, {Wycoff}, {Bastian}, {Schwekendiek}, \& {Wicenec}}]{TYCHO}
{H{\o}g}, E., {Fabricius}, C., {Makarov}, V.~V., {et~al.} 2000{\natexlab{a}}, \aap, 355, L27

\bibitem[{{H{\o}g} {et~al.}(2000{\natexlab{b}}){H{\o}g}, {Fabricius}, {Makarov}, {Urban}, {Corbin}, {Wycoff}, {Bastian}, {Schwekendiek}, \& {Wicenec}}]{Hog2000}
{H{\o}g}, E., {Fabricius}, C., {Makarov}, V.~V., {et~al.} 2000{\natexlab{b}}, \aap, 355, L27

\bibitem[{{Holmberg} \& {Madhusudhan}(2024)}]{Holmberg2024}
{Holmberg}, M. \& {Madhusudhan}, N. 2024, \aap, 683, L2

\bibitem[{{Horch} {et~al.}(2011){Horch}, {Gomez}, {Sherry}, {Howell}, {Ciardi}, {Anderson}, \& {van Altena}}]{Horch2011}
{Horch}, E.~P., {Gomez}, S.~C., {Sherry}, W.~H., {et~al.} 2011, \aj, 141, 45

\bibitem[{{Howell} {et~al.}(2011){Howell}, {Everett}, {Sherry}, {Horch}, \& {Ciardi}}]{Howell2011}
{Howell}, S.~B., {Everett}, M.~E., {Sherry}, W., {Horch}, E., \& {Ciardi}, D.~R. 2011, \aj, 142, 19

\bibitem[{{Hu} {et~al.}(2021){Hu}, {Damiano}, {Scheucher}, {Kite}, {Seager}, \& {Rauer}}]{Hu2021}
{Hu}, R., {Damiano}, M., {Scheucher}, M., {et~al.} 2021, \apjl, 921, L8

\bibitem[{{Jenkins}(2002)}]{Jenkins2002}
{Jenkins}, J.~M. 2002, \apj, 575, 493

\bibitem[{{Jenkins} {et~al.}(2010){Jenkins}, {Chandrasekaran}, {McCauliff}, {Caldwell}, {Tenenbaum}, {Li}, {Klaus}, {Cote}, \& {Middour}}]{Jenkins2010}
{Jenkins}, J.~M., {Chandrasekaran}, H., {McCauliff}, S.~D., {et~al.} 2010, in Society of Photo-Optical Instrumentation Engineers (SPIE) Conference Series, Vol. 7740, Software and Cyberinfrastructure for Astronomy, ed. N.~M. {Radziwill} \& A.~{Bridger}, 77400D

\bibitem[{{Jenkins} {et~al.}(2020){Jenkins}, {Tenenbaum}, {Seader}, {Burke}, {McCauliff}, {Smith}, {Twicken}, \& {Chandrasekaran}}]{JenkinsJM2020}
{Jenkins}, J.~M., {Tenenbaum}, P., {Seader}, S., {et~al.} 2020, {Kepler Data Processing Handbook: Transiting Planet Search}, Kepler Science Document KSCI-19081-003

\bibitem[{{Jenkins} {et~al.}(2016){Jenkins}, {Twicken}, {McCauliff}, {Campbell}, {Sanderfer}, {Lung}, {Mansouri-Samani}, {Girouard}, {Tenenbaum}, {Klaus}, {Smith}, {Caldwell}, {Chacon}, {Henze}, {Heiges}, {Latham}, {Morgan}, {Swade}, {Rinehart}, \& {Vanderspek}}]{Jenkins2016}
{Jenkins}, J.~M., {Twicken}, J.~D., {McCauliff}, S., {et~al.} 2016, in Society of Photo-Optical Instrumentation Engineers (SPIE) Conference Series, Vol. 9913, Software and Cyberinfrastructure for Astronomy IV, ed. G.~{Chiozzi} \& J.~C. {Guzman}, 99133E

\bibitem[{{Jensen}(2013)}]{Jensen:2013}
{Jensen}, E. 2013, {Tapir: A web interface for transit/eclipse observability}, Astrophysics Source Code Library

\bibitem[{{Jin} \& {Mordasini}(2018)}]{Jin2018}
{Jin}, S. \& {Mordasini}, C. 2018, \apj, 853, 163

\bibitem[{{Kempton} {et~al.}(2018){Kempton}, {Bean}, {Louie}, {Deming}, {Koll}, {Mansfield}, {Christiansen}, {L{\'o}pez-Morales}, {Swain}, {Zellem}, {Ballard}, {Barclay}, {Barstow}, {Batalha}, {Beatty}, {Berta-Thompson}, {Birkby}, {Buchhave}, {Charbonneau}, {Cowan}, {Crossfield}, {de Val-Borro}, {Doyon}, {Dragomir}, {Gaidos}, {Heng}, {Hu}, {Kane}, {Kreidberg}, {Mallonn}, {Morley}, {Narita}, {Nascimbeni}, {Pall{\'e}}, {Quintana}, {Rauscher}, {Seager}, {Shkolnik}, {Sing}, {Sozzetti}, {Stassun}, {Valenti}, \& {von Essen}}]{Kempton2018}
{Kempton}, E. M.~R., {Bean}, J.~L., {Louie}, D.~R., {et~al.} 2018, Publications of the Astronomical Society of the Pacific, 130, 114401

\bibitem[{{Kipping}(2013)}]{Kipping2013}
{Kipping}, D.~M. 2013, \mnras, 435, 2152

\bibitem[{{Korth} {et~al.}(2023){Korth}, {Gandolfi}, {{\v{S}}ubjak}, {Howard}, {Ataiee}, {Collins}, {Quinn}, {Mustill}, {Guillot}, {Lodieu}, {Smith}, {Esposito}, {Rodler}, {Muresan}, {Abe}, {Albrecht}, {Alqasim}, {Barkaoui}, {Beck}, {Burke}, {Butler}, {Conti}, {Collins}, {Crane}, {Dai}, {Deeg}, {Evans}, {Grziwa}, {Hatzes}, {Hirano}, {Horne}, {Huang}, {Jenkins}, {Kab{\'a}th}, {Kielkopf}, {Knudstrup}, {Latham}, {Livingston}, {Luque}, {Mathur}, {Murgas}, {Osborne}, {Palle}, {Persson}, {Rodriguez}, {Rose}, {Rowden}, {Schwarz}, {Seager}, {Serrano}, {Sha}, {Shectman}, {Shporer}, {Srdoc}, {Stockdale}, {Tan}, {Teske}, {Van Eylen}, {Vanderburg}, {Vanderspek}, {Wang}, \& {Winn}}]{Korth2023}
{Korth}, J., {Gandolfi}, D., {{\v{S}}ubjak}, J., {et~al.} 2023, \aap, 675, A115

\bibitem[{{Kurucz}(1993)}]{Kurucz-93}
{Kurucz}, R.~L. 1993, {SYNTHE spectrum synthesis programs and line data}

\bibitem[{{Lavie} {et~al.}(2023){Lavie}, {Bouchy}, {Lovis}, {Zapatero Osorio}, {Deline}, {Barros}, {Figueira}, {Sozzetti}, {Gonz{\'a}lez Hern{\'a}ndez}, {Lillo-Box}, {Rodrigues}, {Mehner}, {Damasso}, {Adibekyan}, {Alibert}, {Allende Prieto}, {Cristiani}, {D'Odorico}, {Di Marcantonio}, {Ehrenreich}, {G{\'e}nova Santos}, {Lo Curto}, {Martins}, {Micela}, {Molaro}, {Nunes}, {Palle}, {Pepe}, {Poretti}, {Rebolo}, {Santos}, {Sousa}, {Su{\'a}rez Mascare{\~n}o}, {Tabrenero}, \& {Udry}}]{Lavie2023}
{Lavie}, B., {Bouchy}, F., {Lovis}, C., {et~al.} 2023, \aap, 673, A69

\bibitem[{{Lester} {et~al.}(2021){Lester}, {Matson}, {Howell}, {Furlan}, {Gnilka}, {Scott}, {Ciardi}, {Everett}, {Hartman}, \& {Hirsch}}]{Lester2021}
{Lester}, K.~V., {Matson}, R.~A., {Howell}, S.~B., {et~al.} 2021, \aj, 162, 75

\bibitem[{{Li} {et~al.}(2019){Li}, {Tenenbaum}, {Twicken}, {Burke}, {Jenkins}, {Quintana}, {Rowe}, \& {Seader}}]{Li2019}
{Li}, J., {Tenenbaum}, P., {Twicken}, J.~D., {et~al.} 2019, \pasp, 131, 024506

\bibitem[{{Lillo-Box} {et~al.}(2021){Lillo-Box}, {Faria}, {Su{\'a}rez Mascare{\~n}o}, {Figueira}, {Sousa}, {Tabernero}, {Lovis}, {Silva}, {Demangeon}, {Benatti}, {Santos}, {Mehner}, {Pepe}, {Sozzetti}, {Zapatero Osorio}, {Gonz{\'a}lez Hern{\'a}ndez}, {Micela}, {Hojjatpanah}, {Rebolo}, {Cristiani}, {Adibekyan}, {Allart}, {Allende Prieto}, {Cabral}, {Damasso}, {Di Marcantonio}, {Lo Curto}, {Martins}, {Megevand}, {Molaro}, {Nunes}, {Pall{\'e}}, {Pasquini}, {Poretti}, \& {Udry}}]{Lillo2021}
{Lillo-Box}, J., {Faria}, J.~P., {Su{\'a}rez Mascare{\~n}o}, A., {et~al.} 2021, \aap, 654, A60

\bibitem[{{Lomb}(1976)}]{Lomb1976}
{Lomb}, N.~R. 1976, \apss, 39, 447

\bibitem[{{Lucy} \& {Sweeney}(1971)}]{Lucy1971}
{Lucy}, L.~B. \& {Sweeney}, M.~A. 1971, \aj, 76, 544

\bibitem[{{Luque} \& {Pall{\'e}}(2022)}]{Luque2022}
{Luque}, R. \& {Pall{\'e}}, E. 2022, Science, 377, 1211

\bibitem[{{MacKenzie} {et~al.}(2023){MacKenzie}, {Grenfell}, {Baumeister}, {Tosi}, {Cabrera}, \& {Rauer}}]{MacKenzie2023}
{MacKenzie}, J., {Grenfell}, J.~L., {Baumeister}, P., {et~al.} 2023, \aap, 671, A65

\bibitem[{{Madhusudhan} {et~al.}(2021){Madhusudhan}, {Piette}, \& {Constantinou}}]{Madhusudhan2021}
{Madhusudhan}, N., {Piette}, A. A.~A., \& {Constantinou}, S. 2021, \apj, 918, 1

\bibitem[{{Madhusudhan} {et~al.}(2023){Madhusudhan}, {Sarkar}, {Constantinou}, {Holmberg}, {Piette}, \& {Moses}}]{Madhusudhan2023}
{Madhusudhan}, N., {Sarkar}, S., {Constantinou}, S., {et~al.} 2023, \apjl, 956, L13

\bibitem[{{McCully} {et~al.}(2018){McCully}, {Volgenau}, {Harbeck}, {Lister}, {Saunders}, {Turner}, {Siiverd}, \& {Bowman}}]{McCully:2018}
{McCully}, C., {Volgenau}, N.~H., {Harbeck}, D.-R., {et~al.} 2018, in Society of Photo-Optical Instrumentation Engineers (SPIE) Conference Series, Vol. 10707, \procspie, 107070K

\bibitem[{{Moedas} {et~al.}(2022){Moedas}, {Deal}, {Bossini}, \& {Campilho}}]{Moedas22}
{Moedas}, N., {Deal}, M., {Bossini}, D., \& {Campilho}, B. 2022, \aap, 666, A43

\bibitem[{{Morris} {et~al.}(2020){Morris}, {Twicken}, {Smith}, {Clarke}, {Jenkins}, {Bryson}, {Girouard}, \& {Klaus}}]{Morris2020}
{Morris}, R.~L., {Twicken}, J.~D., {Smith}, J.~C., {et~al.} 2020, {Kepler Data Processing Handbook: Photometric Analysis}, Kepler Science Document KSCI-19081-003

\bibitem[{{Murgas} {et~al.}(2023){Murgas}, {Castro-Gonz{\'a}lez}, {Pall{\'e}}, {Pozuelos}, {Millholland}, {Foo}, {Korth}, {Marfil}, {Amado}, {Caballero}, {Christiansen}, {Ciardi}, {Collins}, {Di Sora}, {Fukui}, {Gan}, {Gonzales}, {Henning}, {Herrero}, {Isopi}, {Jenkins}, {Lillo-Box}, {Lodieu}, {Luque}, {Mallia}, {Morales}, {Morello}, {Narita}, {Orell-Miquel}, {Parviainen}, {P{\'e}rez-Torres}, {Quirrenbach}, {Reiners}, {Ribas}, {Safonov}, {Seager}, {Schwarz}, {Schweitzer}, {Schlecker}, {Strakhov}, {Vanaverbeke}, {Watanabe}, {Winn}, \& {Zechmeister}}]{Murgas2023}
{Murgas}, F., {Castro-Gonz{\'a}lez}, A., {Pall{\'e}}, E., {et~al.} 2023, \aap, 677, A182

\bibitem[{{Noyes} {et~al.}(1984){Noyes}, {Hartmann}, {Baliunas}, {Duncan}, \& {Vaughan}}]{Noyes1984}
{Noyes}, R.~W., {Hartmann}, L.~W., {Baliunas}, S.~L., {Duncan}, D.~K., \& {Vaughan}, A.~H. 1984, \apj, 279, 763

\bibitem[{{Otegi} {et~al.}(2020){Otegi}, {Bouchy}, \& {Helled}}]{Otegi2020}
{Otegi}, J.~F., {Bouchy}, F., \& {Helled}, R. 2020, \aap, 634, A43

\bibitem[{{Owen} \& {Wu}(2017)}]{Owen2017}
{Owen}, J.~E. \& {Wu}, Y. 2017, \apj, 847, 29

\bibitem[{{Palle} {et~al.}(2021){Palle}, {Luque}, {Zapatero Osorio}, {Parviainen}, {Ikoma}, {Tabernero}, {Zechmeister}, {Mustill}, {Bejar}, {Narita}, \& {Murgas}}]{Palle2021}
{Palle}, E., {Luque}, R., {Zapatero Osorio}, M.~R., {et~al.} 2021, \aap, 650, A55

\bibitem[{{Parc} {et~al.}(2024){Parc}, {Bouchy}, {Venturini}, {Dorn}, \& {Helled}}]{Parc2024}
{Parc}, L., {Bouchy}, F., {Venturini}, J., {Dorn}, C., \& {Helled}, R. 2024, \aap, 688, A59

\bibitem[{{Parviainen}(2015)}]{Parviainen2015}
{Parviainen}, H. 2015, \mnras, 450, 3233

\bibitem[{{Parviainen} \& {Aigrain}(2015)}]{Parviainen2015b}
{Parviainen}, H. \& {Aigrain}, S. 2015, \mnras, 453, 3821

\bibitem[{{Paxton} {et~al.}(2011){Paxton}, {Bildsten}, {Dotter}, \& et~al.}]{Paxton2011}
{Paxton}, B., {Bildsten}, L., {Dotter}, A., \& et~al. 2011, \apjs, 192, 3

\bibitem[{{Paxton} {et~al.}(2013){Paxton}, {Cantiello}, {Arras}, {Bildsten}, {Brown}, {Dotter}, {Mankovich}, {Montgomery}, {Stello}, {Timmes}, \& {Townsend}}]{Paxton2013}
{Paxton}, B., {Cantiello}, M., {Arras}, P., {et~al.} 2013, \apjs, 208, 4

\bibitem[{{Paxton} {et~al.}(2015){Paxton}, {Marchant}, {Schwab}, {Bauer}, {Bildsten}, {Cantiello}, {Dessart}, {Farmer}, {Hu}, {Langer}, {Townsend}, {Townsley}, \& {Timmes}}]{Paxton2015}
{Paxton}, B., {Marchant}, P., {Schwab}, J., {et~al.} 2015, \apjs, 220, 15

\bibitem[{{Paxton} {et~al.}(2018){Paxton}, {Schwab}, {Bauer}, {Bildsten}, {Blinnikov}, {Duffell}, {Farmer}, {Goldberg}, {Marchant}, {Sorokina}, {Thoul}, {Townsend}, \& {Timmes}}]{Paxton2018}
{Paxton}, B., {Schwab}, J., {Bauer}, E.~B., {et~al.} 2018, \apjs, 234, 34

\bibitem[{{Paxton} {et~al.}(2019){Paxton}, {Smolec}, {Schwab}, {Gautschy}, {Bildsten}, {Cantiello}, {Dotter}, {Farmer}, {Goldberg}, {Jermyn}, {Kanbur}, {Marchant}, {Thoul}, {Townsend}, {Wolf}, {Zhang}, \& {Timmes}}]{Paxton2019}
{Paxton}, B., {Smolec}, R., {Schwab}, J., {et~al.} 2019, \apjs, 243, 10

\bibitem[{{Pepe} {et~al.}(2021){Pepe}, {Cristiani}, {Rebolo}, {Santos}, {Dekker}, {Cabral}, {Di Marcantonio}, {Figueira}, {Lo Curto}, {Lovis}, {Mayor}, {M{\'e}gevand}, {Molaro}, {Riva}, {Zapatero Osorio}, {Amate}, {Manescau}, {Pasquini}, {Zerbi}, {Adibekyan}, {Abreu}, {Affolter}, {Alibert}, {Aliverti}, {Allart}, {Allende Prieto}, {{\'A}lvarez}, {Alves}, {Avila}, {Baldini}, {Bandy}, {Barros}, {Benz}, {Bianco}, {Borsa}, {Bourrier}, {Bouchy}, {Broeg}, {Calderone}, {Cirami}, {Coelho}, {Conconi}, {Coretti}, {Cumani}, {Cupani}, {D'Odorico}, {Damasso}, {Deiries}, {Delabre}, {Demangeon}, {Dumusque}, {Ehrenreich}, {Faria}, {Fragoso}, {Genolet}, {Genoni}, {G{\'e}nova Santos}, {Gonz{\'a}lez Hern{\'a}ndez}, {Hughes}, {Iwert}, {Kerber}, {Knudstrup}, {Landoni}, {Lavie}, {Lillo-Box}, {Lizon}, {Maire}, {Martins}, {Mehner}, {Micela}, {Modigliani}, {Monteiro}, {Monteiro}, {Moschetti}, {Murphy}, {Nunes}, {Oggioni}, {Oliveira}, {Oshagh}, {Pall{\'e}}, {Pariani}, {Poretti}, {Rasilla}, {Rebord{\~a}o}, {Redaelli}, {Santana Tschudi},
  {Santin}, {Santos}, {S{\'e}gransan}, {Schmidt}, {Segovia}, {Sosnowska}, {Sozzetti}, {Sousa}, {Span{\`o}}, {Su{\'a}rez Mascare{\~n}o}, {Tabernero}, {Tenegi}, {Udry}, \& {Zanutta}}]{Pepe2021}
{Pepe}, F., {Cristiani}, S., {Rebolo}, R., {et~al.} 2021, \aap, 645, A96

\bibitem[{{Pepe} {et~al.}(2000){Pepe}, {Mayor}, {Delabre}, {Kohler}, {Lacroix}, {Queloz}, {Udry}, {Benz}, {Bertaux}, \& {Sivan}}]{Pepe2000}
{Pepe}, F., {Mayor}, M., {Delabre}, B., {et~al.} 2000, in Society of Photo-Optical Instrumentation Engineers (SPIE) Conference Series, Vol. 4008, Optical and IR Telescope Instrumentation and Detectors, ed. M.~{Iye} \& A.~F. {Moorwood}, 582--592

\bibitem[{{Pepe} {et~al.}(2014){Pepe}, {Molaro}, {Cristiani}, {Rebolo}, {Santos}, {Dekker}, {M{\'e}gevand}, {Zerbi}, {Cabral}, {Di Marcantonio}, {Abreu}, {Affolter}, {Aliverti}, {Allende Prieto}, {Amate}, {Avila}, {Baldini}, {Bristow}, {Broeg}, {Cirami}, {Coelho}, {Conconi}, {Coretti}, {Cupani}, {D'Odorico}, {De Caprio}, {Delabre}, {Dorn}, {Figueira}, {Fragoso}, {Galeotta}, {Genolet}, {Gomes}, {Gonz{\'a}lez Hern{\'a}ndez}, {Hughes}, {Iwert}, {Kerber}, {Landoni}, {Lizon}, {Lovis}, {Maire}, {Mannetta}, {Martins}, {Monteiro}, {Oliveira}, {Poretti}, {Rasilla}, {Riva}, {Santana Tschudi}, {Santos}, {Sosnowska}, {Sousa}, {Span{\'o}}, {Tenegi}, {Toso}, {Vanzella}, {Viel}, \& {Zapatero Osorio}}]{Pepe2014}
{Pepe}, F., {Molaro}, P., {Cristiani}, S., {et~al.} 2014, Astronomische Nachrichten, 335, 8

\bibitem[{{Pozuelos} {et~al.}(2020){Pozuelos}, {Su{\'a}rez}, {de El{\'\i}a}, {Berdi{\~n}as}, {Bonfanti}, {Dugaro}, {Gillon}, {Jehin}, {G{\"u}nther}, {Van Grootel}, {Garcia}, {Thuillier}, {Delrez}, \& {Rod{\'o}n}}]{pozuelos2020}
{Pozuelos}, F.~J., {Su{\'a}rez}, J.~C., {de El{\'\i}a}, G.~C., {et~al.} 2020, \aap, 641, A23

\bibitem[{{Pozuelos} {et~al.}(2023){Pozuelos}, {Timmermans}, {Rackham}, {Garcia}, {Burgasser}, {Kane}, {G{\"u}nther}, {Stassun}, {Van Grootel}, {D{\'e}vora-Pajares}, {Luque}, {Edwards}, {Niraula}, {Schanche}, {Wells}, {Ducrot}, {Howell}, {Sebastian}, {Barkaoui}, {Waalkes}, {Cadieux}, {Doyon}, {Boyle}, {Dietrich}, {Burdanov}, {Delrez}, {Demory}, {de Wit}, {Dransfield}, {Gillon}, {G{\'o}mez Maqueo Chew}, {Hooton}, {Jehin}, {Murray}, {Pedersen}, {Queloz}, {Thompson}, {Triaud}, {Z{\'u}{\~n}iga-Fern{\'a}ndez}, {Collins}, {Fausnaugh}, {Hedges}, {Hesse}, {Jenkins}, {Kunimoto}, {Latham}, {Shporer}, {Ting}, {Torres}, {Amado}, {Rod{\'o}n}, {Rodr{\'\i}guez-L{\'o}pez}, {Su{\'a}rez}, {Alonso}, {Benkhaldoun}, {Berta-Thompson}, {Chinchilla}, {Ghachoui}, {G{\'o}mez-Mu{\~n}oz}, {Rebolo}, {Sabin}, {Schroffenegger}, {Furlan}, {Gnilka}, {Lester}, {Scott}, {Aganze}, {Gerasimov}, {Hsu}, {Theissen}, {Apai}, {Chen}, {Gabor}, {Henning}, \& {Mancini}}]{pozuelos2023}
{Pozuelos}, F.~J., {Timmermans}, M., {Rackham}, B.~V., {et~al.} 2023, \aap, 672, A70

\bibitem[{{Rasmussen} \& {Williams}(2010)}]{Rasmussen2006}
{Rasmussen}, C. \& {Williams}, C. 2010, the MIT Press, 122, 935

\bibitem[{{Ricker} {et~al.}(2015){Ricker}, {Winn}, {Vanderspek}, {Latham}, {Bakos}, {Bean}, {Berta-Thompson}, {Brown}, {Buchhave}, {Butler}, {Butler}, {Chaplin}, {Charbonneau}, {Christensen-Dalsgaard}, {Clampin}, {Deming}, {Doty}, {De Lee}, {Dressing}, {Dunham}, {Endl}, {Fressin}, {Ge}, {Henning}, {Holman}, {Howard}, {Ida}, {Jenkins}, {Jernigan}, {Johnson}, {Kaltenegger}, {Kawai}, {Kjeldsen}, {Laughlin}, {Levine}, {Lin}, {Lissauer}, {MacQueen}, {Marcy}, {McCullough}, {Morton}, {Narita}, {Paegert}, {Palle}, {Pepe}, {Pepper}, {Quirrenbach}, {Rinehart}, {Sasselov}, {Sato}, {Seager}, {Sozzetti}, {Stassun}, {Sullivan}, {Szentgyorgyi}, {Torres}, {Udry}, \& {Villasenor}}]{Ricker2015}
{Ricker}, G.~R., {Winn}, J.~N., {Vanderspek}, R., {et~al.} 2015, Journal of Astronomical Telescopes, Instruments, and Systems, 1, 014003

\bibitem[{{Rodrigues} {et~al.}(2017){Rodrigues}, {Bossini}, {Miglio}, {Girardi}, {Montalb{\'a}n}, {Noels}, {Trabucchi}, {Coelho}, \& {Marigo}}]{PARAM3}
{Rodrigues}, T.~S., {Bossini}, D., {Miglio}, A., {et~al.} 2017, MNRAS, 467, 1433

\bibitem[{{Rodrigues} {et~al.}(2014){Rodrigues}, {Girardi}, {Miglio}, {Bossini}, {Bovy}, {Epstein}, {Pinsonneault}, {Stello}, {Zasowski}, {Allende Prieto}, {Chaplin}, {Hekker}, {Johnson}, {M{\'e}sz{\'a}ros}, {Mosser}, {Anders}, {Basu}, {Beers}, {Chiappini}, {da Costa}, {Elsworth}, {Garc{\'\i}a}, {Garc{\'\i}a P{\'e}rez}, {Hearty}, {Maia}, {Majewski}, {Mathur}, {Montalb{\'a}n}, {Nidever}, {Santiago}, {Schultheis}, {Serenelli}, \& {Shetrone}}]{PARAM2}
{Rodrigues}, T.~S., {Girardi}, L., {Miglio}, A., {et~al.} 2014, MNRAS, 445, 2758

\bibitem[{{Rogers} \& {Owen}(2021)}]{Rogers21}
{Rogers}, J.~G. \& {Owen}, J.~E. 2021, \mnras, 503, 1526

\bibitem[{{Santos} {et~al.}(2013){Santos}, {Sousa}, {Mortier}, {Neves}, {Adibekyan}, {Tsantaki}, {Delgado Mena}, {Bonfils}, {Israelian}, {Mayor}, \& {Udry}}]{Santos-13}
{Santos}, N.~C., {Sousa}, S.~G., {Mortier}, A., {et~al.} 2013, \aap, 556, A150

\bibitem[{{Savitzky} \& {Golay}(1964)}]{SavitzkyGolay1964}
{Savitzky}, A. \& {Golay}, M.~J.~E. 1964, Analytical Chemistry, 36, 1627

\bibitem[{{Scargle}(1982)}]{Scargle1982}
{Scargle}, J.~D. 1982, \apj, 263, 835

\bibitem[{{Schwarz}(1978)}]{Schwarz1978}
{Schwarz}, G. 1978, Annals of Statistics, 6, 461

\bibitem[{{Scott} {et~al.}(2021){Scott}, {Howell}, {Gnilka}, {Stephens}, {Salinas}, {Matson}, {Furlan}, {Horch}, {Everett}, {Ciardi}, {Mills}, \& {Quigley}}]{Scott2021}
{Scott}, N.~J., {Howell}, S.~B., {Gnilka}, C.~L., {et~al.} 2021, Frontiers in Astronomy and Space Sciences, 8, 138

\bibitem[{{Shappee} {et~al.}(2014){Shappee}, {Prieto}, {Grupe}, {Kochanek}, {Stanek}, {De Rosa}, {Mathur}, {Zu}, {Peterson}, {Pogge}, {Komossa}, {Im}, {Jencson}, {Holoien}, {Basu}, {Beacom}, {Szczygie{\l}}, {Brimacombe}, {Adams}, {Campillay}, {Choi}, {Contreras}, {Dietrich}, {Dubberley}, {Elphick}, {Foale}, {Giustini}, {Gonzalez}, {Hawkins}, {Howell}, {Hsiao}, {Koss}, {Leighly}, {Morrell}, {Mudd}, {Mullins}, {Nugent}, {Parrent}, {Phillips}, {Pojmanski}, {Rosing}, {Ross}, {Sand}, {Terndrup}, {Valenti}, {Walker}, \& {Yoon}}]{Shappee2014}
{Shappee}, B.~J., {Prieto}, J.~L., {Grupe}, D., {et~al.} 2014, \apj, 788, 48

\bibitem[{{Skrutskie} {et~al.}(2006){Skrutskie}, {Cutri}, {Stiening}, {Weinberg}, {Schneider}, {Carpenter}, {Beichman}, {Capps}, {Chester}, {Elias}, {Huchra}, {Liebert}, {Lonsdale}, {Monet}, {Price}, {Seitzer}, {Jarrett}, {Kirkpatrick}, {Gizis}, {Howard}, {Evans}, {Fowler}, {Fullmer}, {Hurt}, {Light}, {Kopan}, {Marsh}, {McCallon}, {Tam}, {Van Dyk}, \& {Wheelock}}]{2MASS}
{Skrutskie}, M.~F., {Cutri}, R.~M., {Stiening}, R., {et~al.} 2006, \aj, 131, 1163

\bibitem[{{Smith} {et~al.}(2012){Smith}, {Stumpe}, {Van Cleve}, {Jenkins}, {Barclay}, {Fanelli}, {Girouard}, {Kolodziejczak}, {McCauliff}, {Morris}, \& {Twicken}}]{Smith2012}
{Smith}, J.~C., {Stumpe}, M.~C., {Van Cleve}, J.~E., {et~al.} 2012, \pasp, 124, 1000

\bibitem[{{Sneden}(1973)}]{Sneden-73}
{Sneden}, C.~A. 1973, PhD thesis, THE UNIVERSITY OF TEXAS AT AUSTIN.

\bibitem[{{Sousa}(2014)}]{Sousa-14}
{Sousa}, S.~G. 2014, [arXiv:1407.5817] [\eprint[arXiv]{1407.5817}]

\bibitem[{{Sousa} {et~al.}(2021){Sousa}, {Adibekyan}, {Delgado-Mena}, {Santos}, {Rojas-Ayala}, {Soares}, {Legoinha}, {Ulmer-Moll}, {Camacho}, {Barros}, {Demangeon}, {Hoyer}, {Israelian}, {Mortier}, {Tsantaki}, \& {Monteiro}}]{Sousa-21}
{Sousa}, S.~G., {Adibekyan}, V., {Delgado-Mena}, E., {et~al.} 2021, \aap, 656, A53

\bibitem[{{Sousa} {et~al.}(2015){Sousa}, {Santos}, {Adibekyan}, {Delgado-Mena}, \& {Israelian}}]{Sousa-15}
{Sousa}, S.~G., {Santos}, N.~C., {Adibekyan}, V., {Delgado-Mena}, E., \& {Israelian}, G. 2015, \aap, 577, A67

\bibitem[{{Sousa} {et~al.}(2007){Sousa}, {Santos}, {Israelian}, {Mayor}, \& {Monteiro}}]{Sousa-07}
{Sousa}, S.~G., {Santos}, N.~C., {Israelian}, G., {Mayor}, M., \& {Monteiro}, M.~J.~P.~F.~G. 2007, A\&A, 469, 783

\bibitem[{{Sousa} {et~al.}(2008){Sousa}, {Santos}, {Mayor}, {Udry}, {Casagrande}, {Israelian}, {Pepe}, {Queloz}, \& {Monteiro}}]{Sousa-08}
{Sousa}, S.~G., {Santos}, N.~C., {Mayor}, M., {et~al.} 2008, \aap, 487, 373

\bibitem[{{Southworth}(2011)}]{Southworth2011}
{Southworth}, J. 2011, \mnras, 417, 2166

\bibitem[{{Sozzetti} {et~al.}(2021){Sozzetti}, {Damasso}, {Bonomo}, {Alibert}, {Sousa}, {Adibekyan}, {Zapatero Osorio}, {Gonz{\'a}lez Hern{\'a}ndez}, {Barros}, {Lillo-Box}, {Stassun}, {Winn}, {Cristiani}, {Pepe}, {Rebolo}, {Santos}, {Allart}, {Barclay}, {Bouchy}, {Cabral}, {Ciardi}, {Di Marcantonio}, {D'Odorico}, {Ehrenreich}, {Fasnaugh}, {Figueira}, {Haldemann}, {Jenkins}, {Latham}, {Lavie}, {Lo Curto}, {Lovis}, {Martins}, {M{\'e}gevand}, {Mehner}, {Micela}, {Molaro}, {Nunes}, {Oshagh}, {Otegi}, {Pall{\'e}}, {Poretti}, {Ricker}, {Rodriguez}, {Seager}, {Su{\'a}rez Mascare{\~n}o}, {Twicken}, \& {Udry}}]{Sozzetti2021}
{Sozzetti}, A., {Damasso}, M., {Bonomo}, A.~S., {et~al.} 2021, \aap, 648, A75

\bibitem[{{Stassun} {et~al.}(2018){Stassun}, {Oelkers}, {Pepper}, {Paegert}, {De Lee}, {Torres}, {Latham}, {Charpinet}, {Dressing}, {Huber}, {Kane}, {L{\'e}pine}, {Mann}, {Muirhead}, {Rojas-Ayala}, {Silvotti}, {Fleming}, {Levine}, \& {Plavchan}}]{Stassun2018}
{Stassun}, K.~G., {Oelkers}, R.~J., {Pepper}, J., {et~al.} 2018, \aj, 156, 102

\bibitem[{{Stumpe} {et~al.}(2014){Stumpe}, {Smith}, {Catanzarite}, {Van Cleve}, {Jenkins}, {Twicken}, \& {Girouard}}]{Stumpe2014}
{Stumpe}, M.~C., {Smith}, J.~C., {Catanzarite}, J.~H., {et~al.} 2014, \pasp, 126, 100

\bibitem[{{Stumpe} {et~al.}(2012){Stumpe}, {Smith}, {Van Cleve}, {Twicken}, {Barclay}, {Fanelli}, {Girouard}, {Jenkins}, {Kolodziejczak}, {McCauliff}, \& {Morris}}]{Stumpe2012}
{Stumpe}, M.~C., {Smith}, J.~C., {Van Cleve}, J.~E., {et~al.} 2012, \pasp, 124, 985

\bibitem[{{Su{\'a}rez Mascare{\~n}o} {et~al.}(2023){Su{\'a}rez Mascare{\~n}o}, {Gonz{\'a}lez-{\'A}lvarez}, {Zapatero Osorio}, {Lillo-Box}, {Faria}, {Passegger}, {Gonz{\'a}lez Hern{\'a}ndez}, {Figueira}, {Sozzetti}, {Rebolo}, {Pepe}, {Santos}, {Cristiani}, {Lovis}, {Silva}, {Ribas}, {Amado}, {Caballero}, {Quirrenbach}, {Reiners}, {Zechmeister}, {Adibekyan}, {Alibert}, {B{\'e}jar}, {Benatti}, {D'Odorico}, {Damasso}, {Delisle}, {Di Marcantonio}, {Dreizler}, {Ehrenreich}, {Hatzes}, {Hara}, {Henning}, {Kaminski}, {L{\'o}pez-Gonz{\'a}lez}, {Martins}, {Micela}, {Montes}, {Pall{\'e}}, {Pedraz}, {Rodr{\'\i}guez}, {Rodr{\'\i}guez-L{\'o}pez}, {Tal-Or}, {Sousa}, \& {Udry}}]{SuarezMascareno2023}
{Su{\'a}rez Mascare{\~n}o}, A., {Gonz{\'a}lez-{\'A}lvarez}, E., {Zapatero Osorio}, M.~R., {et~al.} 2023, \aap, 670, A5

\bibitem[{{Su{\'a}rez Mascare{\~n}o} {et~al.}(2024){Su{\'a}rez Mascare{\~n}o}, {Passegger}, {Gonz{\'a}lez Hern{\'a}ndez}, {Armstrong}, {Nielsen}, {Lovis}, {Lavie}, {Sousa}, {Silva}, {Allart}, {Rebolo}, {Pepe}, {Santos}, {Cristiani}, {Sozzetti}, {Zapatero Osorio}, {Tabernero}, {Dumusque}, {Udry}, {Adibekyan}, {Allende Prieto}, {Alibert}, {Barros}, {Bouchy}, {Castro-Gonz{\'a}lez}, {Collins}, {Damasso}, {D'Odorico}, {Demangeon}, {Di Marcantonio}, {Ehrenreich}, {Hadjigeorghiou}, {Hara}, {Hawthorn}, {Jenkins}, {Lillo-Box}, {Lo Curto}, {Martins}, {Mehner}, {Micela}, {Molaro}, {Nunes}, {Nari}, {Osborn}, {Pall{\'e}}, {Ricker}, {Rodrigues}, {Rowden}, {Seager}, {Stefanov}, {Str{\o}m}, {Villase{\~n}or}, {Watkins}, {Winn}, {Wohler}, \& {Zambelli}}]{SuarezMascareno2024}
{Su{\'a}rez Mascare{\~n}o}, A., {Passegger}, V.~M., {Gonz{\'a}lez Hern{\'a}ndez}, J.~I., {et~al.} 2024, \aap, 685, A56

\bibitem[{{Su{\'a}rez Mascare{\~n}o} {et~al.}(2015){Su{\'a}rez Mascare{\~n}o}, {Rebolo}, {Gonz{\'a}lez Hern{\'a}ndez}, \& {Esposito}}]{SuarezMascareno2015}
{Su{\'a}rez Mascare{\~n}o}, A., {Rebolo}, R., {Gonz{\'a}lez Hern{\'a}ndez}, J.~I., \& {Esposito}, M. 2015, \mnras, 452, 2745

\bibitem[{{Tokovinin}(2018)}]{Tokovinin2018}
{Tokovinin}, A. 2018, \pasp, 130, 035002

\bibitem[{{Tsai} {et~al.}(2021){Tsai}, {Innes}, {Lichtenberg}, {Taylor}, {Malik}, {Chubb}, \& {Pierrehumbert}}]{Tsai2021}
{Tsai}, S.-M., {Innes}, H., {Lichtenberg}, T., {et~al.} 2021, \apjl, 922, L27

\bibitem[{{Twicken} {et~al.}(2018){Twicken}, {Catanzarite}, {Clarke}, {Girouard}, {Jenkins}, {Klaus}, {Li}, {McCauliff}, {Seader}, {Tenenbaum}, {Wohler}, {Bryson}, {Burke}, {Caldwell}, {Haas}, {Henze}, \& {Sanderfer}}]{Twicken2018}
{Twicken}, J.~D., {Catanzarite}, J.~H., {Clarke}, B.~D., {et~al.} 2018, \pasp, 130, 064502

\bibitem[{{Van Eylen} {et~al.}(2018){Van Eylen}, {Agentoft}, {Lundkvist}, {Kjeldsen}, {Owen}, {Fulton}, {Petigura}, \& {Snellen}}]{VanEylen2018}
{Van Eylen}, V., {Agentoft}, C., {Lundkvist}, M.~S., {et~al.} 2018, \mnras, 479, 4786

\bibitem[{{Van Grootel} {et~al.}(2021){Van Grootel}, {Pozuelos}, {Thuillier}, {Charpinet}, {Delrez}, {Beck}, {Fortier}, {Hoyer}, {Sousa}, {Barlow}, {Billot}, {D{\'e}vora-Pajares}, {{\O}stensen}, {Alibert}, {Alonso}, {Anglada Escud{\'e}}, {Asquier}, {Barrado}, {Barros}, {Baumjohann}, {Beck}, {Bekkelien}, {Benz}, {Bonfils}, {Brandeker}, {Broeg}, {Bruno}, {B{\'a}rczy}, {Cabrera}, {Cameron}, {Charnoz}, {Davies}, {Deleuil}, {Demangeon}, {Demory}, {Ehrenreich}, {Erikson}, {Fossati}, {Fridlund}, {Futyan}, {Gandolfi}, {Gillon}, {Guedel}, {Heng}, {Isaak}, {Kiss}, {Laskar}, {Lecavelier des Etangs}, {Lendl}, {Lovis}, {Magrin}, {Maxted}, {Mecina}, {Mustill}, {Nascimbeni}, {Olofsson}, {Ottensamer}, {Pagano}, {Pall{\'e}}, {Peter}, {Piotto}, {Plesseria}, {Pollacco}, {Queloz}, {Ragazzoni}, {Rando}, {Rauer}, {Ribas}, {Santos}, {Scandariato}, {S{\'e}gransan}, {Silvotti}, {Simon}, {Smith}, {Steller}, {Szab{\'o}}, {Thomas}, {Udry}, {Viotto}, {Walton}, {Westerdorff}, \& {Wilson}}]{vangrootel2021}
{Van Grootel}, V., {Pozuelos}, F.~J., {Thuillier}, A., {et~al.} 2021, \aap, 650, A205

\bibitem[{{Vaughan} {et~al.}(1978){Vaughan}, {Preston}, \& {Wilson}}]{Vaughan1978}
{Vaughan}, A.~H., {Preston}, G.~W., \& {Wilson}, O.~C. 1978, \pasp, 90, 267

\bibitem[{{Venturini} {et~al.}(2020){Venturini}, {Guilera}, {Haldemann}, {Ronco}, \& {Mordasini}}]{Venturini2020}
{Venturini}, J., {Guilera}, O.~M., {Haldemann}, J., {Ronco}, M.~P., \& {Mordasini}, C. 2020, \aap, 643, L1

\bibitem[{{Venturini} {et~al.}(2024){Venturini}, {Ronco}, {Guilera}, {Haldemann}, {Mordasini}, \& {Miller Bertolami}}]{Venturini2024}
{Venturini}, J., {Ronco}, M.~P., {Guilera}, O.~M., {et~al.} 2024, \aap, 686, L9

\bibitem[{{Zacharias} {et~al.}(2013){Zacharias}, {Finch}, {Girard}, {Henden}, {Bartlett}, {Monet}, \& {Zacharias}}]{UCAC4}
{Zacharias}, N., {Finch}, C.~T., {Girard}, T.~M., {et~al.} 2013, \aj, 145, 44

\bibitem[{{Zechmeister} \& {K{\"u}rster}(2009)}]{Zechmeister2009}
{Zechmeister}, M. \& {K{\"u}rster}, M. 2009, \aap, 496, 577

\bibitem[{{Zeng} {et~al.}(2019){Zeng}, {Jacobsen}, {Sasselov}, {Petaev}, {Vanderburg}, {Lopez-Morales}, {Perez-Mercader}, {Mattsson}, {Li}, {Heising}, {Bonomo}, {Damasso}, {Berger}, {Cao}, {Levi}, \& {Wordsworth}}]{Zeng2019}
{Zeng}, L., {Jacobsen}, S.~B., {Sasselov}, D.~D., {et~al.} 2019, Proceedings of the National Academy of Science, 116, 9723

\bibitem[{{Zeng} {et~al.}(2016){Zeng}, {Sasselov}, \& {Jacobsen}}]{Zeng2016}
{Zeng}, L., {Sasselov}, D.~D., \& {Jacobsen}, S.~B. 2016, \apj, 819, 127

\bibitem[{{Ziegler} {et~al.}(2020){Ziegler}, {Tokovinin}, {Brice{\~n}o}, {Mang}, {Law}, \& {Mann}}]{Ziegler2020}
{Ziegler}, C., {Tokovinin}, A., {Brice{\~n}o}, C., {et~al.} 2020, \aj, 159, 19

\end{thebibliography}

\begin{appendix}

\onecolumn
\section{TESS data}
\label{Sec:Appendix_TESS}

\begin{figure*}[h!]
    \centering
    \includegraphics[width=\hsize]{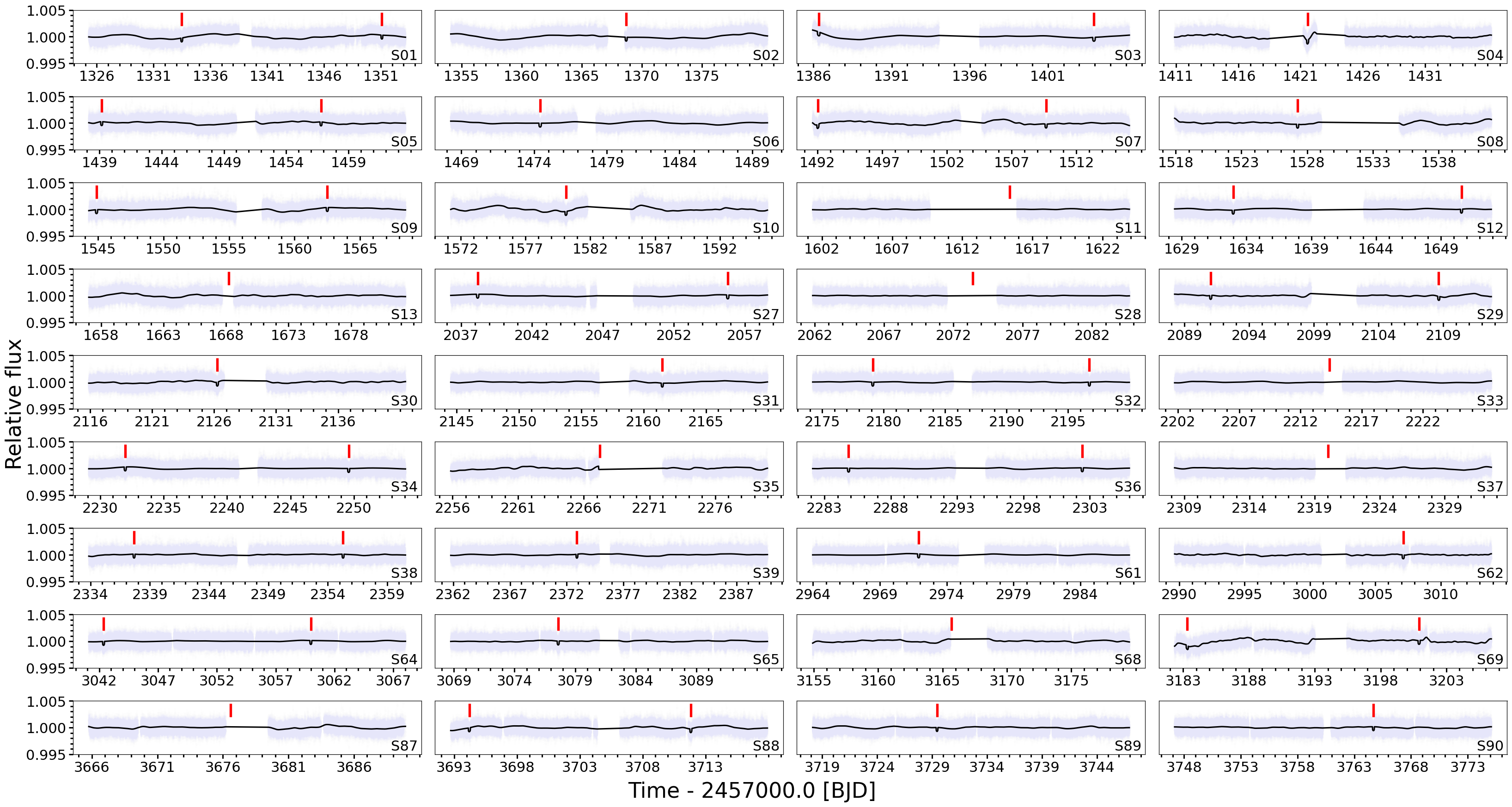}
    \caption{TOI-283\,b TESS light curves analyzed in this work. The black line shows our best model fit. The vertical red lines represent the individual transit events of TOI~283b.}
    \label{Fig:AllTESS_LightCurves}
\end{figure*}

\begin{figure*}[h!]
    \centering
    \includegraphics[width=0.48\textwidth]{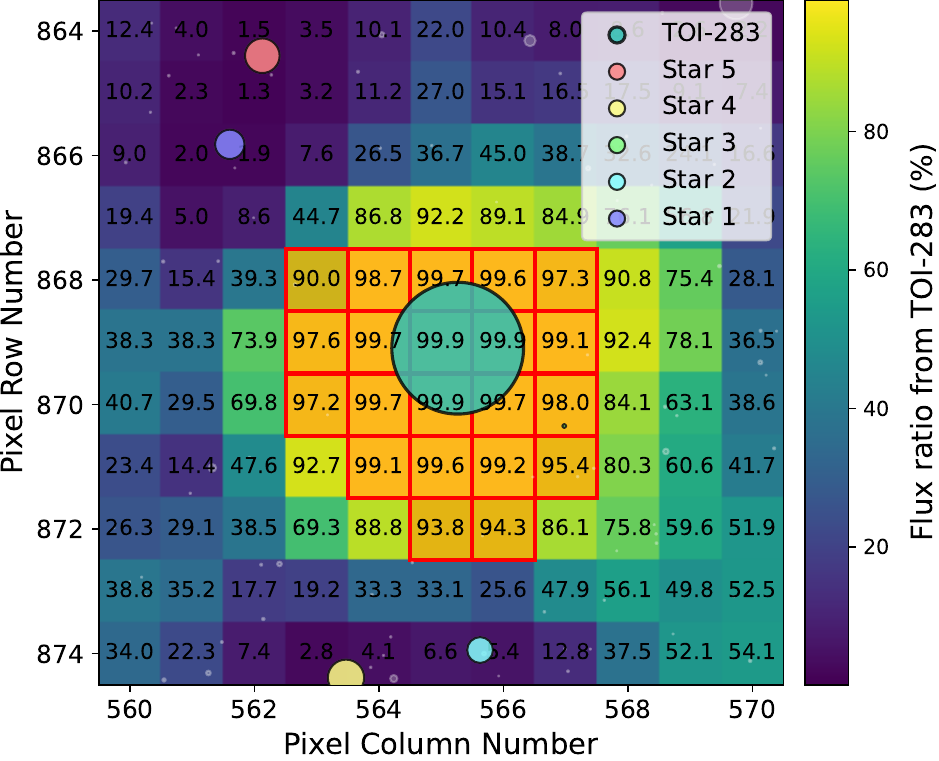}
    \includegraphics[width=0.49\textwidth]{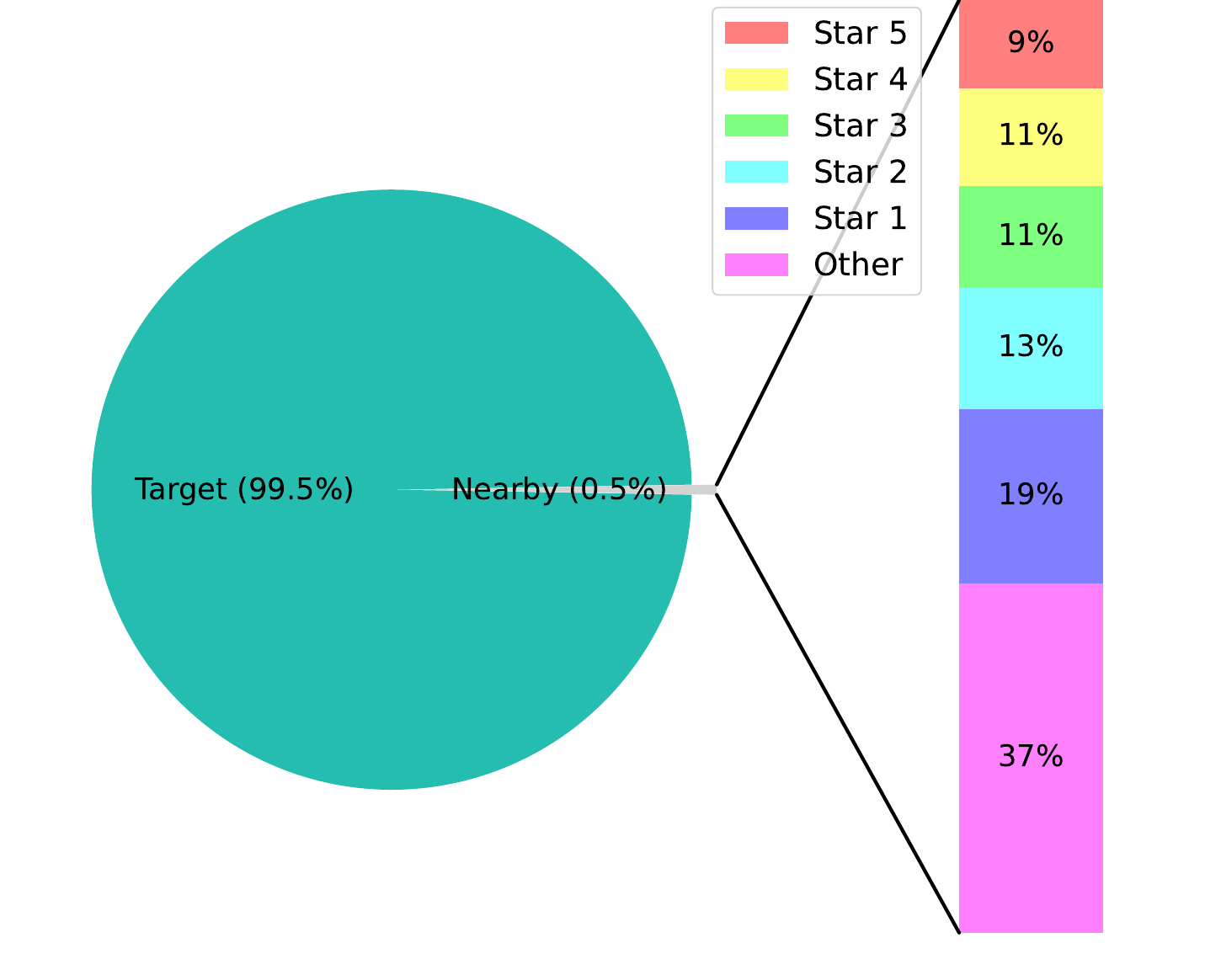}
    \caption{Nearby sources contributing to the TESS photometry TOI-283. Left: TPF-shaped heatmap with the pixel-by-pixel flux fraction from TOI-283 in S1. The red grid is the SPOC aperture. The white disks represent all the \textit{Gaia} sources, and the five sources that most contribute to the aperture flux are highlighted in different colours. Disk areas scale with the emitted fluxes. Right: Flux contributions to the SPOC aperture. ‘Other’ refers to the total contamination from nearby Gaia DR2 sources that contribute less than Star 5.}
    \label{Fig:TESS-cont}
\end{figure*}

\begin{figure}[h!]
    \centering
    \includegraphics[width=\hsize]{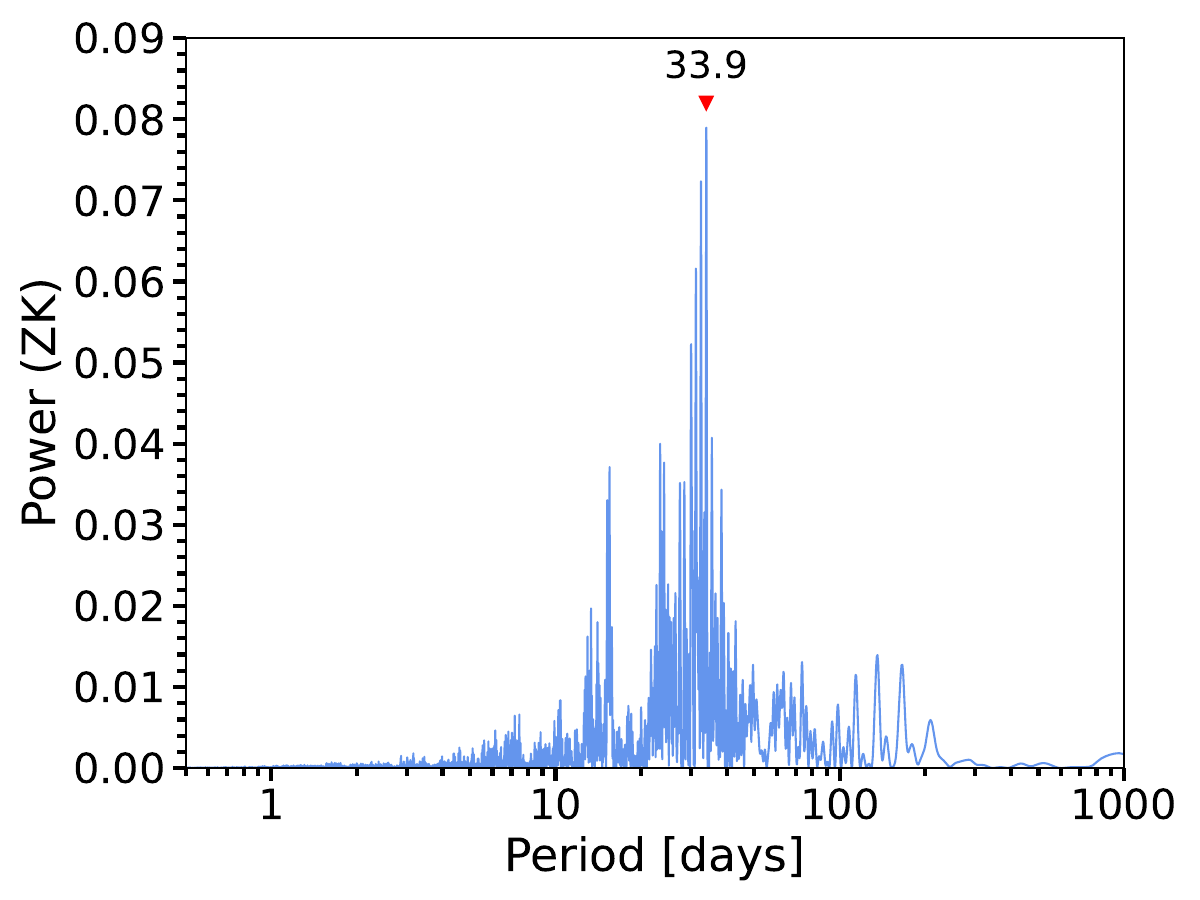}
    \caption{GLS periodogram of the TESS SAP photometry for the 36 sectors analyzed in this paper. The highest power peak is at a period of $\sim$33 days.}
    \label{Fig:TESS_SAP_Periodogram}
\end{figure}

Figure~\ref{Fig:AllTESS_LightCurves} shows the TESS time series and the best fit model from our joint fit analysis (see Sect. \ref{Sec:JointFit}).  Figure~\ref{Fig:TESS_SAP_Periodogram} shows the GLS periodogram of the TESS SAP photometry. Figure~\ref{Fig:TESS-cont} presents the results of the contamination analysis using TESS.

\twocolumn

\section{LCOGT transit}
\label{Sec:Appendix_LCO_transit}
\begin{figure}[h!]
    \centering
    \includegraphics[width=\hsize]{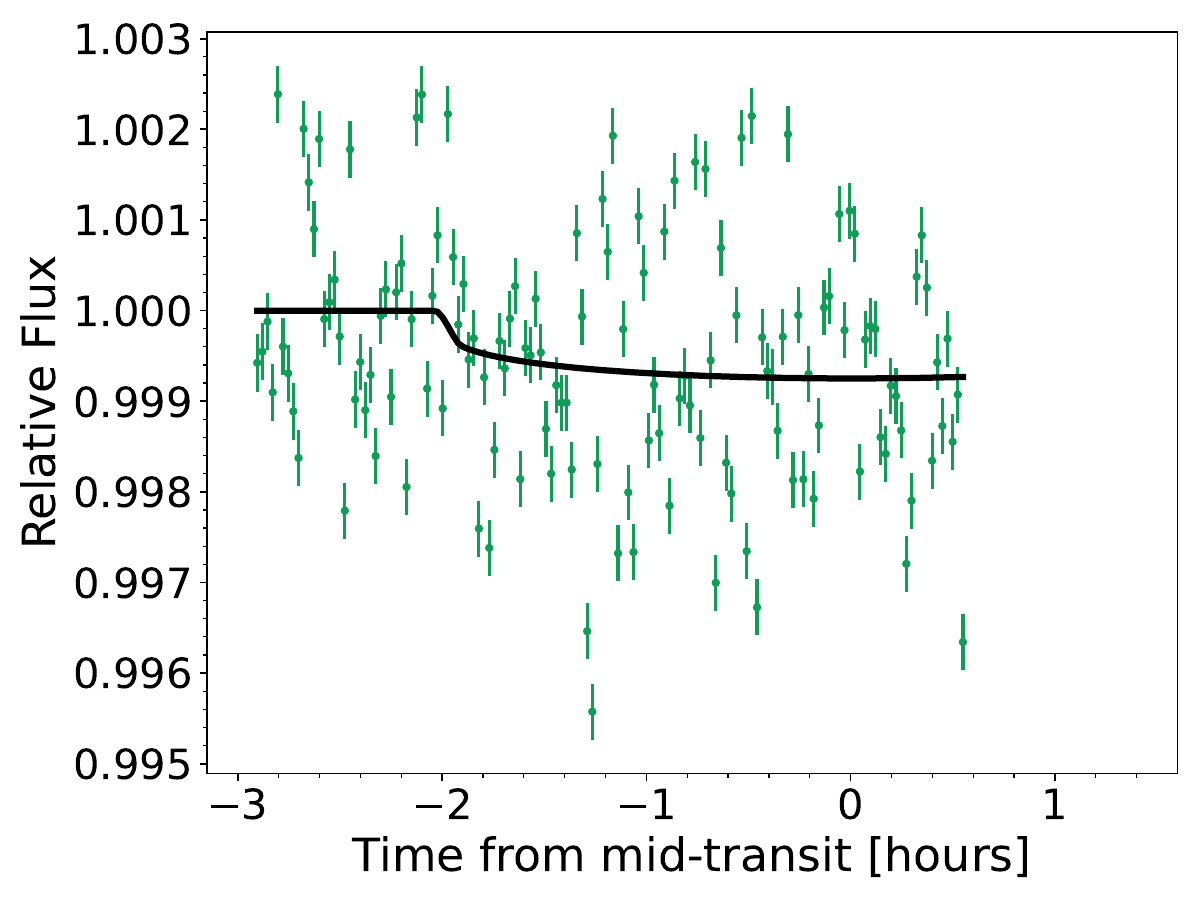}
    \caption{LCOGT transit observation taken on January 26, 2019. The data were collected with the 1 m telescope at CTIO using a Sloan $r'$ filter. The black line shows the best-fit model.}
    \label{Fig:LCO_LightCurve}
\end{figure}

\begin{figure}[h!]
    \centering
    \includegraphics[width=\hsize]{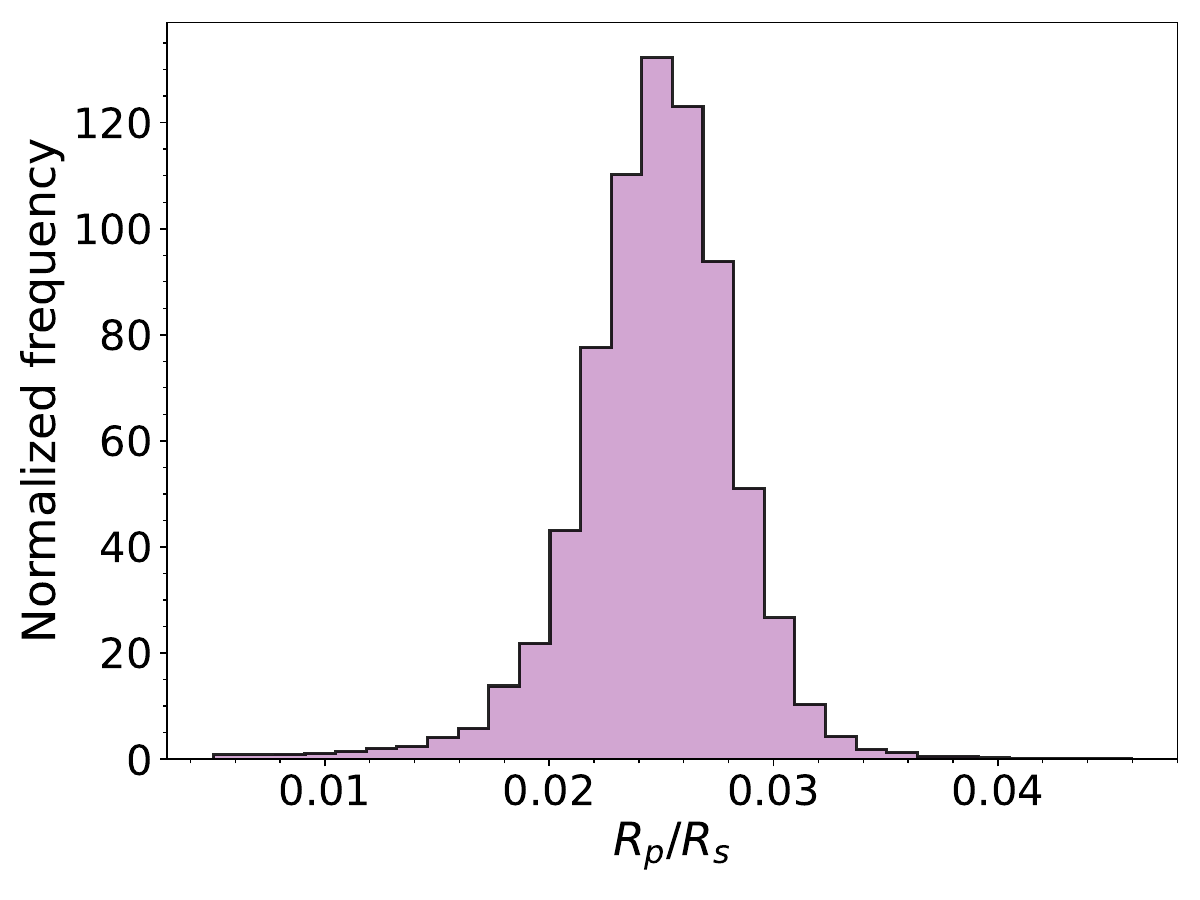}
    \caption{Posterior distribution of the planet-to-star radius ratio ($\mathrm{R}_\mathrm{p}/\mathrm{R}_\star$) for the LCOGT transit observation.}
    \label{Fig:LCO_RpRs_Ditri}
\end{figure}

Here we describe our fitting process for the LCOGT transit light curve taken on 26 January 2019. For the fit we used a detrended light curve provided by \texttt{EXOFOP}, where the time series was detrended using the FWHM as a proxy for seeing changes during the observations. Similar to our joint fit analysis (see Sect. \ref{Sec:JointFit}), we set as free parameters the planet-to-star radius ratio ($\mathrm{R}_\mathrm{p}/\mathrm{R}_\star$), the LD coefficients ($q_1$, $q_2$), the central time of the transit ($\mathrm{T}_{\mathrm{c}}$), the planetary orbital period ($P$), the stellar density ($\rho_\star$), the transit impact parameter ($\mathrm{b}$), a photometric jitter term to model the white noise component ($\sigma_{\mathrm{phot}}$), and the two GP coefficients of the Mat\'ern 3/2 kernel. Because we had a partial transit, we decided to fit the data using normal priors on $\mathrm{T}_{\mathrm{c}}$, $P$, and $\rho_\star$. These priors were derived from the results of a transit fit using only TESS data. After optimizing a posterior probability function with \texttt{PyDE}, we used \texttt{emcee} to sample the posterior with 50 chains and ran an MCMC for 25\,000 iterations. The best-fit parameter values and uncertainties were computed from the percentiles of the posterior distributions. Figure~\ref{Fig:LCO_LightCurve} shows the data and the best-fitting model, with our approach we find a planet-to-star radius ratio of $\mathrm{R}_\mathrm{p}/\mathrm{R}_\star = 0.025 \pm 0.003 $ in agreement with TESS measurements (see Fig.~\ref{Fig:LCO_RpRs_Ditri}). Since this transit event was covered by TESS observations from Sector 7, we decided to omit this dataset from the joint fit.

\onecolumn
\begin{landscape}
\section{Radial velocity data} 
In Table~\ref{Tab:ESPRESSO_Data_Table} we present the ESPRESSO radial velocity measurements and activity indicators. 
\label{Sec:Appendix_ESPRESSO}

\begin{longtable}{ccccccccc}
\caption{ESPRESSO RVs and activity indices.}\\
\label{Tab:ESPRESSO_Data_Table} 

BJD -- 2\,458\,500 & RV [m\,s$^{-1}$] & FWHM [m\,s$^{-1}$] & BIS [m\,s$^{-1}$] & CONT & $S$-Index & Na & H$\alpha$ & Ca \\
\hline
\endfirsthead
\caption{continued.}\\
\hline\hline
BJD -- 2\,458\,500 & RV [m\,s$^{-1}$] & FWHM [m\,s$^{-1}$] & BIS [m\,s$^{-1}$] & CONT & $S$-Index & Na & H$\alpha$ & Ca  \\
\hline
\endhead
\hline
\endfoot

24.772866 & $ -12579.98 \pm 0.57 $ & $ 6454.75 \pm 1.50 $ & $ -51.89 \pm 1.50 $ & $ 61.93 \pm 0.01 $ & $ 0.1680 \pm 0.0005 $ & $ 0.2073 \pm 0.0001 $ & $ 0.2516 \pm 0.0001 $ & $ 0.1313 \pm 0.0004 $ \\
33.658071 & $ -12584.76 \pm 0.47 $ & $ 6463.39 \pm 1.14 $ & $ -51.89 \pm 1.14 $ & $ 61.81 \pm 0.01 $ & $ 0.1959 \pm 0.0003 $ & $ 0.2098 \pm 0.0000 $ & $ 0.2550 \pm 0.0001 $ & $ 0.1580 \pm 0.0003 $ \\
35.752610 & $ -12580.07 \pm 0.54 $ & $ 6477.28 \pm 1.17 $ & $ -55.90 \pm 1.17 $ & $ 61.75 \pm 0.01 $ & $ 0.2080 \pm 0.0003 $ & $ 0.2086 \pm 0.0000 $ & $ 0.2528 \pm 0.0001 $ & $ 0.1696 \pm 0.0003 $ \\
36.615973 & $ -12579.83 \pm 0.45 $ & $ 6483.69 \pm 1.06 $ & $ -53.93 \pm 1.06 $ & $ 61.72 \pm 0.01 $ & $ 0.2387 \pm 0.0003 $ & $ 0.2093 \pm 0.0000 $ & $ 0.2545 \pm 0.0001 $ & $ 0.1990 \pm 0.0002 $ \\
37.682623 & $ -12581.80 \pm 0.68 $ & $ 6497.13 \pm 1.88 $ & $ -43.70 \pm 1.88 $ & $ 61.58 \pm 0.02 $ & $ 0.1796 \pm 0.0007 $ & $ 0.2070 \pm 0.0001 $ & $ 0.2587 \pm 0.0001 $ & $ 0.1424 \pm 0.0006 $ \\
38.658689 & $ -12578.49 \pm 0.63 $ & $ 6494.61 \pm 1.71 $ & $ -41.36 \pm 1.71 $ & $ 61.61 \pm 0.02 $ & $ 0.1611 \pm 0.0006 $ & $ 0.2094 \pm 0.0001 $ & $ 0.2576 \pm 0.0001 $ & $ 0.1247 \pm 0.0005 $ \\
40.732162 & $ -12585.28 \pm 0.64 $ & $ 6492.51 \pm 1.64 $ & $ -45.21 \pm 1.64 $ & $ 61.59 \pm 0.02 $ & $ 0.1618 \pm 0.0005 $ & $ 0.2092 \pm 0.0001 $ & $ 0.2599 \pm 0.0001 $ & $ 0.1253 \pm 0.0005 $ \\
41.602512 & $ -12588.11 \pm 0.66 $ & $ 6485.30 \pm 1.86 $ & $ -33.32 \pm 1.86 $ & $ 61.58 \pm 0.02 $ & $ 0.1573 \pm 0.0006 $ & $ 0.2085 \pm 0.0001 $ & $ 0.2589 \pm 0.0001 $ & $ 0.1210 \pm 0.0006 $ \\
43.608362 & $ -12592.46 \pm 0.47 $ & $ 6476.76 \pm 1.15 $ & $ -35.50 \pm 1.15 $ & $ 61.81 \pm 0.01 $ & $ 0.2070 \pm 0.0003 $ & $ 0.2109 \pm 0.0000 $ & $ 0.2527 \pm 0.0001 $ & $ 0.1686 \pm 0.0003 $ \\
44.682224 & $ -12594.41 \pm 0.52 $ & $ 6466.74 \pm 1.29 $ & $ -41.64 \pm 1.29 $ & $ 61.83 \pm 0.01 $ & $ 0.2024 \pm 0.0004 $ & $ 0.2101 \pm 0.0001 $ & $ 0.2551 \pm 0.0001 $ & $ 0.1641 \pm 0.0003 $ \\
50.742186 & $ -12596.62 \pm 0.49 $ & $ 6437.46 \pm 1.03 $ & $ -46.07 \pm 1.03 $ & $ 61.99 \pm 0.01 $ & $ 0.2043 \pm 0.0003 $ & $ 0.2105 \pm 0.0000 $ & $ 0.2472 \pm 0.0001 $ & $ 0.1660 \pm 0.0003 $ \\
55.610865 & $ -12583.82 \pm 0.59 $ & $ 6460.29 \pm 1.56 $ & $ -51.47 \pm 1.56 $ & $ 61.72 \pm 0.01 $ & $ 0.2002 \pm 0.0005 $ & $ 0.2102 \pm 0.0001 $ & $ 0.2528 \pm 0.0001 $ & $ 0.1621 \pm 0.0005 $ \\
61.551690 & $ -12599.47 \pm 0.53 $ & $ 6455.38 \pm 1.29 $ & $ -37.74 \pm 1.29 $ & $ 61.78 \pm 0.01 $ & $ 0.2304 \pm 0.0004 $ & $ 0.2099 \pm 0.0001 $ & $ 0.2537 \pm 0.0001 $ & $ 0.1910 \pm 0.0003 $ \\
62.631877 & $ -12603.01 \pm 0.56 $ & $ 6458.06 \pm 1.47 $ & $ -41.67 \pm 1.47 $ & $ 61.94 \pm 0.01 $ & $ 0.2296 \pm 0.0005 $ & $ 0.2063 \pm 0.0001 $ & $ 0.2542 \pm 0.0001 $ & $ 0.1902 \pm 0.0004 $ \\
68.670319 & $ -12584.62 \pm 0.60 $ & $ 6472.84 \pm 1.31 $ & $ -42.15 \pm 1.31 $ & $ 61.73 \pm 0.01 $ & $ 0.2088 \pm 0.0004 $ & $ 0.2117 \pm 0.0001 $ & $ 0.2544 \pm 0.0001 $ & $ 0.1703 \pm 0.0004 $ \\
73.610850 & $ -12592.89 \pm 0.46 $ & $ 6454.03 \pm 1.05 $ & $ -39.85 \pm 1.05 $ & $ 61.94 \pm 0.01 $ & $ 0.2467 \pm 0.0003 $ & $ 0.2089 \pm 0.0000 $ & $ 0.2493 \pm 0.0001 $ & $ 0.2066 \pm 0.0002 $ \\
109.529639 & $ -12580.24 \pm 0.79 $ & $ 6454.65 \pm 2.42 $ & $ -57.33 \pm 2.42 $ & $ 61.94 \pm 0.02 $ & $ 0.1911 \pm 0.0010 $ & $ 0.2124 \pm 0.0001 $ & $ 0.2503 \pm 0.0002 $ & $ 0.1534 \pm 0.0009 $ \\
124.503722 & $ -12592.68 \pm 0.52 $ & $ 6433.89 \pm 1.26 $ & $ -41.49 \pm 1.26 $ & $ 62.06 \pm 0.01 $ & $ 0.2447 \pm 0.0004 $ & $ 0.2129 \pm 0.0001 $ & $ 0.2467 \pm 0.0001 $ & $ 0.2047 \pm 0.0003 $ \\
138.475054 & $ -12595.52 \pm 0.53 $ & $ 6418.67 \pm 1.18 $ & $ -48.52 \pm 1.18 $ & $ 62.16 \pm 0.01 $ & $ 0.2166 \pm 0.0004 $ & $ 0.2067 \pm 0.0000 $ & $ 0.2437 \pm 0.0001 $ & $ 0.1778 \pm 0.0003 $ \\
139.497928 & $ -12596.07 \pm 0.77 $ & $ 6412.98 \pm 1.88 $ & $ -53.85 \pm 1.88 $ & $ 62.26 \pm 0.02 $ & $ 0.1738 \pm 0.0007 $ & $ 0.2115 \pm 0.0001 $ & $ 0.2455 \pm 0.0001 $ & $ 0.1368 \pm 0.0007 $ \\
145.481360 & $ -12597.04 \pm 1.11 $ & $ 6448.70 \pm 3.50 $ & $ -46.39 \pm 3.50 $ & $ 61.90 \pm 0.03 $ & $ 0.1232 \pm 0.0018 $ & $ 0.2142 \pm 0.0002 $ & $ 0.2543 \pm 0.0002 $ & $ 0.0883 \pm 0.0017 $ \\
231.910463 & $ -12597.23 \pm 0.48 $ & $ 6417.74 \pm 1.09 $ & $ -51.98 \pm 1.09 $ & $ 62.15 \pm 0.01 $ & $ 0.1685 \pm 0.0003 $ & $ 0.2115 \pm 0.0000 $ & $ 0.2393 \pm 0.0001 $ & $ 0.1317 \pm 0.0003 $ \\
237.911701 & $ -12600.34 \pm 0.48 $ & $ 6412.65 \pm 1.09 $ & $ -51.57 \pm 1.09 $ & $ 62.22 \pm 0.01 $ & $ 0.1896 \pm 0.0003 $ & $ 0.2131 \pm 0.0000 $ & $ 0.2388 \pm 0.0001 $ & $ 0.1519 \pm 0.0003 $ \\
241.870284 & $ -12594.54 \pm 0.75 $ & $ 6418.93 \pm 1.93 $ & $ -49.93 \pm 1.93 $ & $ 62.36 \pm 0.02 $ & $ 0.1181 \pm 0.0008 $ & $ 0.2152 \pm 0.0001 $ & $ 0.2441 \pm 0.0001 $ & $ 0.0834 \pm 0.0008 $ \\
288.741464 & $ -12589.70 \pm 0.84 $ & $ 6421.35 \pm 2.28 $ & $ -54.15 \pm 2.28 $ & $ 62.45 \pm 0.02 $ & $ 0.0465 \pm 0.0011 $ & $ 0.2073 \pm 0.0001 $ & $ 0.2353 \pm 0.0001 $ & $ 0.0149 \pm 0.0010 $ \\
303.682177 & $ -12599.28 \pm 0.58 $ & $ 6434.36 \pm 1.36 $ & $ -48.63 \pm 1.36 $ & $ 62.26 \pm 0.01 $ & $ 0.1635 \pm 0.0005 $ & $ 0.2082 \pm 0.0001 $ & $ 0.2399 \pm 0.0001 $ & $ 0.1269 \pm 0.0005 $ \\
306.818370 & $ -12596.93 \pm 0.60 $ & $ 6415.86 \pm 1.47 $ & $ -56.50 \pm 1.47 $ & $ 62.44 \pm 0.01 $ & $ 0.1656 \pm 0.0005 $ & $ 0.2070 \pm 0.0001 $ & $ 0.2386 \pm 0.0001 $ & $ 0.1289 \pm 0.0005 $ \\
313.722435 & $ -12603.66 \pm 0.38 $ & $ 6411.43 \pm 0.85 $ & $ -53.79 \pm 0.85 $ & $ 62.38 \pm 0.01 $ & $ 0.2012 \pm 0.0002 $ & $ 0.2087 \pm 0.0000 $ & $ 0.2393 \pm 0.0000 $ & $ 0.1631 \pm 0.0002 $ \\
316.658195 & $ -12599.07 \pm 0.48 $ & $ 6402.76 \pm 1.06 $ & $ -59.62 \pm 1.06 $ & $ 62.59 \pm 0.01 $ & $ 0.1883 \pm 0.0003 $ & $ 0.2089 \pm 0.0000 $ & $ 0.2386 \pm 0.0001 $ & $ 0.1507 \pm 0.0003 $ \\
319.672042 & $ -12592.57 \pm 0.55 $ & $ 6420.24 \pm 1.28 $ & $ -60.06 \pm 1.28 $ & $ 62.50 \pm 0.01 $ & $ 0.2308 \pm 0.0004 $ & $ 0.2068 \pm 0.0001 $ & $ 0.2346 \pm 0.0001 $ & $ 0.1914 \pm 0.0004 $ \\
322.837238 & $ -12597.47 \pm 0.54 $ & $ 6422.53 \pm 1.33 $ & $ -49.65 \pm 1.33 $ & $ 62.36 \pm 0.01 $ & $ 0.1656 \pm 0.0004 $ & $ 0.2087 \pm 0.0001 $ & $ 0.2385 \pm 0.0001 $ & $ 0.1289 \pm 0.0004 $ \\
326.824972 & $ -12592.00 \pm 0.39 $ & $ 6430.18 \pm 0.90 $ & $ -60.25 \pm 0.90 $ & $ 62.28 \pm 0.01 $ & $ 0.2082 \pm 0.0002 $ & $ 0.2093 \pm 0.0000 $ & $ 0.2410 \pm 0.0001 $ & $ 0.1697 \pm 0.0002 $ \\
333.599419 & $ -12603.08 \pm 0.69 $ & $ 6437.86 \pm 1.64 $ & $ -46.03 \pm 1.64 $ & $ 62.46 \pm 0.02 $ & $ 0.2165 \pm 0.0006 $ & $ 0.2086 \pm 0.0001 $ & $ 0.2440 \pm 0.0001 $ & $ 0.1777 \pm 0.0006 $ \\
336.763841 & $ -12599.90 \pm 0.42 $ & $ 6424.67 \pm 0.96 $ & $ -48.44 \pm 0.96 $ & $ 62.23 \pm 0.01 $ & $ 0.2094 \pm 0.0003 $ & $ 0.2092 \pm 0.0000 $ & $ 0.2403 \pm 0.0001 $ & $ 0.1709 \pm 0.0002 $ \\
339.598466 & $ -12598.37 \pm 0.54 $ & $ 6419.22 \pm 1.23 $ & $ -50.80 \pm 1.23 $ & $ 62.28 \pm 0.01 $ & $ 0.1962 \pm 0.0004 $ & $ 0.2093 \pm 0.0001 $ & $ 0.2390 \pm 0.0001 $ & $ 0.1582 \pm 0.0004 $ \\
342.607427 & $ -12601.16 \pm 0.52 $ & $ 6406.60 \pm 1.18 $ & $ -54.11 \pm 1.18 $ & $ 62.37 \pm 0.01 $ & $ 0.1877 \pm 0.0004 $ & $ 0.2080 \pm 0.0000 $ & $ 0.2382 \pm 0.0001 $ & $ 0.1501 \pm 0.0004 $ \\
346.713959 & $ -12600.32 \pm 0.46 $ & $ 6409.40 \pm 1.09 $ & $ -59.81 \pm 1.09 $ & $ 62.32 \pm 0.01 $ & $ 0.2171 \pm 0.0003 $ & $ 0.2066 \pm 0.0000 $ & $ 0.2348 \pm 0.0001 $ & $ 0.1783 \pm 0.0003 $ \\
349.580427 & $ -12597.08 \pm 0.46 $ & $ 6414.25 \pm 1.04 $ & $ -58.32 \pm 1.04 $ & $ 62.26 \pm 0.01 $ & $ 0.2242 \pm 0.0003 $ & $ 0.2091 \pm 0.0000 $ & $ 0.2374 \pm 0.0001 $ & $ 0.1851 \pm 0.0003 $ \\
352.680248 & $ -12599.73 \pm 0.43 $ & $ 6416.40 \pm 0.99 $ & $ -51.51 \pm 0.99 $ & $ 62.26 \pm 0.01 $ & $ 0.1927 \pm 0.0003 $ & $ 0.2094 \pm 0.0000 $ & $ 0.2402 \pm 0.0001 $ & $ 0.1548 \pm 0.0003 $ \\
356.625087 & $ -12591.83 \pm 0.43 $ & $ 6423.09 \pm 0.99 $ & $ -56.50 \pm 0.99 $ & $ 62.18 \pm 0.01 $ & $ 0.2103 \pm 0.0003 $ & $ 0.2096 \pm 0.0000 $ & $ 0.2405 \pm 0.0001 $ & $ 0.1718 \pm 0.0003 $ \\
362.658889 & $ -12601.83 \pm 0.50 $ & $ 6444.43 \pm 1.20 $ & $ -41.89 \pm 1.20 $ & $ 62.21 \pm 0.01 $ & $ 0.2139 \pm 0.0004 $ & $ 0.2057 \pm 0.0001 $ & $ 0.2452 \pm 0.0001 $ & $ 0.1752 \pm 0.0004 $ \\
365.626002 & $ -12603.27 \pm 0.65 $ & $ 6431.60 \pm 1.69 $ & $ -45.10 \pm 1.69 $ & $ 62.19 \pm 0.02 $ & $ 0.1722 \pm 0.0007 $ & $ 0.2049 \pm 0.0001 $ & $ 0.2415 \pm 0.0001 $ & $ 0.1353 \pm 0.0006 $ \\
368.575436 & $ -12599.98 \pm 0.46 $ & $ 6416.99 \pm 1.08 $ & $ -51.35 \pm 1.08 $ & $ 62.18 \pm 0.01 $ & $ 0.1958 \pm 0.0003 $ & $ 0.2045 \pm 0.0000 $ & $ 0.2358 \pm 0.0001 $ & $ 0.1578 \pm 0.0003 $ \\
378.659708 & $ -12594.60 \pm 0.48 $ & $ 6417.98 \pm 1.13 $ & $ -52.41 \pm 1.13 $ & $ 62.42 \pm 0.01 $ & $ 0.1906 \pm 0.0004 $ & $ 0.2041 \pm 0.0000 $ & $ 0.2381 \pm 0.0001 $ & $ 0.1529 \pm 0.0003 $ \\
381.559471 & $ -12602.34 \pm 0.59 $ & $ 6414.51 \pm 1.49 $ & $ -47.67 \pm 1.49 $ & $ 62.57 \pm 0.01 $ & $ 0.2400 \pm 0.0006 $ & $ 0.2038 \pm 0.0001 $ & $ 0.2350 \pm 0.0001 $ & $ 0.2002 \pm 0.0005 $ \\
384.567201 & $ -12601.95 \pm 0.65 $ & $ 6422.37 \pm 1.64 $ & $ -63.39 \pm 1.64 $ & $ 62.75 \pm 0.02 $ & $ 0.2386 \pm 0.0006 $ & $ 0.2062 \pm 0.0001 $ & $ 0.2367 \pm 0.0001 $ & $ 0.1989 \pm 0.0006 $ \\
408.739091 & $ -12599.11 \pm 0.47 $ & $ 6405.00 \pm 1.08 $ & $ -53.98 \pm 1.08 $ & $ 62.36 \pm 0.01 $ & $ 0.2119 \pm 0.0003 $ & $ 0.2071 \pm 0.0000 $ & $ 0.2336 \pm 0.0001 $ & $ 0.1733 \pm 0.0003 $ \\
412.729699 & $ -12601.89 \pm 0.55 $ & $ 6402.53 \pm 1.32 $ & $ -55.04 \pm 1.32 $ & $ 62.43 \pm 0.01 $ & $ 0.2258 \pm 0.0005 $ & $ 0.2076 \pm 0.0001 $ & $ 0.2329 \pm 0.0001 $ & $ 0.1866 \pm 0.0004 $ \\
427.712710 & $ -12607.18 \pm 0.57 $ & $ 6422.18 \pm 1.33 $ & $ -50.66 \pm 1.33 $ & $ 62.37 \pm 0.01 $ & $ 0.2207 \pm 0.0005 $ & $ 0.2085 \pm 0.0001 $ & $ 0.2391 \pm 0.0001 $ & $ 0.1817 \pm 0.0005 $ \\
430.591419 & $ -12598.25 \pm 0.40 $ & $ 6428.84 \pm 0.92 $ & $ -58.50 \pm 0.92 $ & $ 62.26 \pm 0.01 $ & $ 0.2231 \pm 0.0002 $ & $ 0.2076 \pm 0.0000 $ & $ 0.2404 \pm 0.0000 $ & $ 0.1840 \pm 0.0002 $ \\
710.725500 & $ -12586.08 \pm 0.50 $ & $ 6409.96 \pm 1.20 $ & $ -56.41 \pm 1.20 $ & $ 62.14 \pm 0.01 $ & $ 0.2291 \pm 0.0003 $ & $ 0.2031 \pm 0.0001 $ & $ 0.2316 \pm 0.0001 $ & $ 0.1898 \pm 0.0003 $ \\
771.581705 & $ -12587.63 \pm 0.56 $ & $ 6403.10 \pm 1.37 $ & $ -57.09 \pm 1.37 $ & $ 62.43 \pm 0.01 $ & $ 0.1279 \pm 0.0004 $ & $ 0.2022 \pm 0.0001 $ & $ 0.2293 \pm 0.0001 $ & $ 0.0928 \pm 0.0004 $ \\
792.550057 & $ -12589.64 \pm 0.45 $ & $ 6399.52 \pm 1.06 $ & $ -54.52 \pm 1.06 $ & $ 62.51 \pm 0.01 $ & $ 0.1479 \pm 0.0003 $ & $ 0.2033 \pm 0.0000 $ & $ 0.2307 \pm 0.0001 $ & $ 0.1119 \pm 0.0003 $ \\
794.657766 & $ -12586.16 \pm 0.59 $ & $ 6391.19 \pm 1.52 $ & $ -57.23 \pm 1.52 $ & $ 62.72 \pm 0.01 $ & $ 0.0889 \pm 0.0005 $ & $ 0.2033 \pm 0.0001 $ & $ 0.2275 \pm 0.0001 $ & $ 0.0554 \pm 0.0005 $ \\
797.558869 & $ -12586.03 \pm 0.46 $ & $ 6395.10 \pm 1.12 $ & $ -57.89 \pm 1.12 $ & $ 62.32 \pm 0.01 $ & $ 0.1301 \pm 0.0003 $ & $ 0.2092 \pm 0.0000 $ & $ 0.2271 \pm 0.0001 $ & $ 0.0950 \pm 0.0003 $ \\
808.673735 & $ -12593.59 \pm 0.69 $ & $ 6413.18 \pm 1.75 $ & $ -68.14 \pm 1.75 $ & $ 62.46 \pm 0.02 $ & $ 0.0805 \pm 0.0006 $ & $ 0.2150 \pm 0.0001 $ & $ 0.2282 \pm 0.0001 $ & $ 0.0474 \pm 0.0006 $ \\
817.549677 & $ -12586.10 \pm 0.54 $ & $ 6411.83 \pm 1.34 $ & $ -56.67 \pm 1.34 $ & $ 62.51 \pm 0.01 $ & $ 0.1273 \pm 0.0004 $ & $ 0.2070 \pm 0.0001 $ & $ 0.2315 \pm 0.0001 $ & $ 0.0923 \pm 0.0004 $ \\
819.542902 & $ -12590.36 \pm 0.48 $ & $ 6402.17 \pm 1.14 $ & $ -52.16 \pm 1.14 $ & $ 62.49 \pm 0.01 $ & $ 0.1631 \pm 0.0003 $ & $ 0.2090 \pm 0.0000 $ & $ 0.2303 \pm 0.0001 $ & $ 0.1266 \pm 0.0003 $ \\
829.510798 & $ -12587.62 \pm 0.66 $ & $ 6389.46 \pm 1.69 $ & $ -62.21 \pm 1.69 $ & $ 62.64 \pm 0.02 $ & $ 0.1357 \pm 0.0006 $ & $ 0.2198 \pm 0.0001 $ & $ 0.2237 \pm 0.0001 $ & $ 0.1003 \pm 0.0005 $ \\
832.541183 & $ -12584.95 \pm 0.47 $ & $ 6396.90 \pm 1.14 $ & $ -59.53 \pm 1.14 $ & $ 62.42 \pm 0.01 $ & $ 0.1583 \pm 0.0003 $ & $ 0.2158 \pm 0.0000 $ & $ 0.2291 \pm 0.0001 $ & $ 0.1220 \pm 0.0003 $ \\
837.504657 & $ -12589.66 \pm 0.62 $ & $ 6386.88 \pm 1.61 $ & $ -53.96 \pm 1.61 $ & $ 62.71 \pm 0.02 $ & $ 0.0913 \pm 0.0005 $ & $ 0.2158 \pm 0.0001 $ & $ 0.2266 \pm 0.0001 $ & $ 0.0578 \pm 0.0005 $ \\
838.514477 & $ -12585.45 \pm 0.71 $ & $ 6388.96 \pm 1.92 $ & $ -60.95 \pm 1.92 $ & $ 62.76 \pm 0.02 $ & $ 0.0674 \pm 0.0007 $ & $ 0.2137 \pm 0.0001 $ & $ 0.2259 \pm 0.0001 $ & $ 0.0348 \pm 0.0007 $ \\
839.497951 & $ -12590.89 \pm 0.51 $ & $ 6395.99 \pm 1.27 $ & $ -62.23 \pm 1.27 $ & $ 62.56 \pm 0.01 $ & $ 0.1469 \pm 0.0004 $ & $ 0.2140 \pm 0.0001 $ & $ 0.2267 \pm 0.0001 $ & $ 0.1110 \pm 0.0003 $ \\
840.476390 & $ -12591.55 \pm 0.43 $ & $ 6393.36 \pm 1.01 $ & $ -64.49 \pm 1.01 $ & $ 62.31 \pm 0.01 $ & $ 0.1632 \pm 0.0002 $ & $ 0.2089 \pm 0.0000 $ & $ 0.2281 \pm 0.0001 $ & $ 0.1266 \pm 0.0002 $ \\
841.496534 & $ -12591.01 \pm 0.69 $ & $ 6405.08 \pm 1.85 $ & $ -60.02 \pm 1.85 $ & $ 62.68 \pm 0.02 $ & $ 0.0931 \pm 0.0007 $ & $ 0.2168 \pm 0.0001 $ & $ 0.2270 \pm 0.0001 $ & $ 0.0595 \pm 0.0006 $ \\
843.495535 & $ -12580.21 \pm 0.44 $ & $ 6406.74 \pm 1.05 $ & $ -54.77 \pm 1.05 $ & $ 62.49 \pm 0.01 $ & $ 0.1827 \pm 0.0003 $ & $ 0.2056 \pm 0.0000 $ & $ 0.2368 \pm 0.0001 $ & $ 0.1453 \pm 0.0003 $ \\
846.491896 & $ -12589.96 \pm 0.66 $ & $ 6407.26 \pm 1.78 $ & $ -50.02 \pm 1.78 $ & $ 62.48 \pm 0.02 $ & $ 0.0935 \pm 0.0006 $ & $ 0.2079 \pm 0.0001 $ & $ 0.2329 \pm 0.0001 $ & $ 0.0599 \pm 0.0006 $ \\
849.508439 & $ -12579.75 \pm 0.49 $ & $ 6404.14 \pm 1.19 $ & $ -49.59 \pm 1.19 $ & $ 62.44 \pm 0.01 $ & $ 0.1213 \pm 0.0004 $ & $ 0.2142 \pm 0.0001 $ & $ 0.2293 \pm 0.0001 $ & $ 0.0865 \pm 0.0003 $ \\
851.499306 & $ -12587.21 \pm 0.46 $ & $ 6405.27 \pm 1.05 $ & $ -51.60 \pm 1.05 $ & $ 62.49 \pm 0.01 $ & $ 0.1621 \pm 0.0003 $ & $ 0.2052 \pm 0.0000 $ & $ 0.2298 \pm 0.0001 $ & $ 0.1256 \pm 0.0003 $ \\
873.483521 & $ -12584.69 \pm 0.51 $ & $ 6402.86 \pm 1.16 $ & $ -61.88 \pm 1.16 $ & $ 62.60 \pm 0.01 $ & $ 0.2007 \pm 0.0003 $ & $ 0.2058 \pm 0.0000 $ & $ 0.2324 \pm 0.0001 $ & $ 0.1626 \pm 0.0003 $ \\
875.483186 & $ -12582.40 \pm 0.54 $ & $ 6407.63 \pm 1.30 $ & $ -55.53 \pm 1.30 $ & $ 62.49 \pm 0.01 $ & $ 0.2362 \pm 0.0004 $ & $ 0.2061 \pm 0.0001 $ & $ 0.2328 \pm 0.0001 $ & $ 0.1966 \pm 0.0004 $ \\
1096.842355 & $ -12576.47 \pm 0.71 $ & $ 6413.12 \pm 1.86 $ & $ -60.70 \pm 1.86 $ & $ 62.45 \pm 0.02 $ & $ 0.0825 \pm 0.0007 $ & $ 0.2017 \pm 0.0001 $ & $ 0.2286 \pm 0.0001 $ & $ 0.0493 \pm 0.0007 $ \\
1123.593646 & $ -12584.18 \pm 0.45 $ & $ 6400.37 \pm 1.07 $ & $ -58.68 \pm 1.07 $ & $ 62.46 \pm 0.01 $ & $ 0.2036 \pm 0.0003 $ & $ 0.2055 \pm 0.0000 $ & $ 0.2315 \pm 0.0001 $ & $ 0.1653 \pm 0.0003 $ \\
1123.810797 & $ -12585.20 \pm 0.63 $ & $ 6403.36 \pm 1.53 $ & $ -57.80 \pm 1.53 $ & $ 62.57 \pm 0.01 $ & $ 0.2060 \pm 0.0005 $ & $ 0.2043 \pm 0.0001 $ & $ 0.2314 \pm 0.0001 $ & $ 0.1677 \pm 0.0005 $ \\
1172.631470 & $ -12583.43 \pm 0.45 $ & $ 6392.38 \pm 1.08 $ & $ -58.50 \pm 1.08 $ & $ 62.47 \pm 0.01 $ & $ 0.1963 \pm 0.0003 $ & $ 0.2084 \pm 0.0000 $ & $ 0.2326 \pm 0.0001 $ & $ 0.1583 \pm 0.0003 $ \\
1174.528400 & $ -12584.35 \pm 0.50 $ & $ 6391.99 \pm 1.22 $ & $ -57.93 \pm 1.22 $ & $ 62.47 \pm 0.01 $ & $ 0.2002 \pm 0.0004 $ & $ 0.2087 \pm 0.0001 $ & $ 0.2311 \pm 0.0001 $ & $ 0.1621 \pm 0.0003 $ \\
1176.649341 & $ -12589.25 \pm 0.72 $ & $ 6401.35 \pm 1.78 $ & $ -54.42 \pm 1.78 $ & $ 62.59 \pm 0.02 $ & $ 0.1845 \pm 0.0007 $ & $ 0.2072 \pm 0.0001 $ & $ 0.2328 \pm 0.0001 $ & $ 0.1470 \pm 0.0007 $ \\
1178.583493 & $ -12586.14 \pm 0.63 $ & $ 6400.00 \pm 1.62 $ & $ -54.15 \pm 1.62 $ & $ 62.63 \pm 0.02 $ & $ 0.1851 \pm 0.0006 $ & $ 0.2098 \pm 0.0001 $ & $ 0.2336 \pm 0.0001 $ & $ 0.1476 \pm 0.0006 $ \\
1180.543249 & $ -12583.25 \pm 0.40 $ & $ 6399.41 \pm 0.93 $ & $ -56.33 \pm 0.93 $ & $ 62.44 \pm 0.01 $ & $ 0.2093 \pm 0.0002 $ & $ 0.2070 \pm 0.0000 $ & $ 0.2332 \pm 0.0001 $ & $ 0.1708 \pm 0.0002 $ \\
1182.552397 & $ -12583.31 \pm 0.48 $ & $ 6397.65 \pm 1.14 $ & $ -54.58 \pm 1.14 $ & $ 62.44 \pm 0.01 $ & $ 0.2046 \pm 0.0003 $ & $ 0.2091 \pm 0.0000 $ & $ 0.2326 \pm 0.0001 $ & $ 0.1663 \pm 0.0003 $ \\
1185.524801 & $ -12584.82 \pm 0.60 $ & $ 6392.69 \pm 1.51 $ & $ -55.85 \pm 1.51 $ & $ 62.33 \pm 0.01 $ & $ 0.2188 \pm 0.0005 $ & $ 0.2102 \pm 0.0001 $ & $ 0.2333 \pm 0.0001 $ & $ 0.1798 \pm 0.0005 $ \\
1187.513957 & $ -12585.26 \pm 0.47 $ & $ 6392.03 \pm 1.11 $ & $ -55.56 \pm 1.11 $ & $ 62.52 \pm 0.01 $ & $ 0.1831 \pm 0.0003 $ & $ 0.2112 \pm 0.0000 $ & $ 0.2345 \pm 0.0001 $ & $ 0.1456 \pm 0.0003 $ \\
1194.552724 & $ -12584.89 \pm 0.64 $ & $ 6394.72 \pm 1.63 $ & $ -59.42 \pm 1.63 $ & $ 62.76 \pm 0.02 $ & $ 0.1490 \pm 0.0006 $ & $ 0.2120 \pm 0.0001 $ & $ 0.2328 \pm 0.0001 $ & $ 0.1130 \pm 0.0006 $ \\
1197.521415 & $ -12582.83 \pm 0.58 $ & $ 6391.07 \pm 1.42 $ & $ -60.21 \pm 1.42 $ & $ 62.44 \pm 0.01 $ & $ 0.1296 \pm 0.0005 $ & $ 0.2135 \pm 0.0001 $ & $ 0.2309 \pm 0.0001 $ & $ 0.0945 \pm 0.0005 $ \\


\end{longtable}
\end{landscape}

\FloatBarrier 
\clearpage

\onecolumn
\section{Radial velocities and activity indicators}
\begin{figure*}[h!]
    \centering
    \includegraphics[width=\hsize]{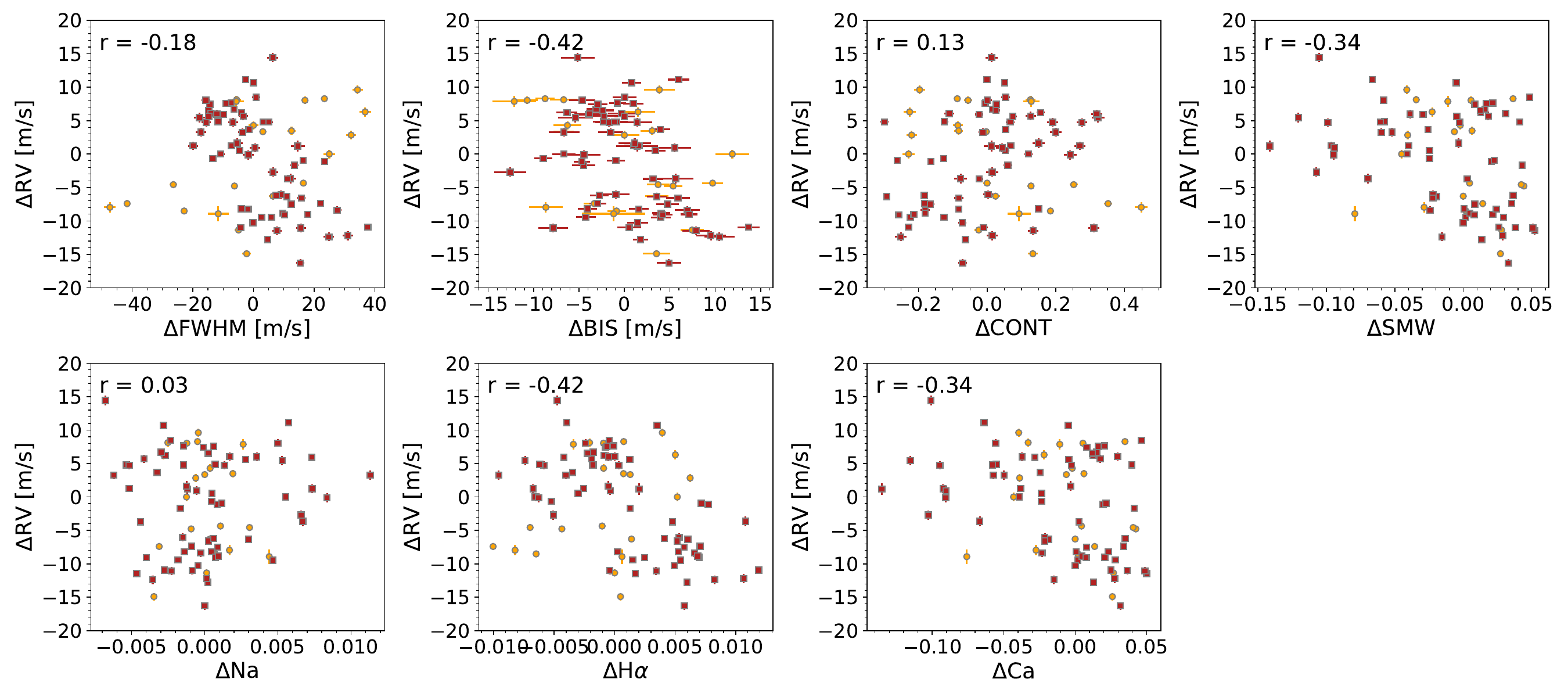}
    \caption{ESPRESSO RV measurements versus activity indices. In each panel, the Pearson product-moment correlation coefficient $r$ is shown. The RVs and activity indices were median-subtracted separately for the pre- (E18) and post- (E19) intervention datasets. E18 data are shown as orange circles, and E19 data as red squares.}
    \label{Fig:RV_vs_ActivityInd}
\end{figure*}

Figure~\ref{Fig:RV_vs_ActivityInd} shows the RV measurements versus the activity indices, after subtracting the median from each dataset (E18 and E19) separately, along with the Pearson product-moment correlation coefficient $r$ between these quantities.

\onecolumn

\section{Joint fit of TESS and ESPRESSO data -- results}
\label{Sec:Appendix_JointFit}
\begin{figure*}[h!]
   \centering
   \includegraphics[width=\hsize]{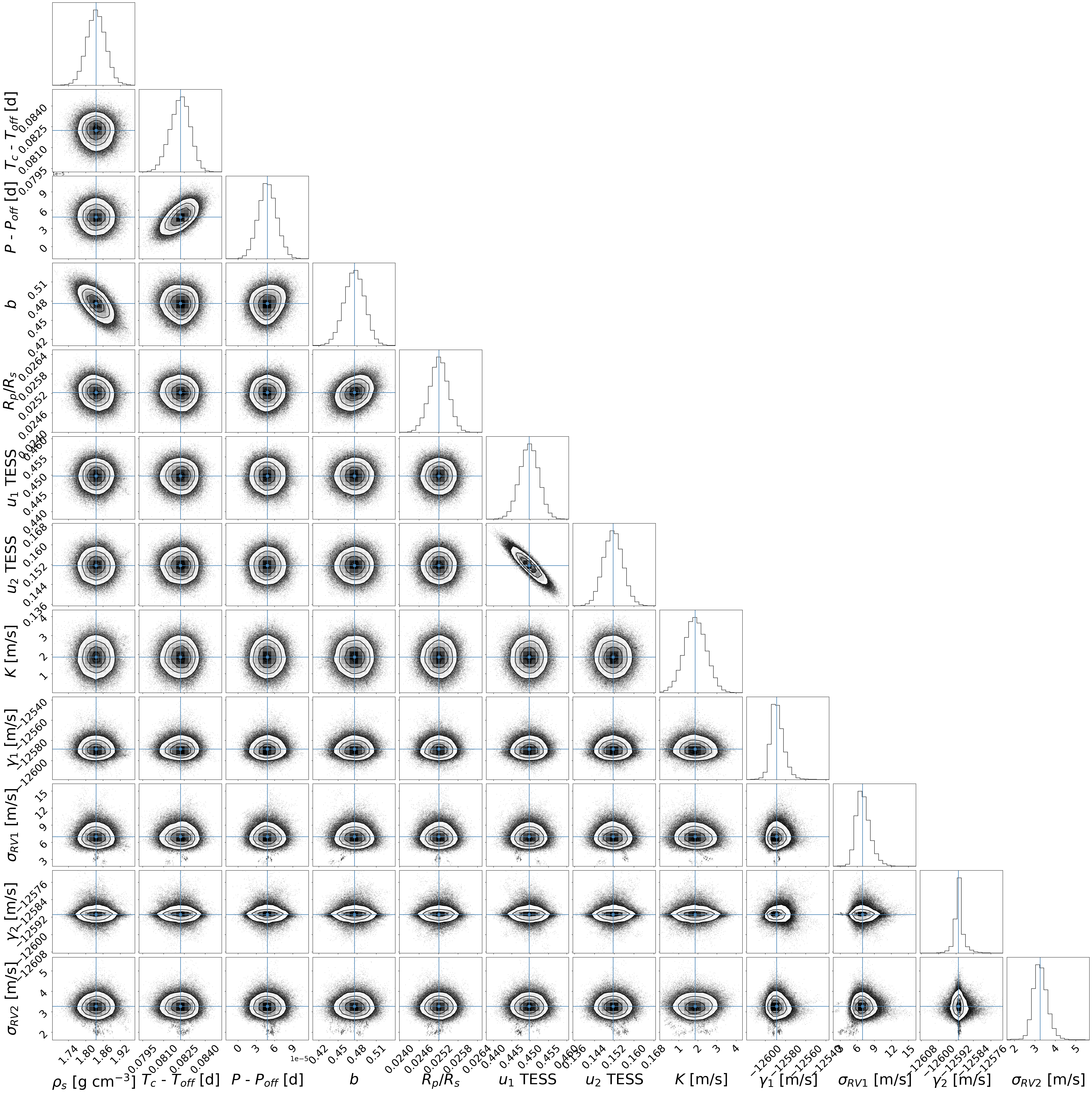}
   \caption{Corner plot of the posterior distributions of the fit transit and orbital parameters of TOI-283\,b. The parameters used to model the systematic effects using GPs have been intentionally left out for ease of viewing. The blue lines mark the median of the distribution. For plotting purposes, the distributions for the central time of the transit and the orbital period have been shifted by $T_{\mathrm{off}} = 2459549.0$ days and $P_{\mathrm{off}} = 17.6174$ days, respectively.}
    \label{Fig:MCMC_CornerPlot}
\end{figure*}

Table~\ref{Tab:GPs_coeffs_jointfit} shows the prior ranges and values of the GP hyperparameters for the circular and eccentric orbit fits. Figure~\ref{Fig:MCMC_CornerPlot} shows the posterior parameter distributions of the joint fit, with the GP parameters for the transit and RV fits intentionally omitted for clarity.

\begin{longtable}{l c c c}

\caption{The GP parameters, priors, and fit values for the joint fit.}\\
\label{Tab:GPs_coeffs_jointfit}

Parameter & Prior & Circular model & Eccentric model  \\
\hline
\endfirsthead
\caption{continued.}\\
\hline
Parameter & Prior & Circular model & Eccentric model \\
\hline
\endhead
\hline
\endfoot

\noalign{\smallskip}
\multicolumn{4}{c}{TESS parameters} \\
\noalign{\smallskip}

$\log c_{01}$ & $\mathcal{U}(-12.0,-1.0)$ & $-8.10^{+0.22}_{-0.18}$ & $-8.10^{+0.22}_{-0.18}$ \\
$\log \tau_{01}$ & $\mathcal{U}(-3.178,2.565)$ & $0.45^{+0.22}_{-0.20}$ & $0.45^{+0.23}_{-0.20}$ \\
$\log c_{02}$ & $\mathcal{U}(-12.0,-1.0)$ & $-8.03^{+0.25}_{-0.20}$ & $-8.03^{+0.25}_{-0.19}$ \\
$\log \tau_{02}$ & $\mathcal{U}(-3.178,2.565)$ & $0.58^{+0.30}_{-0.25}$ & $0.58^{+0.29}_{-0.25}$ \\
$\log c_{03}$ & $\mathcal{U}(-12.0,-1.0)$ & $-7.20^{+0.67}_{-0.48}$ & $-7.20^{+0.68}_{-0.48}$ \\
$\log \tau_{03}$ & $\mathcal{U}(-3.178,2.565)$ & $1.27^{+0.59}_{-0.48}$ & $1.28^{+0.58}_{-0.48}$ \\
$\log c_{04}$ & $\mathcal{U}(-12.0,-1.0)$ & $-8.33^{+0.13}_{-0.11}$ & $-8.34^{+0.12}_{-0.11}$ \\
$\log \tau_{04}$ & $\mathcal{U}(-3.178,2.565)$ & $-1.00^{+0.18}_{-0.17}$ & $-1.00^{+0.17}_{-0.17}$ \\
$\log c_{05}$ & $\mathcal{U}(-12.0,-1.0)$ & $-8.57^{+0.16}_{-0.14}$ & $-8.57^{+0.17}_{-0.14}$ \\
$\log \tau_{05}$ & $\mathcal{U}(-3.178,2.565)$ & $-0.32^{+0.22}_{-0.21}$ & $-0.33^{+0.24}_{-0.21}$ \\
$\log c_{06}$ & $\mathcal{U}(-12.0,-1.0)$ & $-8.50^{+0.26}_{-0.20}$ & $-8.50^{+0.26}_{-0.20}$ \\
$\log \tau_{06}$ & $\mathcal{U}(-3.178,2.565)$ & $0.32^{+0.32}_{-0.30}$ & $0.31^{+0.32}_{-0.30}$ \\
$\log c_{07}$ & $\mathcal{U}(-12.0,-1.0)$ & $-8.20^{+0.16}_{-0.14}$ & $-8.20^{+0.17}_{-0.14}$ \\
$\log \tau_{07}$ & $\mathcal{U}(-3.178,2.565)$ & $-0.40^{+0.35}_{-0.27}$ & $-0.40^{+0.35}_{-0.27}$ \\
$\log c_{08}$ & $\mathcal{U}(-12.0,-1.0)$ & $-8.10^{+0.23}_{-0.18}$ & $-8.10^{+0.24}_{-0.18}$ \\
$\log \tau_{08}$ & $\mathcal{U}(-3.178,2.565)$ & $-0.40^{+0.36}_{-0.31}$ & $-0.39^{+0.36}_{-0.31}$ \\
$\log c_{09}$ & $\mathcal{U}(-12.0,-1.0)$ & $-8.29^{+0.23}_{-0.19}$ & $-8.30^{+0.24}_{-0.19}$ \\
$\log \tau_{09}$ & $\mathcal{U}(-3.178,2.565)$ & $0.30^{+0.26}_{-0.24}$ & $0.29^{+0.26}_{-0.25}$ \\
$\log c_{10}$ & $\mathcal{U}(-12.0,-1.0)$ & $-8.08^{+0.17}_{-0.15}$ & $-8.07^{+0.18}_{-0.15}$ \\
$\log \tau_{10}$ & $\mathcal{U}(-3.178,2.565)$ & $-0.23^{+0.22}_{-0.21}$ & $-0.22^{+0.22}_{-0.21}$ \\
$\log c_{11}$ & $\mathcal{U}(-12.0,-1.0)$ & $-9.48^{+0.19}_{-0.18}$ & $-9.48^{+0.19}_{-0.19}$ \\
$\log \tau_{11}$ & $\mathcal{U}(-3.178,2.565)$ & $-0.74^{+0.34}_{-0.34}$ & $-0.75^{+0.33}_{-0.35}$ \\
$\log c_{12}$ & $\mathcal{U}(-12.0,-1.0)$ & $-9.12^{+0.23}_{-0.19}$ & $-9.13^{+0.22}_{-0.19}$ \\
$\log \tau_{12}$ & $\mathcal{U}(-3.178,2.565)$ & $0.05^{+0.32}_{-0.30}$ & $0.05^{+0.30}_{-0.29}$ \\
$\log c_{13}$ & $\mathcal{U}(-12.0,-1.0)$ & $-8.56^{+0.15}_{-0.14}$ & $-8.56^{+0.16}_{-0.14}$ \\
$\log \tau_{13}$ & $\mathcal{U}(-3.178,2.565)$ & $-0.31^{+0.27}_{-0.25}$ & $-0.31^{+0.28}_{-0.25}$ \\
$\log c_{27}$ & $\mathcal{U}(-12.0,-1.0)$ & $-8.95^{+0.23}_{-0.19}$ & $-8.94^{+0.22}_{-0.19}$ \\
$\log \tau_{27}$ & $\mathcal{U}(-3.178,2.565)$ & $0.10^{+0.34}_{-0.33}$ & $0.09^{+0.34}_{-0.33}$ \\
$\log c_{28}$ & $\mathcal{U}(-12.0,-1.0)$ & $-9.87^{+0.21}_{-0.26}$ & $-9.88^{+0.21}_{-0.27}$ \\
$\log \tau_{28}$ & $\mathcal{U}(-3.178,2.565)$ & $-1.42^{+0.48}_{-0.55}$ & $-1.42^{+0.49}_{-0.57}$ \\
$\log c_{29}$ & $\mathcal{U}(-12.0,-1.0)$ & $-8.73^{+0.17}_{-0.14}$ & $-8.73^{+0.17}_{-0.14}$ \\
$\log \tau_{29}$ & $\mathcal{U}(-3.178,2.565)$ & $-0.71^{+0.25}_{-0.23}$ & $-0.72^{+0.24}_{-0.22}$ \\
$\log c_{30}$ & $\mathcal{U}(-12.0,-1.0)$ & $-8.71^{+0.16}_{-0.13}$ & $-8.71^{+0.15}_{-0.13}$ \\
$\log \tau_{30}$ & $\mathcal{U}(-3.178,2.565)$ & $-0.63^{+0.24}_{-0.23}$ & $-0.64^{+0.24}_{-0.23}$ \\
$\log c_{31}$ & $\mathcal{U}(-12.0,-1.0)$ & $-9.01^{+0.18}_{-0.15}$ & $-9.01^{+0.18}_{-0.15}$ \\
$\log \tau_{31}$ & $\mathcal{U}(-3.178,2.565)$ & $-0.23^{+0.33}_{-0.32}$ & $-0.24^{+0.33}_{-0.31}$ \\
$\log c_{32}$ & $\mathcal{U}(-12.0,-1.0)$ & $-9.41^{+0.19}_{-0.17}$ & $-9.41^{+0.19}_{-0.17}$ \\
$\log \tau_{32}$ & $\mathcal{U}(-3.178,2.565)$ & $-0.22^{+0.43}_{-0.45}$ & $-0.20^{+0.41}_{-0.45}$ \\
$\log c_{33}$ & $\mathcal{U}(-12.0,-1.0)$ & $-9.05^{+0.24}_{-0.18}$ & $-9.06^{+0.22}_{-0.18}$ \\
$\log \tau_{33}$ & $\mathcal{U}(-3.178,2.565)$ & $0.05^{+0.39}_{-0.37}$ & $0.03^{+0.38}_{-0.36}$ \\
$\log c_{34}$ & $\mathcal{U}(-12.0,-1.0)$ & $-9.06^{+0.29}_{-0.22}$ & $-9.05^{+0.30}_{-0.23}$ \\
$\log \tau_{34}$ & $\mathcal{U}(-3.178,2.565)$ & $0.52^{+0.33}_{-0.30}$ & $0.52^{+0.34}_{-0.30}$ \\
$\log c_{35}$ & $\mathcal{U}(-12.0,-1.0)$ & $-8.42^{+0.18}_{-0.15}$ & $-8.42^{+0.18}_{-0.14}$ \\
$\log \tau_{35}$ & $\mathcal{U}(-3.178,2.565)$ & $-0.49^{+0.24}_{-0.23}$ & $-0.51^{+0.24}_{-0.23}$ \\
$\log c_{36}$ & $\mathcal{U}(-12.0,-1.0)$ & $-9.39^{+0.20}_{-0.17}$ & $-9.39^{+0.20}_{-0.18}$ \\
$\log \tau_{36}$ & $\mathcal{U}(-3.178,2.565)$ & $-0.21^{+0.38}_{-0.39}$ & $-0.21^{+0.37}_{-0.38}$ \\
$\log c_{37}$ & $\mathcal{U}(-12.0,-1.0)$ & $-8.95^{+0.18}_{-0.15}$ & $-8.95^{+0.18}_{-0.15}$ \\
$\log \tau_{37}$ & $\mathcal{U}(-3.178,2.565)$ & $-0.39^{+0.29}_{-0.27}$ & $-0.39^{+0.28}_{-0.27}$ \\
$\log c_{38}$ & $\mathcal{U}(-12.0,-1.0)$ & $-9.14^{+0.16}_{-0.14}$ & $-9.14^{+0.16}_{-0.14}$ \\
$\log \tau_{38}$ & $\mathcal{U}(-3.178,2.565)$ & $-0.47^{+0.47}_{-0.40}$ & $-0.46^{+0.48}_{-0.40}$ \\
$\log c_{39}$ & $\mathcal{U}(-12.0,-1.0)$ & $-9.06^{+0.18}_{-0.15}$ & $-9.05^{+0.18}_{-0.15}$ \\
$\log \tau_{39}$ & $\mathcal{U}(-3.178,2.565)$ & $-0.07^{+0.31}_{-0.29}$ & $-0.06^{+0.31}_{-0.30}$ \\
$\log c_{61}$ & $\mathcal{U}(-12.0,-1.0)$ & $-9.17^{+0.26}_{-0.20}$ & $-9.18^{+0.24}_{-0.20}$ \\
$\log \tau_{61}$ & $\mathcal{U}(-3.178,2.565)$ & $0.26^{+0.30}_{-0.29}$ & $0.24^{+0.30}_{-0.28}$ \\
$\log c_{62}$ & $\mathcal{U}(-12.0,-1.0)$ & $-9.03^{+0.12}_{-0.13}$ & $-9.04^{+0.13}_{-0.13}$ \\
$\log \tau_{62}$ & $\mathcal{U}(-3.178,2.565)$ & $-1.31^{+0.58}_{-0.67}$ & $-1.24^{+0.53}_{-0.69}$ \\
$\log c_{64}$ & $\mathcal{U}(-12.0,-1.0)$ & $-9.38^{+0.21}_{-0.18}$ & $-9.38^{+0.21}_{-0.17}$ \\
$\log \tau_{64}$ & $\mathcal{U}(-3.178,2.565)$ & $0.07^{+0.34}_{-0.35}$ & $0.06^{+0.34}_{-0.35}$ \\
$\log c_{65}$ & $\mathcal{U}(-12.0,-1.0)$ & $-9.21^{+0.15}_{-0.15}$ & $-9.22^{+0.15}_{-0.15}$ \\
$\log \tau_{65}$ & $\mathcal{U}(-3.178,2.565)$ & $-0.62^{+0.45}_{-0.38}$ & $-0.64^{+0.45}_{-0.37}$ \\
$\log c_{68}$ & $\mathcal{U}(-12.0,-1.0)$ & $-8.57^{+0.18}_{-0.14}$ & $-8.57^{+0.18}_{-0.14}$ \\
$\log \tau_{68}$ & $\mathcal{U}(-3.178,2.565)$ & $-0.64^{+0.32}_{-0.31}$ & $-0.65^{+0.32}_{-0.32}$ \\
$\log c_{69}$ & $\mathcal{U}(-12.0,-1.0)$ & $-7.89^{+0.14}_{-0.12}$ & $-7.90^{+0.14}_{-0.12}$ \\
$\log \tau_{69}$ & $\mathcal{U}(-3.178,2.565)$ & $-0.69^{+0.17}_{-0.16}$ & $-0.70^{+0.18}_{-0.16}$ \\
$\log c_{87}$ & $\mathcal{U}(-12.0,-1.0)$ & $-8.50^{+0.19}_{-0.15}$ & $-8.49^{+0.19}_{-0.16}$ \\
$\log \tau_{87}$ & $\mathcal{U}(-3.178,2.565)$ & $-0.28^{+0.23}_{-0.22}$ & $-0.28^{+0.23}_{-0.22}$ \\
$\log c_{88}$ & $\mathcal{U}(-12.0,-1.0)$ & $-8.40^{+0.19}_{-0.16}$ & $-8.40^{+0.20}_{-0.16}$ \\
$\log \tau_{88}$ & $\mathcal{U}(-3.178,2.565)$ & $-0.10^{+0.25}_{-0.22}$ & $-0.09^{+0.25}_{-0.22}$ \\
$\log c_{89}$ & $\mathcal{U}(-12.0,-1.0)$ & $-8.84^{+0.18}_{-0.15}$ & $-8.83^{+0.18}_{-0.15}$ \\
$\log \tau_{89}$ & $\mathcal{U}(-3.178,2.565)$ & $-0.11^{+0.24}_{-0.23}$ & $-0.10^{+0.24}_{-0.23}$ \\
$\log c_{90}$ & $\mathcal{U}(-12.0,-1.0)$ & $-9.16^{+0.15}_{-0.13}$ & $-9.16^{+0.14}_{-0.13}$ \\
$\log \tau_{90}$ & $\mathcal{U}(-3.178,2.565)$ & $-0.53^{+0.26}_{-0.24}$ & $-0.54^{+0.26}_{-0.24}$ \\

\multicolumn{4}{c}{RV parameters} \\
\noalign{\smallskip}

$\eta_1$ & $\mathcal{U}(0, 50)$ & $2.61^{+2.93}_{-1.71}$ & $2.33^{+2.54}_{-1.54}$ \\
$\eta_2$ & $\mathcal{U}(0.01, 200.0)$ & $125.12^{+51.56}_{-70.22}$ & $116.81^{+56.29}_{-69.05}$ \\
$\eta_3$ & $\mathcal{N}(30, 6)$ & $29.90^{+6.11}_{-5.88}$ & $30.06^{+5.95}_{-6.03}$ \\
$\eta_4$ & $\mathcal{U}(0.01, 200.0)$ & $96.53^{+69.68}_{-67.59}$ & $96.86^{+70.74}_{-68.45}$ \\

\end{longtable}

\FloatBarrier
\twocolumn

\end{appendix}

\end{document}